\documentclass[]{JHEP3}

\pdfoutput=1

\usepackage{amsmath}
\usepackage{epsfig,multicol,bbm}
\usepackage{graphicx}
\usepackage{bm}
\usepackage{multirow}

\renewcommand{\d}[0]{\textrm{ d}}

\newcommand{\be}[0]{\begin{equation}}
\newcommand{\ee}[0]{\end{equation}}
\newcommand{\sfrac}[2]{\textstyle{\frac{#1}{#2}}}

\newcommand{\kt}[0]{\text{k}_{\text{T}}}

\def\Dslash{D\!\!\!\!\slash}

\def\nslash{n\!\!\!\slash}
\def\bnslash{\bar n\!\!\!\slash}

\newcommand{\nn}{\nonumber}
\newcommand{\bn}{{\bar n}}
\newcommand{\mcdot}{\!\cdot\!}

\newcommand{\vect}[1]{\mathbf{#1}}
\newcommand{\abs}[1]{\left\lvert #1\right\rvert}
\newcommand{\bra}[1]{\left\langle #1\right\rvert}
\newcommand{\ket}[1]{\left\lvert #1\right\rangle}

\newcommand{\GeV}{\text{ GeV}}

\newcommand{\eq}[1]{Eq.~\eqref{#1}}
\newcommand{\eqs}[2]{Eqs.~\eqref{#1} and \eqref{#2}}
\newcommand{\eqss}[3]{Eqs.~\eqref{#1}, \eqref{#2}, and \eqref{#3}}
\renewcommand{\sec}[1]{Sec.~\ref{#1}}
\newcommand{\ssec}[1]{Sec.~\ref{ssec:#1}}
\newcommand{\fig}[1]{Fig.~\ref{#1}}
\newcommand{\tab}[1]{Table~\ref{tab:#1}}
\DeclareMathOperator{\Tr}{Tr}
\DeclareMathOperator{\Real}{Re}

\DeclareMathOperator{\sgn}{sgn}
\DeclareMathOperator{\Li}{Li}
\newcommand{\CF}{C_F}
\newcommand{\TR}{T_R}
\newcommand{\CA}{C_A}
\newcommand{\NF}{N_f}

\newcommand{\as}{\alpha_s}

\newcommand{\cO}{\mathcal{O}}

\newcommand{\cusp}{\! \rm cusp}

\newcommand{\xcone}{x_{\rm cone}}

\newcommand{\meas}{{\rm meas}}
\newcommand{\unmeas}{ {\rm unmeas}}

\newcommand{\incl}{ {\rm incl}}

\newcommand{\ijpair}{{i \neq j}}
\newcommand{\qjetnaive}{\tilde{J}}
\newcommand{\gjetnaive}{\tilde{J}}
\newcommand{\tabrule}{\rule{-5pt}{3.5ex} \rule[-2.5ex]{0pt}{0pt}}
\renewcommand{\dotplus}[0]{+}

\title{Jet Shapes and Jet Algorithms in SCET}

\author{Stephen D. Ellis, Christopher K. Vermilion, and Jonathan R. Walsh\\
University of Washington, Seattle, WA  98195-1560, USA
 \\ E-mail: \email{sdellis@u.washington.edu}, \email{verm@uw.edu}, \email{jrwalsh@u.washington.edu} }

\author{Andrew Hornig and Christopher Lee\\
Theoretical Physics Group, Lawrence Berkeley National Laboratory,\\
  and Center for Theoretical Physics, University of California, Berkeley, CA 94720, USA
 \\ E-mail: \email{ahornig@uw.edu}, \email{clee137@mit.edu} }

\preprint{
UCB-PTH-10/01 \\
arXiv:1001.0014
}	

\abstract{Jet shapes are weighted sums over the four-momenta of the constituents of a jet and reveal details of its internal structure, potentially allowing discrimination of its partonic origin. In this work we make predictions for quark and gluon jet shape distributions in $N$-jet final states in $e^+e^-$ collisions, defined with a cone or recombination algorithm, where we measure some jet shape observable on a subset of these jets. Using the framework of Soft-Collinear Effective Theory, we prove  a factorization theorem for jet shape distributions
and demonstrate the consistent renormalization-group running of the functions in the factorization theorem for any number of measured and unmeasured jets, any number of quark and gluon jets, and any angular size $R$ of the jets, as long as $R$ is much smaller than the angular separation between jets.  We calculate the jet and soft functions for angularity jet shapes $\tau_a$ to one-loop order ($\mathcal{O}(\as)$) and resum a subset of the large logarithms of $\tau_a$ needed for next-to-leading logarithmic (NLL) accuracy for both cone and $\kt$-type jets.   We compare our predictions for the resummed $\tau_a$ distribution of a quark or a gluon jet produced in a 3-jet final state in $e^+ e^-$ annihilation to the output of a Monte Carlo event generator  and find that the dependence on $a$ and $R$ is very similar.
}

\keywords{Jets, Factorization, Resummation, Effective Field Theory}

\begin{document}


\section{Introduction}
\label{sec:intro}

\subsection{Motivation and Objectives}

Jets provide troves of information about physics within and beyond the Standard Model of particle physics. On the one hand, jets display the behavior of Quantum Chromodynamics (QCD) over a wide range of energy scales, from the energy of the hard scattering, through intermediate scales of branching and showering, to the lowest scale of hadronization. On the other hand, jets contain signatures of exotic physics when produced by the decays of heavy, strongly-interacting particles such as top quarks or particles beyond the Standard Model.

Recently, several groups have explored  strategies to probe jet substructure to distinguish jets produced by light partons in QCD from those produced by heavier particles \cite{Seymour:1993mx, Butterworth:2002tt, Butterworth:2007ke, Brooijmans:2008zz,  Butterworth:2008iy,Butterworth:2009qa, Plehn:2009rk,Kribs:2009yh}, and methods to ``clean'' jets of soft radiation to more easily identify their origin, such as ``filtering'' or ``pruning'' for jets from heavy particles  \cite{Butterworth:2008iy, Ellis:2009me,Ellis:2009su} or ``trimming'' for jets from light partons \cite{Krohn:2009th}. Another type of strategy, explored in  \cite{Almeida:2008yp}, to probe jet substructure is the use of \emph{jet shapes}, which are modifications of event shapes \cite{Dasgupta:2003iq} such as thrust. Jet shapes are continuous variables constructed by taking a weighted sum over the four-momenta of all particles constituting a jet. Different choices of weighting functions produce different jet shapes, and can be designed to probe regions closer to or further from the jet axis with greater sensitivity.\footnote{The original ``jet shape,'' to which the name properly belongs, is the quantity $\Psi(r/R)$, the fraction of the total energy of a jet of radius $R$ that is contained in a subjet of radius $r$ \cite{Ellis:1991vr,Ellis:1992qq,Abe:1992wv} . This observable falls into the larger class of jet shapes we have described here and for which we have hijacked the name.} While such jet shapes may integrate over some of the detailed substructure for which some other methods search, they are better suited to analytical calculation and understanding from the underlying theory of QCD.

In this paper, we consider measuring the shape of one or more jets in an $e^+e^-$ collision at center-of-mass energy $Q$ producing $N$ jets with an angular size $R$ according to a cone or recombination jet algorithm, with an energy cut $\Lambda$ on the radiation allowed outside of jets. We use this exclusive characterization of an $N$-jet final state looking forward to extension of our results to a hadron collider environment, where such a final state definition is more typical. For the jet shape observable we choose the angularity $\tau_a$ of a jet, defined by (cf. \cite{Almeida:2008yp,Berger:2003iw}),
\be
\label{angularitydefn}
\tau_a \equiv \frac{1}{2E_J}\sum_{i\in J}\left|\vect{p}_T^i\right|e^{-\eta_i(1-a)} \, ,
\ee
where $a$ is a parameter taking values $-\infty<a<2$ (for IR safety, although factorizability will require $a<1$), the sum is over all particles in the jet, $E_J$ is the jet energy, $\vect{p}_T$ is the transverse momentum relative to the jet direction, and $\eta=-\ln\tan(\theta/2)$ is the (pseudo)rapidity measured from the jet direction. The jet is defined by a jet algorithm, such as a cone algorithm, the details of which we will discuss below. We complete the calculation for the jet shape $\tau_a$ for jets defined by cone or recombination algorithms, but our logic and methods  could be applied to a wider spectrum of jet shapes and jet algorithms. We have organized our results in such a way that the pieces independent of the choice of jet shape and dependent only on the jet algorithm are easily identifiable, requiring recalculation only of the observable-dependent pieces to extend our results to other choices of jet shapes.

Reliable theoretical prediction of jet observables in the presence of jet algorithms is made challenging by the presence of many scales. Logarithms of ratios of these scales can become large and spoil the behavior of perturbative expansions predicting these quantities. These scales are determined by the jet energy $\omega$, the cut on the angular size of a jet $R$, the measured value of the jet shape such as $\tau_a$, and any other cut or selection parameters introduced by the jet algorithm.

Precisely this separation of scales, however, allows us to take advantage of the powerful tools of factorization and effective field theory. Factorization separates the calculation of a hard scattering cross section into hard, jet and soft functions each depending only on physics at a single scale \cite{Collins:1989gx,Sterman:1995fz}. Renormalization group (RG) evolution of these functions between scales resums logarithms of these scales to all orders in $\as$, with the logarithmic accuracy determined by the order to which the anomalous dimensions in the running are calculated  \cite{Contopanagos:1996nh}. Effective field theory organizes these concepts and tools into a conceptually simple framework unifying many ingredients going into traditional methods, such as power counting, gauge invariance,  and resummation through RG evolution. The rules of effective theory facilitate proofs of factorization and achievement of logarithmic resummation at leading order in the power counting and make straightforward the improvement of results order-by-order in power counting and logarithmic accuracy of resummation.

\subsection{Soft-Collinear Effective Theory and Factorization}

Soft-Collinear Effective Theory (SCET)  \cite{Bauer:2000ew,Bauer:2000yr,Bauer:2001ct,Bauer:2001yt} has been successfully applied to the analysis of many hard scattering cross sections \cite{Bauer:2002nz} including the production of jets. SCET is constructed by integrating out of QCD all degrees of freedom except those collinear to a lightlike direction $n$ and those which are soft, that is, have much lower energy than the energy of the hard scattering or of the jets. Using this formalism, the factorization and calculation of two-jet cross sections and event shape distributions  in SCET were developed in \cite{Bauer:2002ie,Bauer:2003di,Fleming:2007qr,Bauer:2008dt}. Later, these techniques were extended to the factorization of jet cross sections and observables using jet algorithms in  \cite{Bauer:2008jx}. Calculations in SCET of two-jet rates using jet algorithms have been performed in  \cite{Bauer:2003di,Trott:2006bk}, and more recently in \cite{Cheung:2009sg}.
Calculations of cross sections with more than two jet directions have been given in \cite{Bauer:2006mk,Bauer:2006qp,Becher:2009th}.

Building on many of the ideas in these previous studies, in this paper, we will demonstrate a factorization theorem for jet shape distributions in $e^+e^-\to N$ jet events,
\begin{align}
\label{factorizationtheorem}
\frac{d\sigma(\vect{P}_1,\dots,\vect{P}_N)}{d\tau_1\cdots d\tau_M} &= \sigma^{(0)}(\vect{P}_1,\dots,\vect{P}_N)H(n_1,\omega_1;\cdots n_N,\omega_N;\mu) \\
&\quad\times \Bigl[J_{n_1,\omega_1} (\tau_1;\mu)\cdots J_{n_M,\omega_M}(\tau_M;\mu)\Bigr]\otimes S_{n_1\cdots n_N}(\tau_1,\dots,\tau_M;R,\Lambda;\mu) \nn \\
&\quad \times  J_{n_{M+1},\omega_{M+1}}(R; \mu)\cdots J_{n_N,\omega_N}(R;\mu) \,, \nn
\end{align}
where the $N$ jets have three-momenta $\vect{P}_i$, and $M\leq N$ of the jets' shapes $\tau_1,\dots,\tau_M$ are measured. $\sigma^{(0)}$ is the Born cross-section, $H$ is a hard function dependent on the directions $n_i$ and energies $\omega_i$ of the $N$ jets, $J_{n,\omega}(\tau)$ is the jet function for a jet whose shape is measured to be $\tau$, $J_{n,\omega}(R)$ is the jet function for a jet with size $R$ whose shape is not measured, and $S$ is the soft function connecting all $N$ jets, dependent on all jets' shapes $\tau_i$, sizes $R$, and total energy $\Lambda$ that is left outside of all jets.  The symbol ``$\otimes$'' stands for a set of convolution integrals in the variables $\tau_i$ between the measured jet functions and the soft function. All terms in the factorization theorem depend on the factorization scale $\mu$.

SCET is typically constructed as a power expansion in a small parameter $\lambda$ formed by the ratio of soft to collinear or collinear to hard scales, determined by the kinematics of the process under study. $\lambda$ is roughly the typical transverse momentum $p_T$ of the constituent of a jet (relative to the jet direction) divided by the jet energy $E_J$. This is set either by the measured value of the jet shape $\tau_a$ for a measured jet or the algorithm measure $R$ for an unmeasured jet. Thus we encounter in this work the new twist that the size of $\lambda$ may be different for different jets. We will comment on further implications of this in subsequent sections. Still, in each separate collinear sector, the momentum $p_n$ of collinear modes in the light-cone direction $n$ in SCET is separated into a large ``label'' momentum $\tilde p_n$ containing $\mathcal{O}(E_J)$ and $\mathcal{O}(\lambda E_J)$ components and a ``residual'' component of $\mathcal{O}(\lambda^2 E_J)$, the same size as soft momenta. Effective theory fields have dynamical momenta only of this soft or residual scale. This fact, along with the fact that soft quarks and soft gluons can be shown to decouple from collinear modes at the level of the Lagrangian \cite{Bauer:2001yt}, makes possible the factorization of a jet shape distribution into hard, jet, and soft functions depending only on the dynamics at those respective scales.

In using SCET for jets in multiple directions and using jet algorithms to define the jets, we will encounter the need for several additional criteria to ensure the validity of the $N$-jet factorization theorem.
\begin{itemize}
\item First, to ensure that the algorithm does not group final-state particles into fewer than $N$ jets, the jets must be ``well separated.'' This allows us to use as the effective theory Lagrangian a sum of $N$ copies of the collinear part of the SCET Lagrangian for a single direction $n$  and a soft part, and to construct a basis of $N$-jet operators built from fields from each of these sectors to produce the final state. Our calculations will reveal the precise quantitative condition that jets must satisfy to be ``well separated''.

\item Second, to ensure that the jet algorithm does not find  more than $N$ jets, we place an energy cut $\Lambda$ on the total energy outside of the observed jets. We will take this energy $\Lambda$ to scale as a soft momentum so that we will be able to identify the total energy of each  jet with the ``label'' momentum on the SCET collinear jet field producing the jet. Corrections to this identification are subleading in the SCET power counting.

\item Third (and related to the above two), we will assume that the $N$-jet restriction on the final state can itself be factorized into a product of $N$ 1-jet restrictions, one in each collinear sector, and a $0$-jet restriction in the soft sector. We represent the energetic particles in the $i$th jet by collinear fields in the SCET Lagrangian in the $n_i$ collinear sector  and soft particles everywhere with fields in the soft part of the Lagrangian. We then stipulate that the jet algorithm acting on states in the $n_i$ collinear sector find exactly one jet in that sector, and when acting on the soft final state  find no additional jet in that sector.

\item Fourth, the way in which a jet algorithm combines particles in the process of finding a jet must respect the order of steps envisioned by factorization.  In particular, factorization requires that the jet directions and energies be determined by the collinear particles alone, so that the soft function knows only about the directions and colors of the jets, not the details of any collinear recombinations. Ideally, all energetic collinear particles should be recombined first, with soft particles within a radius $R$ of the jet axis being recombined into the jet only afterwards. Jet algorithms in use at experiments do not have this precise behavior, but we will discuss in \sec{sec:power} the extent to which common algorithms meet this requirement and estimate the size of the power corrections due to their failure to do so. In general, we will find that for sufficiently large $R$, infrared-safe cone algorithms and $\kt$-type recombination algorithms satisfy the requirements of factorizability, with anti-$\kt$ allowing smaller values of $R$ than $\kt$.
\end{itemize}

After enforcing the above requirements, a key test of the consistency of \eq{factorizationtheorem} will be the independence of the physical cross section on the factorization scale $\mu$. This requires the anomalous dimensions of the hard, jet, and soft functions to sum to zero,
\begin{equation}
\label{consistencycondition}
\begin{split}
0 &= [\gamma_H(\mu) + \gamma_{J_{M+1}}(R;\mu) + \cdots + \gamma_{J_{N}}(R;\mu)]\delta(\tau_1)\cdots \delta(\tau_M) \\
&\quad + \gamma_{J_1}(\tau_1;\mu)\delta(\tau_2)\cdots \delta(\tau_N) + \cdots + \delta(\tau_1)\cdots \delta(\tau_{M-1})\gamma_{J_M}(\tau_M;\mu) \\
&\quad + \gamma_S(\tau_1,\dots,\tau_M,R;\mu)
\,.
\end{split}
\end{equation}
It seems highly nontrivial that this condition would be satisfied for any number, size, and flavors of jets (and that the soft anomalous dimension be independent of $\Lambda$), but we will demonstrate that it does hold at $\mathcal{O}(\as)$, up to corrections of $\cO(1/t^2)$ which violate \eq{consistencycondition}, where $t$ is a measure of the separation between jets.  In particular, for a pair of jets, $i,j$, with 3-vector directions separated by a polar angle $\psi_{ij}$, the separation $t_{ij}$ is given by
\begin{equation}
\label{tdef}
t_{k,l} = \frac{\tan(\psi_{k,l}/2)}{\tan(R/2)}\,.
\end{equation}
Now define $t$ (no indices) as the minimum of $t_{ij}$ over all jet pairs.  This quantifies the qualitative condition of jets being well-separated, $t  \gg 1$, that is required to justify the factorization theorem \eq{factorizationtheorem}. The factorization theorem is valid up to corrections of $\mathcal{O}(\lambda)$ in the  SCET power expansion parameter and corrections of $\mathcal{O}(1/t^2)$ in the separation parameter. As an example of the magnitude of $t$, for three jets in a Mercedes-Benz configuration ($\psi=2\pi/3$ for all pairs of jets),  $1/t^2= 0.04$ for $R=0.7$ and $1/t^2 = 0.1$ for $R=1$, so these corrections are indeed small.  More generally, for non-overlapping jets, $\psi > 2R$, we have $1/t^2 < 1/4$.

Notice that for back-to-back jets ($\psi=\pi$), $t\to\infty$. Thus, for all cases previously considered in the literature, the jets are infinitely separated according to this measure, and no additional criterion regarding jet separation is required for consistency of the factorization and running. A key insight of our work is that for an $N$-jet cross-section described by  \eq{factorizationtheorem}, the factorization theorem receives corrections not only in the usual SCET power counting parameter $\lambda$, but also corrections due to jet separation beginning at $\cO(1/t^2)$.

\subsection{Power Corrections to Factorized Jet Shape Distributions}

As always, there are power corrections to the factorization theorem which we must ensure are small. One class of power corrections arises from approximating the jet axis of the measured jet with the collinear direction $n_i$, which labels that jet in the SCET Lagrangian. This direction $n_i$ is the direction of the parent parton initiating the jet. The jet observable must be such that the difference between the parent parton direction and the jet axis identified by the algorithm makes a subleading correction to the  calculated value of the jet observable. In the context of angularity event shapes, such corrections were estimated in
\cite{Berger:2003iw,Bauer:2008dt} and found to be negligible for $a<1$, and we find the same condition for jet shapes.

In the presence of algorithms, however, there are additional power corrections due to the difference in the soft particles that are included or excluded in a jet by the actual algorithm and in its approximated form in the factorization theorem. We study the effect of this difference on the measurement of jet shapes, and find that for sufficiently large $R$ the power corrections due to the action of the algorithm on soft particles remain small enough not to spoil the factorization for infrared-safe cone and $\kt$-type recombination algorithms.
Algorithm-related power corrections to jet momenta were studied more quantitatively in   \cite{Dasgupta:2007wa}, and their estimated $R$ dependence is consistent with our observations.

  We do not address in this work the issue of power corrections to jet shapes due to hadronization. Event shape distributions are known to receive power corrections of the order $1/(\tau_aQ)$, enhanced in the endpoint region but suppressed by large energy. The endpoints of our jet shape distribution near $\tau_a\to 0$, therefore, will have to be corrected by a nonperturbative shape function. Such functions have been constructed for event shapes in \cite{Korchemsky:2000kp,Hoang:2007vb}. The shift in the first moment of event shape distributions induced by these shape functions was postulated to take a universal form in \cite{Dokshitzer:1995zt,Dokshitzer:1997ew} based on the behavior of single soft gluon emission, and the universality was proven to all orders in soft gluon emission at leading order in the SCET power counting in \cite{Lee:2006nr,Lee:2006fn}. This universality relied on the boost invariance of the soft function describing soft gluon radiation from two back-to-back collinear jets. The extent to which such universality may survive for jet shapes with multiple jets in arbitrary directions is an open question that must be addressed in order to construct appropriate soft shape function models to deal adequately with the power corrections to jet shapes from hadronization. Nonperturbative power corrections  to jet observables  from hadronization and the underlying event in hadron collisions  were also studied in \cite{Dasgupta:2007wa}, and hadronization corrections were found to scale like $1/R$.   In this work, we focus only on the perturbative calculation and resummation of large logarithms of jet shapes, and leave inclusion of nonperturbative power corrections for future work.

\subsection{Resummation and Logarithmic Accuracy}

Knowing the anomalous dimensions of the hard, jet, and soft functions in the factorization theorem allows us to resum logarithms of ratios of the hard, jet, and soft scales. We take this opportunity to explain the order of accuracy to which we are able to resum these logarithms. For an event shape distribution $d\sigma/d\tau$ (i.e. \eq{factorizationtheorem} with two jets and integrated against $\delta(\tau-\tau_1-\tau_2)$), the accuracy of logarithmic resummation \cite{Catani:1992ua} is typically characterized by counting logs in the exponent $\ln R(\tau)$ of the ``radiator,''
\begin{equation}
\label{radiator}
R(\tau) = \frac{1}{\sigma_0} \int_0^\tau d\tau'\frac{d\sigma}{d\tau'}\,,
\end{equation}
where they appear in the form $\alpha_s^n \ln^m \tau$ with $m\leq n+1$. At leading-logarithmic (LL) accuracy all the terms with $m=n+1$ are summed; next-to-leading-logarithmic (NLL) accuracy sums also the $m=n$ terms, and so on. In more traditional methods in QCD, event shapes that have been resummed include NLL resummation of  thrust in \cite{Catani:1992ua,Catani:1991kz}, jet masses in \cite{Catani:1992ua,Catani:1991bd,Dasgupta:2001sh,Burby:2001uz},  jet broadening in \cite{Catani:1992jc,Dokshitzer:1998kz}, the $C$-parameter in \cite{Catani:1998sf}, and angularities in  \cite{Berger:2003iw}.  Resummation of an event shape distribution using the modern SCET method was first illustrated with  the thrust distribution to LL accuracy in \cite{Schwartz:2007ib}. Heavy quark jet mass distributions were resummed in SCET to NLL accuracy as part of a proposed method to extract the top quark mass in \cite{Fleming:2007xt}. The N$^3$LL resummed thrust distribution in SCET was compared to LEP data to extract a value for the strong coupling $\alpha_s$ to high precision in \cite{Becher:2008cf}. Angularities were resummed to NLL accuracy  in SCET in \cite{Hornig:2009vb} directly in $\tau_a$-space instead of in moment space as in \cite{Berger:2003iw}.

Summation of logarithms in effective field theory is achieved by RG evolution. In the factorized radiator of the thrust distribution \eq{radiator}, one finds that the hard function contains logarithms of $\mu/Q$, the jet functions contain logarithms of $\mu/(Q\sqrt{\tau})$, and the soft function contains logarithms of $\mu/(Q\tau)$. Thus, evaluating these functions respectively at the hard scale $\mu_H=Q$, jet scale $\mu_J = Q\sqrt{\tau}$ and soft scale $\mu_S = Q\tau$ eliminates large logarithms in each function. They can then be RG-evolved to the common factorization scale $\mu$ after calculating their anomalous dimensions. The solutions of the RG evolution equations are of the form that logarithms of $\tau$ are resummed to all orders in $\alpha_s$ to a logarithmic accuracy determined by the order in $\alpha_s$ to which the anomalous dimensions and hard/jet/soft functions are known. This underlying hierarchy of scales is illustrated \fig{fig:scales} [in this case, with only one (measured) jet scale and soft scale and $\omega=Q$] along with a table that lists the order in $\as$ to which various quantities must be known in order to achieve a given N$^k$LL accuracy in the exponent of the radiator \eq{radiator}. The power of the EFT framework is to organize of the logs of $\tau$ arising in \eq{radiator} into those that arise from ratios of the jet to the hard scale and those that arise from ratios of the soft to the hard scale, which then allows RG evolution to resum them.

\FIGURE[t]{
\includegraphics[width = \textwidth]{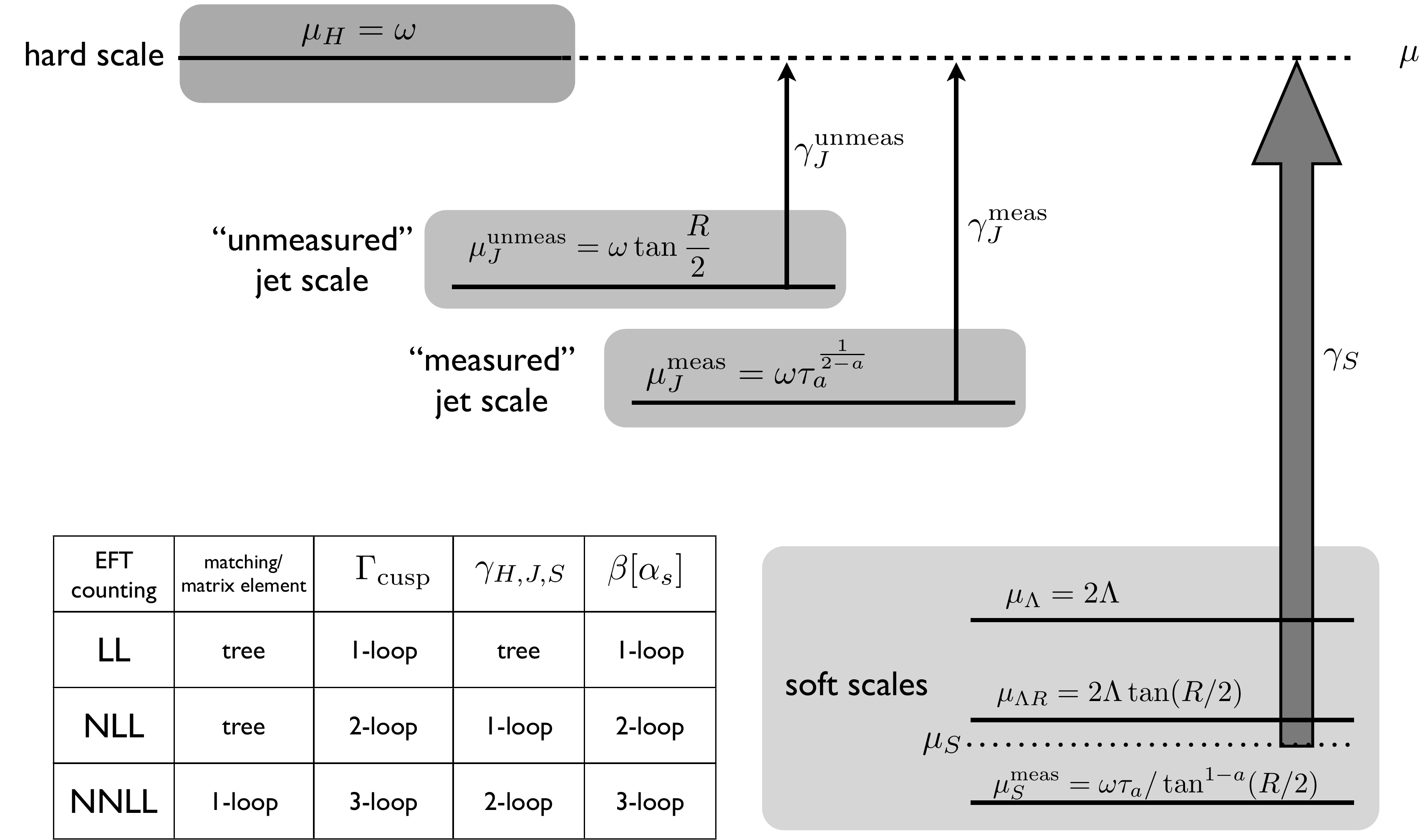}

{ \caption[1]{An illustration of generic scales along with a table of log-accuracy versus perturbative order. A cross section with jets of energy $\sim\omega$, radius $R$, and energy $\Lambda$ outside the jets, with some jets' shapes $\tau_a$ being measured and others' shapes left unmeasured, induces measured and unmeasured jet scales at $\mu_J^{\text{meas}}$ and $\mu_J^{\text{unmeas}}$. Dynamics at these scales are described by separate collinear modes in SCET. Soft dynamics occur at several soft scales, $\mu_\Lambda$ and $\mu_{\Lambda R}$ induced by the soft out-of-jet energy cut $\Lambda$ and jet radius $R$, and $\mu_S^{\text{meas}}$ induced by the measured jet shape $\tau_a$. RG evolution in SCET resums logs of ratios of jet scales to the hard scale $\mu_H$ individually, and logs of the ratio of a ``common'' soft scale $\mu_S$ to the hard scale. Remaining logs of ratios of soft scales to one another are not resummed in current formulations of SCET. The accuracy of logarithmic resummation of these ratios of scales is determined by the order to which anomalous dimensions and matching coefficients or matrix elements (i.e. hard/jet/soft functions) are calculated in perturbation theory. In this paper we perform the RG evolution indicated by the arrows to NLL accuracy.}
\label{fig:scales}}
}

For the multijet shape distribution in \eq{factorizationtheorem}, the strategy to sum logs is the same, but is complicated by the presence of additional scales. This also makes trickier the classification of logarithmic accuracy into the standard N$^k$LL scheme.  Our aim will be to sum as many logs of the jet shapes $\tau_a$ as possible, while not worrying about any others. For instance, phase space cuts induce logs of $R$ and $\Lambda/\omega$ (where $\omega$ is a typical hard jet energy), and the presence of multiple jets induces logs of jet separations $n_i\cdot n_j$ or ratios of jet energies $\omega_i/\omega_j$. We will not aim to sum these types logs systematically in this paper, only those of $\tau_a$ (though we sum subsets of the others incidentally). In particular, resumming the phase space logs of $R$ or $\Lambda/\omega$ is complicated by how the phase space cuts act order-by-order in perturbation theory\footnote{The JADE algorithm is one well-known example in which resummability even of leading logarithms of  the jet mass cut $y$ is spoiled by the  differences in the jet phase space at different orders in perturbation theory \cite{Brown:1990nm}. Another example that will not work is using a $\kt$-type algorithm with $R$ randomly chosen for each recombination. This is  clearly such that resummation of logarithms of $R$ cannot be achieved.}, and the fact that a simple angular cut $R$ is less restrictive than a small jet mass or angularity on how collimated a jet must be. That is, an angular cut allows particles in a jet to be anywhere within an angle $R$ of the jet axis regardless of their energy, while a small jet mass or angularity forces harder particles to be closer to the jet axis. The former allows hard particles to lie along the edges of a jet, and soft radiation from such configurations that escapes the jets can lead to logs of $\Lambda/\omega$ that are not captured in our treatment.   These are not issues we solve in this paper, in which we focus on resumming logs of jet shapes $\tau_a$. (Some exploration of phase space logarithms in SCET was carried out in \cite{Trott:2006bk,Cheung:2009sg}.)

A way to understand how we sum logs and which ones we capture is presented in \fig{fig:scales}. The factorization theorem \eq{factorizationtheorem} organizes logs in the multijet cross section into those in the hard function, those in measured jet functions, those in unmeasured jet functions, and those in the soft function, much like for the simple thrust distribution. The difference is the presence now of multiple jet and soft scales. Logarithms in jet functions can still be minimized by choices of individual jet scales, $\mu_J^{\text{meas}}\sim \omega\tau_a^{1/(2-a)}$ for a jet whose shape $\tau_a$ is measured, and $\mu_J^{\text{unmeas}}\sim\omega\tan(R/2)$ for a jet whose shape is not measured but has a radius $R$. Thus logs arising from ratios of these scales to the hard scale can be summed completely to a desired N$^k$LL order. The complication is in the soft function. The soft function is sensitive to soft radiation into measured and unmeasured jets and outside of all jets. As we will see by explicit calculation, this induces logs of $\mu\tan^{1-a}(R/2)/(\omega\tau_a)$ from radiation into measured jets, and  logs of  $\mu/(2\Lambda)$ and  $\mu/(2\Lambda\tan\frac{R}{2})$ from  radiation from unmeasured jets  cut off by the energy $\Lambda$. In addition, though not illustrated in \fig{fig:scales}, there can be logs of multiple jet shapes to one another, $\tau_a^i/\tau_a^j$. No single choice of a soft scale $\mu_S$ will minimize all of these logs.

In the present work, we will start with the simple strategy of choosing a single soft scale $\mu_S\sim \omega\tau_a/\tan^{1-a}(R/2)$ for a jet whose shape $\tau_a$ we are measuring and logs of which we aim to resum. We will calculate hard/jet/soft functions and anomalous dimensions corresponding to ``NLL'' accuracy listed in \fig{fig:scales}. By this we do not mean all potentially large logs in \eq{factorizationtheorem} are resummed to NLL, but only those logs of ratios of a jet scale to the hard scale or of the (common) soft scale to the hard scale. We do not attempt to sum logs of ratios of soft scales to one another completely to NLL accuracy (which can contain $\tau_a$).  In the case that all jets' shapes are measured and are similar to one another, $\tau_a^i\sim\tau_a^j$, our resummation of large logs of $\tau_a^i$ would be complete to NLL accuracy.

We will nevertheless venture to propose a framework to ``refactorize'' the soft function into further pieces dependent on only a single soft scale at a time and perform additional RG running between these scales to resum the additional logarithms, and will implement it at the level of the $\mathcal{O}(\as)$ soft functions we calculate. However, one cannot really address mixed logarithms such as $\log(\tau_a^i/\tau_a^j)$ that arise for multiple jets until $\cO(\as^2)$, the first order at which two soft gluons can probe two different physical regions. This lies beyond the scope of the present work. (We note, however, that our implementation of refactorization using the one-loop soft function does already seem to tame logarithmic dependence on $\Lambda$ in our numerical studies of jet shape distributions.)

These issues are related to some types of ``non-global'' logarithms described by
\cite{Dasgupta:2001sh,Banfi:2002hw,Appleby:2003ai,Dokshitzer:2003uw} that spoil the simple characterization of NLL accuracy.
In \cite{Banfi:2010pa}  these were identified as next-to-leading logs of  $\Lambda R^2/(\omega_i\tau_a)$ and $\Lambda/Q$ (when $R \ll 1$) that appear at $\cO(\as^2)$  in jet shape distributions. These authors organized the radiator for a single jet shape distribution into a ``global'' and ``non-global'' part \cite{Dokshitzer:2003uw,Banfi:2010pa},
\begin{equation}
R(\tau_a^i,R,\Lambda;\omega_i,Q) = R_{gl}\left(\tau_a^i,R,\frac{\Lambda}{Q}\right) R_{ng} \left(\frac{\Lambda R^2}{\omega_i\tau_a^i},\frac{\Lambda}{Q}\right)\,.
\end{equation}
In this language, the calculations we undertake in this paper resum logs in the global part to NLL accuracy but not in the non-global part.  The first argument in $R_{ng}$ is related to ratios of soft scales illustrated in \fig{fig:scales}, and the second argument arises when there are unmeasured jets. In the case that all jets are measured, $R\sim 1$, and $\Lambda\sim\omega_i\tau_a^i$, the non-global logs vanish.

While summing all global and non-global logs to at least NLL accuracy will be important for precision jet phenomenology, what we contribute in this paper are key developments and calculations necessary to resum even global logs of jet shapes for jets defined with algorithms. We also believe the  effective theory approach and the idea of  refactorizing the soft function will help us understand  and resum many types of non-global logarithms.

\subsection{Detailed Outline of This Work}

In this paper,
we will formulate and prove a factorization theorem for distributions in the jet shape variables we introduced above, calculate the jet and soft functions appearing in the factorization theorem to $\cO(\as)$ in SCET, and use the renormalization group evolution of these functions to sum global logs of $\tau_a$ to NLL accuracy. We consider $N$ jets (defined with a cone or $\kt$ algorithm) produced in an $e^+e^-$ collision, with $M$ of the jets' shapes (angularities) being measured. The key formal result is our demonstration of \eq{consistencycondition}, the consistency of the anomalous dimensions of hard, jet and soft functions to $\cO(\as)$  for any number of total jets, any numbers of quark and gluon jets, any number of these jets whose shapes are measured, and any value of the distance measure $R$ in cone or $\kt$-type algorithms (as long as $t\gg 1$). 
These results lead to accurate predictions for the shape of the $\tau_a$  distribution near the peak, but not necessarily the endpoints for very small $\tau_a$ (where hadronization corrections dominate) and very large $\tau_a$ (where fixed-order NLO QCD corrections take over, which are not yet calculated and not captured by NLL resummation).\footnote{Jet shapes were also studied in the QCD factorization approach in  \cite{Almeida:2008tp}. In that work QCD jet functions for quark and gluon jets defined with an algorithm and whose jet masses $m_J^2$ are measured were calculated to $\mathcal{O}(\alpha_s)$. The jet mass$^2$ corresponds to $\tau_a$ for $a=0$, $\tau_0 = m^2/\omega^2$ ($\tau_0 \ll 1$). A fixed-order QCD jet function as defined in \cite{Almeida:2008tp} is given by the convolution of our fixed-order SCET jet function and soft function for a measured jet away from $\tau_a=0$.}

In \sec{sec:shapes} we describe in detail the jet shapes and jet algorithms that we use. We describe features of an ``ideal'' jet algorithm that would respect exactly the order of operations envisioned in the factorization theorem derived in SCET, and the extent to which cone and recombination algorithms actually in use resemble this idealization.

In \sec{sec:fact},  using the tools of SCET, we will derive in detail a factorization theorem for exclusive $3$-jet production where we measure the angularity jet shape of one of the jets, and then perform the straightforward extension  to $N$-jet production with $M \le N$ measured jets. We will give a review of the necessary technical details of SCET in \ssec{SCET}. In the process of justifying the factorization theorem, we identify the new requirements listed above on $N$-jet final states and jet algorithms that must be satisfied for factorization to hold. In \sec{sec:power} we explore in some detail the power corrections to the factorization theorem due to soft radiation and the action of jet algorithms that cause tension with these requirements, and argue that for sufficiently large $R$ in infrared-safe cone and recombination algorithms, these corrections are sufficiently small.

Next we calculate to $\cO(\as)$ the jet and soft functions corresponding to $N$ cone or $\kt$-type jets, with $M$ jets' shapes measured.

In \sec{sec:jet} we calculate the jet functions for measured quark jets, $J_\omega^q(\tau_a)$, unmeasured quark jets, $J_\omega^q$, measured gluon jets, $J_\omega^g(\tau_a)$, and unmeasured gluon jets, $J_\omega^g$, where $\omega = 2E_J$ is the label momentum of the collinear jet field in each jet function. We find that in collinear sectors for measured jets, the collinear scale (and thus the SCET power counting parameter in that sector $\lambda_i$) is given by $\omega_i\tau_a^{1/(2-a)}$, and in unmeasured jet sectors, $\lambda_i\sim\tan(R/2)$. In studying power corrections, however, as mentioned above, we find that $R$ must be parametrically larger than $\tau_a$.  So, in collinear sectors for measured jets, $\lambda_i$ is set by the shape $\tau_a$ with $R\sim \lambda_i^0$, while in unmeasured jet sectors, $\lambda_i\sim \tan(R/2)$. Thus one should understand $\tan(R/2)$ to be significantly less than 1 but much larger than any jet shape $\tau_a$.

In \sec{sec:soft} we calculate the soft function. To do this, we split the soft function into several contributions from different parts of phase space in order to facilitate the calculation and elucidate its intuitive structure. We find it most convenient to split the soft function into an observable-independent part that arises from soft emission out of the jets, $S^\unmeas$, and a part that depends on our choice of angularities as the observable that arises from soft emission into measured jet $i$, $S^\meas(\tau^i_a)$. $S^\unmeas$ is hence sensitive to the scale $\Lambda$ while $S^\meas(\tau_a^i)$ is sensitive to the scale $\omega_i \tau_a^i$.

In \sec{sec:cons}, having calculated all the jet and soft function contributions to $\cO(\as)$, we extract the anomalous dimensions and perform renormalization-group (RG) evolution.  We find the hard anomalous dimension from existing results in the literature.
The hard, jet, and soft anomalous dimensions have to satisfy the consistency condition \eq{consistencycondition} in order for the physical cross section to be independent of the arbitrary factorization scale $\mu$. Our calculations reveal that, as long as the jet separation parameter $t$ \eq{tdef} between all pairs of jets is much larger than 1, the condition is satisfied.

Even after requiring $t \gg 1$, the satisfaction of the consistency condition is non-trivial. The hard function knows only about the direction of each jet and the jet function knows only the jet size $R$; the soft function knows about both. Furthermore, it is not sufficient only to include regions of phase space where radiation enters the measured jets. We learn from our results in this Section that it is crucial to include
soft radiation outside of all jets with an upper energy cutoff of $\Lambda$. Only after including all of these contributions from the various parts of phase space do the jet, hard, and soft anomalous dimensions cancel and we arrive at a consistent factorization theorem.

We conclude  \sec{sec:cons} by proposing in \ssec{refactorization} a strategy to sum logs due to a hierarchy of scales in the soft function, by  ``refactorizing'' it into multiple pieces, each sensitive to a single scale, as suggested by the discussion surrounding \fig{fig:scales}. Our current implementation of this procedure does tame the logarithmic dependence of jet shape distributions on the ratio $\Lambda/\omega$ in our numerical studies, but we leave for further development the resummation of all ``non-global'' logs of ratios of multiple soft scales that begin at NLL and $\cO(\as^2)$.

To help the reader find the results of the calculations in \sec{sec:jet} through \sec{sec:cons} just outlined, \tab{master-table} provides a summary with equation numbers.
\TABLE[t]{
\begin{tabular}{ | c | c | c | c | }
\hline
Category &  Contribution & Symbol & Location  \\
 \hline
 \hline
& measured quark jet function & $J_\omega^q(\tau_a)$ & \eq{Jtotal} \\
\cline{ 2-4}
& unmeas. quark jet function & $J_\omega^q$ & \eq{Jq} \\
 \cline{ 2-4}
& measured gluon jet function & $J_\omega^g(\tau_a)$ & \eq{Gjet} \\
\cline{ 2-4}
& unmeas. gluon jet function & $J_\omega^g$ & \eq{G2result} \\
\cline{ 2-4}
NLO contributions & summary of divergent & \multirow{2}{*}{--- }  &  \multirow{2}{*}{\tab{soft} } \\
 before resummation: & parts of soft func. (any $t$) &  & \\
\cline{ 2-4}
& total universal &  \multirow{2}{*}{$S^\unmeas$} & \multirow{2}{*}{ \eq{eq:Sunmeas-result}} \\
& soft func. (large $t$) & & \\
\cline{ 2-4}
& total measured &  \multirow{2}{*}{$S^\meas(\tau_a^i)$} &  \multirow{2}{*}{\eq{eq:Smeas-result}} \\
& soft func. (large $t$)  & & \\
 \hline
  anomalous dimensions: & --- & --- & \tab{anomalous-coeff} \\
 \hline
& measured jet function & $f_J^i(\tau_a^i; \mu_J^i)$ & \eq{measuredlogs-jet} \\
\cline{ 2-4}
NLO contributions & measured soft function & $f_S(\tau_a^i; \mu_J^i)$ & \eq{measuredlogs-soft}\\
\cline{ 2-4}
 after resummation:  & unmeas. jet function & $J_\omega^i(\mu_J)$ & \eq{unmeasuredlogs-jet} \\
\cline{ 2-4}
 & universal soft function & $S^\unmeas(\mu_S^\Lambda)$ & \eq{unmeasuredlogs-soft} \\
 \hline
 Total NLL Distribution: & --- & --- & \eq{NjetcsRG}   \\
 \hline
\end{tabular}
{ \caption[1]{Directory of main results: the fixed-order NLO quark and gluon jet functions for jets whose shapes $\tau_a$ are measured or not; the fixed-order NLO contributions to the soft functions from parts of phase space where a soft gluon enters a measured jet, $S^\meas(\tau_a)$, or not, $S^\unmeas$; their anomalous dimensions; the contributions the jet and soft functions make to the finite part of the NLL resummed distributions; and the full NLL resummed jet shape distribution itself.}
\label{tab:master-table} }}

In \sec{sec:plots} we compare our resummed perturbative predictions for jet shape distributions to the output of a Monte Carlo event generator. We test both the accuracy of these predictions  and assess the extent to which hadronization corrections affect jet shapes. We will illustrate our results in the case of $e^+ e^-\to \text{3 jets}$, with the jets constrained to be in a configuration where each has equal energy and are maximally separated. In both the effective theory and Monte Carlo, we can take the jets to have been produced by an underlying hard process $e^+ e^-\to q\bar q g$. After placing cuts on jets to ensure each parton corresponds to a nearby jet, we measure the angularity jet shape of one of the jets. We compare our resummed theoretical predictions with the Monte Carlo output for quark and gluon jet shapes with various values of $a$ and $R$.  We find that the dependencies on $a$ and $R$ of the shapes of the distribution and the peak value of $\tau_a$  agree well between the theory and Monte Carlo, with small but noticeable corrections due to hadronization.  We can estimate these corrections by comparing output with hadronization turned on or off in Monte Carlo.

In \sec{sec:conclusions}, we give our conclusions and outlook. We also collect a number of technical details and results for $\cO(\as)$ finite pieces of jet and soft functions in the Appendices.

Our work is, to our knowledge, the first achieving factorization and resummation of a jet observable distribution in an exclusive $N$-jet final state defined by a non-hemisphere jet algorithm.\footnote{Dijet cross sections for cone jets were factorized and resummed in \cite{Kidonakis:1998ur}.}  Having demonstrated the consistency of this factorization for any number of quark and gluon jets, measured and unmeasured jets, and phase space cuts in cone and $\kt$-type algorithms, and having constructed a framework to resum logarithms of jet shapes in the presence of these phase space cuts, we hope to have provided a starting point for future precision calculations of many jet observables both in $e^+ e^-$ and hadron-hadron collisions. The case of $pp$ collisions will require a number of modifications, including turning two of our outgoing jet functions into  incoming ``beam functions'' introduced in \cite{Stewart:2009yx}.  We leave this generalization for future work.

The reader wishing to follow the general structure of our ideas and logic and understand the basis of the final results of the paper without working through all the technical details may read Secs.~\ref{sec:intro} and \ref{sec:shapes}, and then skip to \sec{sec:plots}. Some short less technical discussion also appears in \sec{sec:power}.

\section{Jet Shapes and Jet Algorithms}
\label{sec:shapes}

\subsection{Jet Shapes}

Event shapes, such as thrust, characterize events based on the distribution of energy in the final state by assigning differing weights to events with differing energy distributions.
Events that are two-jet like, with two very collimated back-to-back jets, produce values of the observable at one end of the distribution, while spherical events with a broad energy distribution produce values of the observable at the other end of the distribution.  While event shapes can quantify the global geometry of events, they are not sensitive to the detailed structure of jets in the event.  Two classes of events may have similar values of an event shape but characteristically different structure in terms of number of jets and the energy distribution within those jets.

Jet shapes, which are event shape-like observables applied to single jets, are an effective tool to measure the structure of individual jets.  These observables can be used to not only quantify QCD-like events, but study more complex, non-QCD topologies, as illustrated  for light quark vs. top quark and $Z$ jets in \cite{Almeida:2008yp,Almeida:2008tp}.
Broad jets, with wide-angle energy depositions, and very collimated jets, with a narrow energy profile, take on distinct values for jet shape observables.  In this work, we consider the example of the class of jet shapes called angularities, defined in \eq{angularitydefn} and denoted $\tau_a$.  Every value of $a$ corresponds to a different jet shape.  As $a$ decreases, the angularity weights particles at the periphery of the jet more, and is therefore more sensitive to wide-angle radiation.  Simultaneous measurements of the angularity of a jet for different values of $a$ can be an additional probe of the structure of the jet.

\subsection{Jet Algorithms}

A key component of the distribution of jet shapes is the jet algorithm, which builds jets from the final state particles in an event. (We are using the term ``particle'' generically here to refer to actual individual tracks, to cells/towers in a calorimeter, to partons in a perturbative calculation, and to combinations of these objects within a jet.)  Since the underlying jet is not intrinsically well defined, there is no unique jet algorithm and a wide variety of jet algorithms have been proposed and implemented in experiments.  The details of each algorithm are motivated by particular properties desired of jets, and different algorithms have different strengths and weaknesses.  In this work we will calculate angularity distributions for jets coming from a variety of algorithms.  Because we calculate  (only)  at next-to-leading order, there are at most 2 particles in a jet, and jet algorithms that implement the same phase space cuts at NLO simplify to the same algorithm.  At this order the two standard classes of algorithms, cone algorithms and recombination algorithms, each simplify to a generic jet algorithm at NLO.  At NLO the cone algorithms place a constraint on the separation between each particle and the jet axis, while the recombination algorithms place a constraint on the separation between the two particles.

Cone algorithms build jets by grouping particles within a fixed geometric shape, the cone, and finding ``stable'' cones.  A cone contains all of the particles within an angle $R$ of the cone axis, and the angular parameter $R$ sets the size of the jet.  In found jets (stable cones), the direction of the total three-momentum of particles in the cone equals the cone (jet) axis.  Different cone algorithms employ different methods to find stable cones and deal with differently the ``split/merge'' problem of overlapping stable cones.  The SISCone algorithm \cite{Salam:2007xv} is a modern implementation of the cone algorithm that finds all stable cones and is free of infrared unsafety issues.  In the next-to-leading order calculation we perform, there are at most two particles in a jet, and we only consider configurations where all jets are well-separated.  Therefore, it is straightforward to find all stable cones, there are no issues with overlapping stable cones, and the phase space cuts of all cone algorithms are equivalent.  This simplifies all standard cone algorithms to a generic cone-type algorithm, in which each particle is constrained to be within an angle $R$ of the jet axis.  For a two-particle jet, if we label the particles $1$ and $2$ and the jet axis $\vect{n}$, then the cone-like constraints for the two particles to be in a jet are
\be
\textrm{cone jet: } \theta_{1\vect{n}} < R \textrm{ and } \theta_{2\vect{n}} < R \, .
\ee
This defines our cone-type algorithm.

Recombination algorithms build jets by recursively merging pairs of particles.  Two distance metrics, defined by the algorithm, determine when particles are merged and when jets are formed.  A pairwise metric $\rho_{\rm pair}$ (the recombination metric) defines a distance between pairs of particles, and a single particle metric $\rho_{\rm jet}$ (the beam, or promotion, metric) defines a distance for each single particle.  Using these metrics, a recombination algorithm builds jets with the following procedure:\footnote{This defines an \emph{inclusive} recombination algorithm more typically applied to hadron-hadron colliders.  We are applying it here to the simpler case of $e^+ e^-$ collisions in order to facilitate the eventual transition to LHC studies.  Exclusive recombination algorithms, more typical of $e^+ e^-$ collisions, are described along with other jet algorithms in \cite{Salam:2009jx} and their description in SCET is given in \cite{Cheung:2009sg}.}

\begin{itemize}
\item[0.] Begin with a list $L$ of particles.
\item[1.] Find the smallest distance for all pairs of particles (using $\rho_{\rm pair}$) and all single particles (using $\rho_{\rm jet}$).
\item[2a.] If the smallest distance is from a pair, merge those particles together by adding their four momenta.  Replace the pair in $L$ with the new particle.
\item[2b.] If the smallest distance is from a single particle, promote that particle to a jet and remove it from $L$.
\item[3.] Loop back to step 1 until all particles have been merged into jets.
\end{itemize}

The $\kt$, Cambridge-Aachen, and anti-$\kt$ algorithms are common recombination algorithms, and their distance metrics are part of a general class of recombination algorithms. For $e^+e^-$ colliders, a class of recombination algorithms can be defined by the parameter $\alpha$:
\begin{align}
\rho_{\rm pair}(i,j) &= \min\left(E_i^{\alpha},E_j^{\alpha}\right)\frac{\theta_{ij}}{R} \nn \\
\rho_{\rm jet}(i) &= E_i^{\alpha} \, ,
\end{align}
where $\alpha = 1$ for $\kt$, $0$ for Cambridge-Aachen, and $-1$ for the anti-$\kt$ algorithm.  The parameter $R$ sets the maximum angle between two particles for a single recombination.\footnote{We use $R$ for both cone and $\kt$ algorithms for ease of notation. For $\kt$, this parameter is sometimes referred to as $D$. We emphasize that having the same size $R$ for different algorithms does not in general guarantee the same sized jets.}  In the multijet configurations we consider the jets are separated by an angle larger than $R$, so that only the pairwise metric is relevant for describing the phase space constraints for particles in each jet.  For a two-particle (NLO) jet, the only phase space constraint imposed by this class of recombination algorithms is that the two particles be separated by an angle less than $R$:
\be
\textrm{$\kt$ jet: } \theta_{12} < R \, .
\ee
This defines a generic recombination algorithm suitable for our calculation.  We will denote this as the $\kt$-type algorithm.

The configurations with two widely separated energetic particles best distinguish cone-type jets from $\kt$-type jets at NLO.  For instance, the case where the two energetic particles are at opposite edges of a cone jet (at an angle $2R$ apart) is not a single $\kt$ jet.  However, it is important to note that these configurations will not be accurately described in this SCET calculation for $R \gg \lambda$, as such configurations are power suppressed in our description of jets.  Our concern is in accurately describing the configurations with narrow jets (small $\tau_a$), and not the wide angle configurations above.

Because jets are reliable degrees of freedom and provide a useful description of an event when they have large  energy, in the description of an event we impose a cut $\Lambda$ on the minimum energy of jets.  An $N$-jet event, therefore, is one where $N$ jets have energy larger than the cutoff $\Lambda$, with any number of jets having energy less than the cutoff.  In our calculation, we impose the same constraint: any jet with energy less than $\Lambda$ is not considered when we count the number of jets in the final state.  This imposes phase space cuts: for a gluon radiated outside of all jets in the event, that gluon is required to have energy $E_g < \Lambda$ to maintain the same number of jets in the event.  The proper division of phase space in calculating the jet and soft functions is a key part of the discussion below, and careful treatment of the phase space cuts is needed.

\subsection{Do Jet Algorithms Respect Factorization?}

The factorization theorem places specific requirements on the structure of jet algorithms used to describe events.  As in \eq{factorizationtheorem}, the factorization theorem divides the cross section into separately calculable hard, jet, and soft functions.  The hard function depends only on the configuration of jets, while the jet and soft functions describe the degrees of freedom in each jet in terms of the observable $\tau$.  While the soft function is global, the jet function depends only on the collinear degrees of freedom in a single jet.  The limited dependence of the hard and jet functions implies constraints on the jet algorithm.

Because jets are built from the long distance degrees of freedom arising from evolution of energetic partons to lower energies, the configuration of jets in an event depends on dynamics across all energy scales.  This naively breaks factorization in SCET, since the configuration of jets in the hard function would depend on dynamics at low energy in the soft function.  However, we can describe a jet algorithm that respects factorization, and in \sec{sec:power} we will parameterize the power corrections that arise from various algorithms.

The primary constraint on the jet algorithm in order to satisfy the factorization theorem is that the phase space cuts on the collinear particles in the jet are determined only by the collinear degrees of freedom.  This essentially ensures that the jet functions are independent of dynamics in the soft function. Correspondingly the soft function can only know about the jet directions and their color representations. The direction of the jet is naturally set by the collinear particles, as soft particles have energy parametrically lower than the collinear ones and change the jet direction by a power suppressed amount. The further restriction that the phase space cuts on the collinear degrees of freedom are independent of the soft degrees of freedom places a strong constraint on the action of the jet algorithm.  Cone algorithms already implement this constraint: the jet boundary (the cone) is determined by the location of the jet axis, which is the direction of the sum of all collinear particles up to a power correction.  Recombination algorithms, however, are constrained by the factorization theorem to operate in a specific way: all collinear particles must be recombined before soft particles.  As discussed in \sec{sec:power}, commonly used algorithms obey this constraint up to power corrections in the observable for measured jets.  Of particular note is the anti-kT algorithm, which exhibits behavior very close to what is required by the factorization theorem (similarly to the way cone algorithms behave).

\section{Factorization of Jet Shape Distributions in $e^+ e^-$ to $N$ Jets}
\label{sec:fact}

In this Section we formulate a factorization theorem for jet shape distributions in $e^+ e^-$ annihilation to $N$ jets. All the formal aspects we need to describe an $N$-jet cross section appear already in the 3-jet cross section, so we will give explicit details only for that case.  We will use the framework of Soft-Collinear Effective Theory (SCET), developed in \cite{Bauer:2000ew,Bauer:2000yr,Bauer:2001ct,Bauer:2001yt},  to formulate the factorization theorem. We begin with a basic review of the relevant aspects of the effective theory.

\subsection{Overview of SCET}
\label{ssec:SCET}

SCET is the effective field theory for QCD with all degrees of freedom integrated out, other than those traveling with large energy but small virtuality along a light-like trajectory $n$, and those with small, or soft, momenta in all components.  A particularly useful set of coordinates is light-cone coordinates, which uses light-like directions $n$ and $\bar{n}$, with $n^2 = \bar{n}^2 = 0$ and $n\cdot\bar{n} = 2$.  In Minkowski coordinates, we take $n = (1,0,0,1)$ and $\bn = (1,0,0,-1)$, corresponding to collinear particles moving in the $+z$ direction. A generic four-vector $p^{\mu}$ can be decomposed into components
\be
p^{\mu} = \bar{n}\cdot p\frac{n^{\mu}}{2} + n\cdot p\frac{\bar{n}^{\mu}}{2} + p_{\perp}^{\mu} \, .
\ee
In terms of these components, $p = (\bar n\cdot p, n\cdot p, p_\perp)$, collinear and soft momenta scale with some small parameter $\lambda$ as
\begin{equation}
p_n = E(1,\lambda^2,\lambda)\,,\quad p_s\sim E(\lambda^2,\lambda^2,\lambda^2)\,,
\end{equation}
where $E$ is a large energy scale, for example, the center-of-mass energy in an $e^+ e^-$ collision.  $\lambda$ is then the ratio of the typical transverse momentum of the constituents of the jet to the total jet energy. Quark and gluon fields in QCD are divided into collinear and soft effective theory fields with these respective momentum scalings:
\begin{equation}
\label{QCDsplit}
q(x) = q_n(x) + q_s(x)\,,\quad A^\mu(x) = A_n^\mu(x) + A_s^\mu(x)\,.
\end{equation}
We factor out a phase containing the largest components of the collinear momentum from the fields $q_n,A_n$. Defining the ``label'' momentum $\tilde p_n^\mu = \bar n\cdot\tilde p_n \frac{n^\mu}{2} + \tilde p_\perp^\mu$, where $\bar n\cdot \tilde p_n$ contains the $\mathcal{O}(1)$ part of the large light-cone component of the collinear momentum $p_n$, and $\tilde p_\perp$ the $\mathcal{O}(\lambda)$ transverse component, we can partition the collinear fields $q_n,A_n$ into their labeled components,
\begin{equation}
q_n(x) = \sum_{\tilde p\not = 0} e^{-i\tilde p\cdot x}q_{n,p}(x)\,,\quad A_{n}^\mu(x) = \sum_{\tilde p \not=0} e^{-i\tilde p\cdot x} A_{n,p}^\mu(x)\,.
\end{equation}
The sums are over a discrete set of $\mathcal{O}(1,\lambda)$ label momenta into which momentum space is partitioned. The bin $\tilde p = 0$ is omitted to avoid double-counting the soft mode in \eq{QCDsplit} \cite{Manohar:2006nz}. The labeled fields $q_{n,p},A_{n,p}$ now have spacetime fluctuations in $x$ which are conjugate to ``residual'' momenta $k$ of the order $E\lambda^2$, describing remaining fluctuations within each labeled momentum partition \cite{Bauer:2001ct,Manohar:2006nz}. It will be convenient to define label operators $\mathcal{P}^\mu = \bn\cdot \mathcal{P} n^\mu/2 + \mathcal{P}_\perp^\mu$  which pick out just the label components of momentum of a collinear field:
\begin{equation}
\mathcal{P}^\mu \phi_{n,p}(x) = \tilde p^\mu \phi_{n,p}(x)\,.
\end{equation}
Ordinary derivatives $\partial^\mu$ acting on effective theory fields $\phi_{n,p}(x)$ are of order $E\lambda^2$.

The final step to construct the effective theory fields is to isolate the two large components of the Dirac spinor $q_{n,p}$ for a fermion with lightlike momentum along $n$. The large components $\xi_{n,p}$ and the small $\Xi_{n,p}$ can be separated by the projections
\begin{equation}
\xi_{n,p} =  \frac{\nslash\bnslash}{4}q_{n,p}\,,\quad \Xi_{n,p} = \frac{\bnslash\nslash}{4}q_{n,p}\,,
\end{equation}
and we have $q_{n,p} = \xi_{n,p} + \Xi_{n,p}$. One can show, substituting these definitions into the QCD Lagrangian, that the fields $\Xi_{n,p}$ have an effective mass of order $E$ and can be integrated out of the theory. The effective theory Lagrangian at leading order in $\lambda$ is \cite{Bauer:2000yr,Bauer:2001ct,Bauer:2001yt}
\begin{equation}
\label{SCETLag}
\mathcal{L}_{\text{SCET}} = \mathcal{L}_{\xi} + \mathcal{L}_{A_n} + \mathcal{L}_s\,,
\end{equation}
where the collinear quark Lagrangian $\mathcal{L}_\xi$ is
\begin{equation}
\label{xiLag}
\mathcal{L}_\xi = \bar\xi_{n} (x) \left[in\cdot D + i\Dslash_\perp^{\; c} W_n(x) \frac{1}{i\bar n\cdot \mathcal{P}} W_n^\dag(x) i\Dslash_\perp^{\; c}\right] \frac{\bnslash}{2}\xi_n(x)\,,
\end{equation}
where $W_n$ is the Wilson line of collinear gluons,
\begin{equation}
\label{Wdef}
W_n(x) = \sum_{\text{perms}} \exp\left[-g\frac{1}{\bar n\cdot{\mathcal{P}}}\bar n\cdot A_n(x)\right]\, ;
\end{equation}
the collinear gluon Lagrangian $\mathcal{L}_{A_n}$ is
\begin{equation}
\label{AnLag}
\begin{split}
\mathcal{L}_{A_n} &= \frac{1}{2g^2}\Tr\biggl\{\Bigl[ i\mathcal{D}^\mu + gA_n^\mu, i\mathcal{D}^\nu + gA_n^\nu \Bigr]\biggr\}^2  \\
& \quad + 2\Tr\biggl\{\bar c_n \Bigl[i\mathcal{D}_\mu, \Bigl[i\mathcal{D}^\mu + g A_n^\mu,c_n\Bigr]\Bigr]\biggl\} + \frac{1}{\alpha} \Tr\biggl\{ \Bigl[i\mathcal{D}_\mu , A_n^\mu\Bigr]\biggr\}\,,
\end{split}
\end{equation}
where $c_n$ is the collinear ghost field and $\alpha$ the gauge-fixing parameter; and the soft Lagrangian $\mathcal{L}_s$ is
\begin{equation}
\mathcal{L}_s = \bar q_s i\Dslash_s q_s(x) - \frac{1}{2}\Tr G_s^{\mu\nu} G_{s\mu\nu}(x)\,,
\end{equation}
which is identical to the form of the full QCD Lagrangian (the usual gauge-fixing terms are implicit). In the collinear Lagrangians, we have defined several covariant derivative operators,
\begin{equation}
D^\mu = \partial^\mu - ig A_n^\mu - ig A_s^\mu\,,\quad iD_c^{\mu} = \mathcal{P}^\mu + g A_n^\mu \,,\quad i\mathcal{D}^\mu  = \mathcal{P}^\mu + in\mcdot D\frac{\bar n^\mu}{2}\,.
\end{equation}
In addition, there is an implicit sum over the label momenta of each collinear field and the requirement that the total label momentum of each term in the Lagrangian be zero.

Note the soft quarks do not couple to collinear particles at leading order in $\lambda$. Meanwhile, the coupling of the soft gluon field to a collinear field is in the component $n\mcdot A_s$ only, according to \eqs{xiLag}{AnLag}, which makes possible the decoupling of  such interactions through a field redefinition of the soft gluon field given in \cite{Bauer:2001yt}. We will utilize this soft-collinear decoupling to simplify the proof of factorization below.

The SCET Lagrangian  \eq{SCETLag} may be extended to include collinear particles in more than one direction \cite{Bauer:2002nz}. One adds multiple copies of the collinear quark and gluon Lagrangians \eqs{xiLag}{AnLag} together. The collinear fields in each direction $n_i$ constitute their own independent set of quark and gluon fields, and are governed in principle by different expansion parameters $\lambda$ associated with the transverse momentum of each jet, set either by the angular cut $R$ in the jet algorithm or by the measured value of the jet shape $\tau_a$. Each collinear sector may be paired with its own associated soft field $A_s$ with momentum of order $E\lambda^2$ with the appropriate $\lambda$. For the purposes of keeping the notation tractable while proving the factorization theorem in this section, we will for simplicity take all $\lambda$'s to be the same, with a single soft gluon field $A_s$ coupling to collinear modes in all sectors. In \ssec{refactorization}, we will discuss how to ``refactorize'' the soft function further into separate soft functions each depending only on one of the various possible soft scales.

The effective theory containing $N$ collinear sectors and the soft sector is appropriate to describe QCD processes with strongly-interacting particles collimated in $N$ well-separated directions. Thus, in addition to the power counting in the small parameter $\lambda$ within each sector, guaranteeing that the particles in each direction are well collimated, we will find in calculating an $N$-jet cross section the need for another parameter that guarantees that the different directions $n_i$ are well separated. This latter condition requires $t \gg 1$, where $t$ is defined in \eq{tdef}.\footnote{This condition is a consequence of our insistence on using operators with exactly $N$ directions to create the final state. We could move away from the large-$t$ limit and account for corrections to it by using a basis of operators with arbitrary numbers of jets and properly accounting for the regions of overlap between an $N$ jet operator and $(N\pm 1)$-jet operators. This is outside the scope of the present work, where we limit ourselves to kinematics well described by an $N$-jet operator, and thus, limit ourselves to the large-$t$ limit.}

\subsection{Jet Shape Distribution in $e^+ e^-\to\text{3 Jets}$}
\label{ssec:jetshapedist3jets}

Consider a 3-jet cross section differential in the jet 3-momenta  $\vect{P}_{1,2,3}$, where we measure the shape $\tau_a^1$ of one of the jets, which we will call jet  1. The full theory cross section for $e^+ e^- \to \gamma^*\to\text{3 jets}$ at center-of-mass energy $Q$ is
\begin{equation}
\begin{split}
\frac{d\sigma}{d\tau_a^1 d^3 \vect{P}_{1,2,3}} &= \frac{1}{2Q^2} \sum_X\abs{\bra{X} j^\mu(0)\ket{0} L_\mu}^2 (2\pi)^4\delta^4(Q - p_X) \delta_{N(\mathcal{J}(X)) , 3} \\
&\quad\times\delta\bigl(\tau_a^1 - \tau_a(\text{jet } 1)\bigr)
\prod_{j=1}^3 \delta^3\bigl(\vect{P}_j-\vect{P}(\text{jet }j)\bigr)\,,
\end{split}
\end{equation}
where the $\mathcal{J}(X)$ is the jet algorithm acting on final state $X$, and $N(\mathcal{J}(X))$ is the number of jets identified by the algorithm \cite{Bauer:2008jx}. $\vect{P}(\text{jet }j)$  is the 3-momentum of jet $j$, and is also a function of the output of the jet algorithm $\mathcal{J}(X)$. $L_\mu$ is the leptonic part of the amplitude for $e^+ e^- \to\gamma^*\to q\bar q g$. The current $j^\mu$ is
\begin{equation}
\label{QCDcurrent}
j^\mu = \sum_{a,f}\bar q^a\gamma^\mu q^a,
\end{equation}
summing over colors $a$ and flavors $f$.

When the three jet directions are well separated, we can match the QCD current $j^\mu(x)$ onto a basis of three-jet operators in SCET \cite{Bauer:2006qp,Marcantonini:2008qn}. We build these operators from quark jet fields $\chi_n$, related to collinear quark fields $\xi_n$ by $\chi_n = W_n^\dag\xi_n$, where $W_n$ is given by \eq{Wdef},
and a gluon jet field ${B}_n^\perp$ related to gluons $A_n$ by
\begin{equation}
{B}_{n}^{\perp} = \frac{1}{g}W_{n}^\dag({\mathcal{P}}_\perp + A_{n}^{\perp})W_n\,.
\end{equation}
The matching relation is
\begin{equation}
\label{3jetmatching}
j^\mu(x) = \sum_{n_1n_2n_3}\sum_{\tilde p_1\tilde p_2\tilde p_3}e^{i(\tilde p_1-\tilde p_2 + \tilde p_3)\cdot x}C^{\mu}_{\alpha\beta\nu}(n_1,\tilde p_1;n_2, \tilde p_2;n_3, \tilde p_3)\bar\chi_{n_1,p_1}^\alpha (g{B}^{\perp\nu}_{n_3,p_3})\chi_{n_2,p_2}^\beta(x)\,,
\end{equation}
with sums over Dirac spinor indices $\alpha,\beta$ and Lorentz index $\nu$, and over label directions $n_{1,2,3}$ and label momenta $\tilde p_{1,2,3}$. Sums over colors and flavors are implied. We have chosen to produce a  quark  in direction $n_1$, antiquark in $n_2$, and gluon in $n_3$. The matching coefficients $C^\mu_{\alpha\beta\nu}$ are found by equating QCD matrix elements of $j^\mu$ to SCET matrix elements of the right-hand side of \eq{3jetmatching}. These coefficients have been found at tree level in \cite{Marcantonini:2008qn}. The number of independent Dirac and Lorentz structures that can actually appear with nonzero coefficients is considerably smaller than suggested by \eq{3jetmatching} due to symmetries.   We will keep the form of these coefficients general to make extension to $N$ jets transparent, which would require the introduction of a basis of $N$ jet fields in \eq{3jetmatching}, with specified numbers of quark, antiquark, and gluon fields. We will not write the details for an $N$-jet cross section here, but the procedures are obvious extensions of all the steps in factorizing the 3-jet cross section below.

As a final step before factorization, we redefine the collinear fields to decouple collinear-soft interactions in the Lagrangian \cite{Bauer:2001yt}:
\begin{subequations}
\label{BPS}
\begin{align}
\chi_{n}(x) &= Y_{n}^\dag(x)\chi_{n}^{(0)}(x) \\
\bar\chi_{n}(x) &= \bar\chi_{n}^{(0)}(x){Y}_{n}(x) \\
A_{n}(x) &= \mathcal{Y}_n(x) A_{n}^{(0)}(x)\,,
\end{align}
\end{subequations}
where $Y_n$ is a Wilson line of soft gluons along the light-cone direction $n$,
\begin{equation}
\label{Yndef}
Y_n(x) = P\exp\left[ig\int_0^\infty ds\,n\cdot A_s(ns + x)\right]\,,
\end{equation}
with $A_s$ in the fundamental representation.\footnote{The path choice (0 to $\infty$) in \eq{Yndef} is convenient for outgoing particles. The physical cross section is independent of whether the path goes to $\pm \infty$ if the transformation of the external states $X$ is also taken into account \cite{Arnesen:2005nk}.
} $\mathcal{Y}_n$ is similar but in the adjoint representation. The new fields $\chi_{n}^{(0)},A_n^{(0)}$ do not have interactions with soft fields in the SCET Lagrangian at leading order in $\lambda$. Henceforth, we use only these redefined fields, but for simplicity drop the $(0)$ superscripts.

The cross section in SCET can now be written,
\begin{align}
\label{SCETcs1}
\frac{d\sigma}{d\tau_a^1 d^3\vect{P}_{1,2,3}} &= \frac{N_F L^2}{6Q^2} \sum_X\delta_{N(\mathcal{J}(X)),3} \delta(\tau_a^1 - \tau_a(\text{jet }1)) \prod_{j=1}^3\delta^3(\vect{P}_j -\vect{ P}(\text{jet }j)) \nn \\
&\quad \times \sum_{n_{1,2,3}}\sum_{\tilde p_{1,2,3}} \int d^4 x\, e^{i(Q-\tilde p_1+\tilde p_2 - \tilde p_3)\cdot x}C^\mu_{\alpha\beta\nu}(n_{1,2,3};\tilde p_{1,2,3})C_{\mu\gamma\delta\rho}^*(n_{1,2,3};\tilde p_{1,2,3}) \nn \\
&\quad\times\bra{0}\bar T\Bigl\{ \bar \chi_{n_2,p_2}^{a\delta}  Y_{n_2}^{ab} \mathcal{Y}_{n_3}^{AB} (g{B}_{n_3,p_3}^{\perp\rho B}) T^A_{bc} Y_{n_1}^{\dag cd}\chi^{d\gamma}_{n_1,p_1}(x)\Bigr\}\ket{X} \nn \\
&\quad \times \bra{X}T\Bigl\{ \bar\chi_{n_1,p_1}^{e \alpha} Y_{n_1}^{ef} \mathcal{Y}_{n_3}^{CD}(g{B}_{n_3,p_3}^{\perp\nu D})T^C_{fg} Y_{n_2}^{\dag gh}\chi_{n_2,p_2}^{h\beta}(0)\Bigr\} \ket{0}\,.
\end{align}
To proceed to factorize this cross section, it is convenient to rewrite the remaining delta functions that depend on the final state $X$ in terms of operators acting on $X$. Those quantities depending on the jet algorithm $\mathcal{J}$ can be rewritten in terms of an operator containing the momentum flow operator,
\begin{equation}
\mathcal{E}_\mu(\vect{n}) = \lim_{R\to\infty}\int_0^\infty dt\,n_i T_{\mu i}(t,R\vect{n}) \,,
\end{equation}
where $T_{\mu\nu}$ is the energy-momentum tensor, evaluated at time $t$ and position $R\vect{n}$. The operator $\mathcal{E}_\mu(\vect{n})$ measures the flow of four-momentum $P_\mu$ in the direction $\vect{n}$ (cf. \cite{Bauer:2008dt,Korchemsky:1997sy,Belitsky:2001ij}), and the jet algorithm $\mathcal{J}$ can be written as an operator $\hat{\mathcal{J}}$ acting on the momentum flow in all directions \cite{Bauer:2008jx}.  Correspondingly we can define an operator for the 3-momentum of the jet, $\hat{\vect{P}}(J_j(\hat{\mathcal{J}}))$. In addition, the event shape $\tau_a(\text{jet 1})$ can also be expressed as an operator $\hat\tau_a(J_1(\hat{\mathcal{J}}))$, built from the momentum flow operator, acting on the state $\ket{X}$ (cf. \cite{Bauer:2008dt}):
\begin{equation}
\hat\tau_a(J_1(\hat{\mathcal{J}})) = \int d\eta\,e^{-\eta(1-a)}\mathcal{E}_T(\eta) \Theta(\eta - \eta_{\text{min}}(J_1(\hat{\mathcal{J}})))\,.
\end{equation}
The operator is constructed to count only particles actually entering the jet in direction $n_1$ determined by the action of the jet algorithm (for simplicity we will suppress the argument $J_1(\hat{\mathcal{J}})$ of $\hat\tau_a$ in the following, but add a superscript for the jet number).
Using these operators, we can eliminate the $X$ dependence in the delta functions in \eq{SCETcs1} and perform the sum over states $X$, obtaining
\begin{align}
\label{SCETcs1.5}
\frac{d\sigma}{d\tau_a^1 d^3\vect{P}_{1,2,3}} &= \frac{L^2}{6Q^2}\sum_{n_{1,2,3}}\sum_{\tilde p_{1,2,3}}\int d^4x \, e^{i(Q-\tilde p_1 + \tilde p_2 - \tilde p_3)\cdot x}C^\mu_{\alpha\beta\nu}(n_{1,2,3};\tilde p_{1,2,3})C_{\mu\gamma\delta\rho}^*(n_{1,2,3};\tilde p_{1,2,3}) \nn \\
&\qquad\times\bra{0}\bar T\Bigl\{ \bar \chi_{n_2,p_2}^{a\delta}  Y_{n_2}^{ab} \mathcal{Y}_{n_3}^{AB} (g{B}_{n_3,p_3}^{\perp\rho B}) T^A_{bc} Y_{n_1}^{\dag cd}\chi^{d\gamma}_{n_1,p_1}(x)\Bigr\} \nn \\
&\qquad\times \delta_{N(\hat{\mathcal{J}}) , 3} \delta(\tau_a^1 - \hat \tau_a^1)\prod_{j=1}^3 \delta^3(\vect{P}_j - \hat{\vect{P}}(J_j(\hat{\mathcal{J}}))) \nn \\
&\qquad \times T\Bigl\{ \bar\chi_{n_1,p_1}^{e \alpha} Y_{n_1}^{ef} \mathcal{Y}_{n_3}^{CD}(g{B}_{n_3,p_3}^{\perp\nu D})T^C_{fg} Y_{n_2}^{\dag gh}\chi_{n_2,p_2}^{h\beta}(0)\Bigr\} \ket{0}\,.
\end{align}
The matrix element can be calculated as the sum over cuts of time-ordered Feynman graphs, with the delta function operators inside the matrix element enforcing phase space restrictions from the jet algorithm and jet shape measurement on the final state created by the cut.

The operators $\hat\tau_a$ and $\hat{\mathcal{J}}$ depend linearly on the energy-momentum tensor, which itself splits linearly in SCET into separate collinear and soft pieces,
\begin{equation}
T_{\mu\nu} = \sum_{i} T_{\mu\nu}^{n_i} + T_{\mu\nu}^s\,,
\end{equation}
which will aid us to factorize the full matrix element in \eq{SCETcs1.5} into separate collinear and soft matrix elements. To achieve this factorization, however, we must make some more approximations:
\begin{enumerate}

\item The contribution of soft particles and residual collinear momenta to the momentum $\vect{P}(\text{jet }j)$ of each jet can be neglected, and the jet momentum  is just given by the label momentum $\tilde p_j$ of the collinear state $\ket{X_j}$. Thus the energy and jet axis of each jet is approximated to be that of the parent collinear parton initiating the jet.  In particular, the squared mass of the jet is order $\lambda^2$ compared to its energy.  So in this approximation we take the jet energy to be equal to the magnitude of its 3-momentum.  On the other hand, we keep the leading non-zero contribution to the angularity even though it is also of order $\lambda^2$.
These approximations also require that we treat the energy of any particles outside all of the jets, and thus the cutoff $\Lambda$,  as a soft or residual energy.

\item The Kronecker delta restricting the total number of jets to 3 can be factored into three separate Kronecker deltas restricting the number of jets in each collinear direction $n_i$ to 1, and one Kronecker delta restricting the soft particles not to create an additional jet. This approximation requires the separation between jets to be much larger than the size of any individual jet so that different jets do not overlap. Factoring the restriction on the number of jets in this way is one reason that the parameter $t_{ij}$ in \eq{tdef} is required to be large.

\end{enumerate}
We describe to what extent the algorithms we consider actually satisfy these two approximations in \sec{sec:power}.  For now we assume these approximations and facts hold, which allows us to factor the cross section \eq{SCETcs1},
\begin{equation}
\label{SCETcs2}
\begin{split}
&\frac{d\sigma}{d\tau_a^1 dE_{1,2,3} d^2\Omega_{1,2,3}} = \frac{L^2}{6Q^2}\sum_{n_{1,2,3}}\sum_{\omega_{1,2,3}}  C^\mu_{\alpha\beta\nu}(n_{1,2,3}; \omega_{1,2,3})C_{\mu\gamma\delta\rho}^*(n_{1,2,3}; \omega_{1,2,3}) \\
&\qquad \times\int d^4 x\,e^{i(Q-\omega_1 n_1/2+ \omega_2 n_2/2 - \omega_3n_3/2)\cdot x} \int d\tau_J d\tau_S\delta(\tau_a^1 - \tau_J - \tau_S) \\
&\qquad \times\bra{0}\chi_{n_1,\omega_1}^{f\gamma}(x)  \delta_{N(\hat{\mathcal{J}}),1}\delta(\tau_J - \hat \tau_a^{n_1})  \bar\chi_{n_1, \omega_1}^{e\alpha}(0)\ket{0} \delta\left(E_1 - \frac{\omega_1}{2}\right) \delta^2(\Omega_1 -  \vect{n}_1) \\
&\qquad\times\bra{0}\bar\chi_{n_2,-\omega_2}^{a\delta}(x)\delta_{N(\hat{\mathcal{J}}),1} \chi_{n_2,-\omega_2}^{h\beta}(0)\ket{0} \delta\left(E_2 - \frac{\omega_2}{2}\right) \delta^2(\Omega_2 -  \vect{n}_2)\\
&\qquad\times\bra{0}(g{B}_{n_3,\omega_3}^{\perp\rho A})(x)  \delta_{N(\hat{\mathcal{J}}),1} (g{B}_{n_3,\omega_3}^{\perp\nu B}(0)\ket{0} \delta\left(E_3 - \frac{\omega_3}{2}\right) \delta^2(\Omega_3 -  \vect{n}_3)\\
&\qquad\times\bra{0}\overline Y_{n_2}^{\dag ab} Y_{n_3}^{\dag bc}T^A_{cd} \overline Y_{n_3}^{\dag de}Y_{n_1}^{\dag ef}(x)   \delta_{N(\hat{\mathcal{J}}),0}  \delta(\tau_S - \hat\tau_a^s)Y_{n_1}^{gh}\overline Y_{n_3}^{hi} T^B_{ij}Y_{n_3}^{jk}\overline Y_{n_2}^{kl} (0)\ket{0}
\end{split}
\end{equation}
We have rewritten the cross section to be differential in $E_i$ (the magnitude of $\vect{P}_i$) and $\Omega_i$ (the direction of $\vect{P}_i$). In the sum over label directions, $\vect{n}_i$ can be chosen to align with $\vect{P}_i$ such that $\tilde{{p}}_i^\perp= 0$. In \eq{SCETcs2} we have written the label momentum as $\omega_i \equiv \bar n_i\cdot\tilde p_i$.
In \eq{angularitydefn} we approximate the jet axis by this $\vect{n}_i$ and the jet energy by $\bar n_i\cdot \tilde p_i/2$, so that they do not depend on soft momenta at all. The operators $\hat\tau_a^{n_1}$ and $\hat\tau_a^s$ are defined to count only particles inside the measured jet identified by the algorithm.

In the soft matrix element in \eq{SCETcs2}, we have introduced the soft Wilson line $\overline Y_n$ in the antifundamental representation to remove the  time- and anti-time-ordering operators $T,\bar T$ in \eq{SCETcs1} \cite{Bauer:2003di}, and related Wilson lines $\mathcal{Y}_n$ in the adjoint representation to those in the fundamental representation by \cite{Bauer:2001yt}
\begin{equation}
\mathcal{Y}_n^{AB} T^B = Y_n^\dag T^A Y_n\,.
\end{equation}

Defining the jet functions by the relations
\begin{subequations}
\label{jetfuncdefs}
\be
\int \!\frac{d^4 k_1}{(2\pi)^4}e^{-ik_1\cdot x}J_{n_1,\omega_1}(\tau_J,n_1\cdot k_1) \left(\!\frac{\nslash_1}{2}\!\right)_{\!\!\gamma\alpha}\!\!\delta^{ef} = \bra{0}\chi_{n_1,\omega_1}^{f\gamma}(x)  \delta_{N(\hat{\mathcal{J}}) , 1} \delta(\tau_J - \hat\tau_a^{n_1}) \bar\chi_{n_1, \omega_1}^{e\alpha}(0)\ket{0}
\ee
\be
\int \frac{d^4 k_2}{(2\pi)^4}e^{-ik_2\cdot x}J_{n_2, \omega_2}(n_2\mcdot k_2) \left(\frac{\nslash_2}{2}\right)_{\beta\delta}\delta^{ah} = \bra{0}\bar\chi_{n_2, \omega_2}^{a\delta}(x) \delta_{N(\hat{\mathcal{J}}) ,1}\chi_{n_2,\omega_2}^{h\beta}(0)\ket{0}
\ee
\be
\int \frac{d^4 k_3}{(2\pi)^4}e^{-ik_3\cdot x}J_{n_3, \omega_3}(n_3\mcdot k_3) g_\perp^{\rho\nu}\delta^{AB} = -\omega_3\bra{0}(g{B}_{n_3, \omega_3}^{\perp\rho A}) \delta_{N(\hat{\mathcal{J}}) ,1}(g{B}_{n_3,\omega_3}^{\perp\nu B})\ket{0}\,,
\ee\end{subequations}
and the soft function by
\begin{equation}
\label{softfuncdef}
\begin{split}
\int\frac{d^4 r}{(2\pi)^4}e^{-ir\cdot x} S(\tau_s,r) =  \frac{1}{N_C C_F}\Tr & \bra{0}\overline Y_{n_2}^{\dag }\overline Y_{n_3}^\dag T^A Y_{n_3}^{\dag}Y_{n_1}^{\dag}(x) \delta_{N(\hat{\mathcal{J}}),0} \delta(\tau_S - \hat\tau_a^s)\\
&\times Y_{n_1}Y_{n_3} T^B Y_{n_3}\overline Y_{n_2} (0)\ket{0}
\end{split}
\end{equation}
we can express the  cross section \eq{SCETcs2} as
\begin{align}
\label{SCETcs3}
\frac{d\sigma}{d\tau_a^1 dE_{1,2,3}d^2\Omega_{1,2,3}} &= \frac{L^2 N_F N_C C_F}{6Q^2} \sum_{n_{1,2,3}}\sum_{\omega_{1,2,3}} \int d^4 x\,e^{i [Q - (\omega_1 n_1 - \omega_2 n_2 + \omega_3 n_3)/2]\cdot x} \\
&\quad\times C^\mu_{\alpha\beta\nu}(n_{1,2,3};\omega_{1,2,3})C^*_{\mu\gamma\delta\rho}(n_{1,2,3},\omega_{1,2,3}) \left(\frac{\nslash_1}{2}\right)_{\gamma\alpha}\left(\frac{\nslash_2}{2}\right)_{\beta\delta}g_\perp^{\nu\rho}\nn  \\
&\quad\times \int d\tau_J d\tau_S\delta(\tau_a^1 - \tau_J - \tau_S) \prod_{i=1}^3\delta\left(E_i - \frac{\omega_i}{2}\right)\delta^2(\Omega_i - \vect{n_i}) \nn \\
&\quad\times \int\frac{d^4 k_1}{(2\pi)^4}e^{-ik_1\cdot x}\int\frac{d^4 k_2}{(2\pi)^4}e^{-ik_2\cdot x}\int\frac{d^4 k_3}{(2\pi)^4}e^{-ik_3\cdot x}\int \frac{d^4 r}{(2\pi)^4}e^{-ir\cdot x} \nn \\
&\quad\times J_{n_1,\omega_1}(\tau_J,n_1\cdot k_1)J_{n_2,-\omega_2}(n_2\cdot k_2) J_{n_3,\omega_3}(n_3\cdot k_3) S(\tau_S,r)  \,, \nn
\end{align}
where now $P_i = E_i(1,\vect{n}_i)$. The quark and antiquark jet functions are now for a single flavor, and we have summed over $N_F$ flavors to obtain the factor in front.  The jet functions depend only on the smallest component of momentum $n_i\cdot k_i$ in each collinear direction.
The residual and soft momenta appearing in the exponentials can be reabsorbed into the label momenta by a series of reparameterizations of the label momenta and directions, under which the SCET Lagrangian is invariant \cite{Manohar:2002fd}. The three-jet operator \eq{3jetmatching} will receive corrections of order $\lambda^2$ (which we can drop) under the reparameterizations we perform below, or can be constructed from the outset to be explicitly reparameterization invariant (RPI) \cite{Marcantonini:2008qn}.

First, collect the residual and soft momenta together by defining $K = k_1 + k_2 + k_3 + r$. We can decompose $x$ in $n_1$ light-cone coordinates, so
\begin{equation}
e^{-iK\cdot x} = e^{-i(\bn_1\cdot K\, n_1\cdot x/2 + n_1\cdot K \, \bn_1\cdot x/2 + K_{\perp1}\cdot x_{\perp 1})}\,.
\end{equation}
Performing a type-A transformation (in the language of \cite{Manohar:2002fd}) on the label momentum $\tilde p_1  = \omega_1 n_1/2$,
\begin{equation}
\omega_1 \to \omega_1 + \bar n_1\cdot K\,,
\end{equation}
and a type-IB transformation on the vector $n_1$ itself,
\begin{equation}
n_1 \to n_1 + \Delta^\perp\,,\quad \Delta^\perp = - \frac{2}{\omega_1} K^\perp\,,
\end{equation}
shifts the label momentum on the jet function 1 by $\omega_1 n_1/2 \to (\omega_1 + \bar n_1\cdot  K) n_1 /2+ K_{\perp 1}$. The summation variables $n_1,\omega_1$ can then be shifted to eliminate $\bar n_1\cdot K$ and $K_{\perp 1}$ from the exponentials entirely. We drop all corrections suppressed by $\lambda^2$ due to these shifts.

It remains to absorb the $n_1\cdot K$ component of residual and soft momentum, appearing in the exponential factor $e^{-i n_1\cdot K\, \bn_1\cdot x/2}$. This cannot be achieved by RPI transformations in the $n_1$ sector since this momentum is purely residual---there is no label momentum in this direction. However, in a multijet cross section, $\bn_1$ can be written as a linear combination of $n_2,n_3$, and, say, $n_{\perp 2}$ (a unit vector transverse to $n_2,\bn_2$), so that RPI transformations on $\omega_2,\omega_3$ and $n_2$ similar to those above can absorb $n_1\cdot K$ into the label momenta also. Then, the $x$-dependent residual and soft exponentials all disappear, and we can combine the nine superfluous $\bn_i\cdot k_i$ and $k_{\perp i}$ integrals with the nine discrete sums over label directions and momenta to give continuous integrals over total momenta. Performing these with the remaining energy and direction delta functions in \eq{SCETcs3} and the $x$ integral with the remaining exponential gives the momentum conservation delta function $\delta^4(Q - E_1 n_1 - E_2 n_2 - E_3n_3)$.

The resulting cross section \eq{SCETcs3} takes the form
\begin{equation}
\label{3jetfinal}
\begin{split}
\frac{d\sigma}{d\tau_a^1 dE_{1,2,3} d^2\vect{n}_{1,2,3}} &= \frac{d\sigma^{(0)}}{dE_{1,2,3}d^2\vect{n}_{1,2,3}} H(n_{1,2,3};E_{1,2,3})\int d\tau_J d\tau_S\delta(\tau_a^1 - \tau_J - \tau_S) \\
&\quad \times\int\frac{dn_1\mcdot k_1}{2\pi} \int\frac{dn_2\mcdot k_2}{2\pi}\int\frac{dn_3\mcdot k_3}{2\pi} \int\frac{d^4 r}{(2\pi)^4}\\
&\quad\times J_{n_1,2E_1}(\tau_J,n_1\cdot k_1)J_{n_2,2E_2}(n_2\cdot k_2)J_{n_3,2E_3}(n_3\cdot k_3)S(\tau_S,r)\,,
\end{split}
\end{equation}
where we used that the matching coefficients $C^\mu_{\alpha\beta\nu}(n_i;\tilde p_i)$ are such that, by construction, the right-hand side at tree-level is simply the Born cross section (denoted by $\sigma^{(0)}$) for $e^+ e^-\to q\bar q g$ times $\delta(\tau_a^1)$. The hard function $H = 1+\mathcal{O}(\alpha_s)$ is determined by calculating the matching coefficients $C$ order-by-order in perturbation theory.

The above may be easily modified to describe the antiquark or gluon jet angularities, by moving the appropriate delta function $\delta(\tau_a^i - \tau_a(\text{jet } i))$ into the $J_2$ or $J_3$ jet functions. In addition, we may choose from among various jet algorithms. The choice determines what $\Theta$-function restrictions must be inserted into the final state phase space integrations created by cutting  the jet and soft diagrams to determine which particles make it  into the jet. As long as the algorithm is such that the approximations enumerated above hold, it will not violate factorization of the jet shape cross section. We will discuss factorization and its potential breakdown in the context of particular jet algorithms in more detail in \sec{sec:power}.

Another check of the validity of the factorization theorem is that the factorized jet and soft functions be separately IR safe, which is a stronger condition than the full cross section being IR safe. If the observable \cite{Hornig:2009vb,Hornig:2009kv} or algorithm \cite{Cheung:2009sg} too strongly weights final states with narrow jets whose invariant masses are the same as the virtuality of soft particles, then the jet and soft functions for the observable in standard SCET with dimensional regularization become IR divergent.
When this occurs it does not necessarily mean that factorization is not possible; but at least not in the standard form derived from the version of SCET we utilized above. It could be, for example, that a scheme to  further separate modes by defining the theory with an explicit cutoff  \cite{Cheung:2009sg} or by factorizing modes by rapidity instead of virtuality \cite{Manohar:2006nz} can restore a version of the factorization theorem. We leave an explicit study of which algorithms and observables give IR safe jet and soft functions in SCET in dimensional regularization for a separate publication. However, we note here that the algorithms and observables ($\tau_a$ for $a<1$) that we consider in this paper, at least at NLO, do give rise to IR safe jet and soft functions.

\subsection{Jet Shapes in $e^+ e^-\to N\text{ jets}$}
 \label{ssec:jetshapedistNjets}

To generalize the result \eq{3jetfinal} to an arbitary number $N$ of jets, we simply add more quark and gluon jet fields to the operator matching in \eq{3jetmatching}, resulting in the corresponding number of additional quark and gluon jet functions in \eq{3jetfinal}, along with a hard coefficient and a soft function for $N$ jet directions. Consider an event with $2N_q$ quark and antiquark jets and $N_g$ gluon jets, where $2N_q+N_g = N$. Furthermore, we can choose to measure the angularity shape for any number of these jets. Achieving a factorization theorem that remains consistent for any of these combinations is a nontrivial task and thus a powerful test of the validity of the effective theory.

For an $N = 2N_q + N_g$ jet event, we generalize the matching of the QCD current in \eq{3jetmatching} to:
\begin{equation}
\label{Njetmatching}
\begin{split}
j^\mu(x) = \sum_{n_1\cdots n_N}\sum_{\tilde p_1\cdots \tilde p_N} & e^{i(\tilde p_1 + \cdots + \tilde p_N)\cdot x} C^{\mu a_1\cdots a_{N_q} b_1\cdots b_{N_q} A_1\cdots A_{N_g}}_{\alpha_1\cdots \alpha_{N_q}\beta_1\cdots \beta_{N_q}\nu_1\cdots \nu_{N_g}}(n_1,\tilde p_1;\cdots;n_N,\tilde p_N) \\
& \times\prod_{i=1}^{N_q}\bar\chi_{n_i,p_i}^{\alpha_i a_i}(x) \prod_{j=1}^{N_q}\chi_{n_j,-\tilde p_j}^{\beta_j b_j}(x)\prod_{k=1}^{N_g}(g B_{n_k,-\tilde p_k}^{\perp\nu_k A_k})(x)\,,
\end{split}
\end{equation}
with implicit sums over Dirac spinor indices $\alpha_i,\beta_j$, Lorentz indices $\nu_k$, (anti-)fundamental color indices $a_i,b_j$, and adjoint color indices $A_k$.
The $N$ jet cross section differential in $M$ jet shapes, with $M<N$, factors in SCET into the form
\begin{align}
\label{Njetcs}
\frac{d\sigma(E_1,\vect{n_1};\cdots;E_N,\vect{n_N})}{d\tau_{a_1}^1\cdots d\tau_{a_M}^M} &= \sigma^{(0)}(E_1,\vect{n}_1;\cdots;E_N,\vect{n}_N)H^{a_i b_j A_k}(n_1,E_1;\cdots n_N,E_N) \nn \\
&\quad\times \prod_{l=1}^M\int d\tau_J^l d\tau_S^l \delta(\tau_a^{l} - \tau_J^{l} - \tau_S^{l}) \int \frac{dn_1\cdot k_1}{2\pi}\cdots \int \frac{dn_N\cdot k_N}{2\pi} \nn \\
&\quad\times J_{n_1,2E_1}^{f_1}(\tau_J^1;n_1\cdot k_1) \cdots J_{n_M,2E_M}^{f_M}(\tau_J^M;n_M\cdot k_M) \nn \\
&\quad\times J_{n_{M+1}, 2E_{M+1}}^{f_{M+1}}(n_{M+1}\cdot k_{M+1})\cdots J_{n_N,2E_N}^{f_N}(n_N\cdot k_{N}) \nn \\
&\quad\times \int\frac{d^4 r}{(2\pi)^4} S^{a_i b_j A_k}(\tau_S^1,\dots,\tau_S^M;r)\,,
\end{align}
where $\sigma^{(0)}$ is the Born cross section for $e^+ e^-$ to $N_q$ quarks, $N_q$ antiquarks, and $N_g$ gluons; the color indices on the hard and soft functions $H$ and $S$ allow for color mixing; and $f_i$ is the flavor of each jet function (quark, antiquark, or gluon). $H$ itself is determined by calculating the matching coefficients $C$ in \eq{Njetmatching}. The jet functions have the same definitions given in \eq{jetfuncdefs}, and the soft function is given by the appropriate generalization of \eq{softfuncdef} with Wilson lines in the directions and color representations corresponding to the choice of fields in \eq{Njetmatching}. We rearrange the order of flavor and color indices in the hard and soft functions to agree with the choices of flavor indices on the jet functions.

\subsection{Do Jet Algorithms Induce Large Power Corrections to Factorization?}
\label{sec:power}

In this section we explore when power corrections to the factorization theorem above become large, in particular those that are due to the action of jet algorithms.
We will argue that power corrections to jet angularities induced by the commonly-used cone and recombination algorithms remain suppressed as long as $R$ is sufficiently large. In particular, we need in general that $R$ satisfies $ R  \gtrsim\lambda$ and, for the case of the $\kt$ algorithm, we require $R \gg \lambda$.\footnote{We noted earlier that there may be different $\lambda$'s for the SCET modes describing different jets. For measured jets, $\lambda^2\sim \tau_a$, while for unmeasured jets, $\lambda\sim \tan(R/2)$, and for soft gluons outside jets, $\lambda^2\sim\Lambda/Q$. In this section, we mean by $\lambda$ the expansion parameter associated with \textit{measured} jets, and ensuring $R$ is much larger than this $\lambda$. But if $R$ is \emph{too} large, the separation parameter $t\propto 1/\tan(R/2)$ becomes too small. We will consider $R\sim 0.4$ to $1$ to be safe.} These power corrections are associated with assumptions we made about the action of the jet algorithm on final states in deriving \eq{Njetcs}.  In general, the size of these power corrections depends both on the algorithm and the observable.

Power corrections to the $p_T$ of a jet arising from perturbative emissions (as well as from hadronization and the underlying event in $pp$ collisions) for various jet algorithms were explored in \cite{Dasgupta:2007wa}.  These power corrections arise for similar reasons as those we discuss below, namely, perturbative emissions changing which partons get combined into the jet. Ref. \cite{Dasgupta:2007wa} finds that such power corrections scale like $\ln R$ for small $R$.  This result is consistent with our qualitative discussion below, where we argue that power corrections to jet angularities arising from the jet algorithms we use are minimized when $R$ is sufficiently large. For us, $R$ should be at least $\cO(\lambda)$ (and for the case of the $\kt$ algorithm, $R\gg \lambda$).

One set of power corrections that is independent of the choice of algorithm arises from the approximation of setting the jet axis equal to the label direction $n$. Since this neglects the effects of soft particles, it is valid up to $\mathcal{O}(\lambda^2)$ corrections. It was argued in Refs.~\cite{Berger:2003iw,Bauer:2008dt,Lee:2006nr} for the case of hemisphere jet algorithms that these corrections in turn induce corrections to the angularity $\tau_a$  of order $\lambda^{2(2-a)}$, which, for $\tau_a \sim \lambda^2$, are subleading for $a<1$. Essentially the same arguments can be applied to all of the algorithms we consider.

Jet algorithm-dependent power corrections arise from the difference in soft particles included in a jet by a given algorithm and those included in the soft function in the factorization theorem. The algorithms also differ amongst themselves in which soft particles they include in a jet. For observables that scale as $\cO(1)$, such as the jet energies and 3-momenta, the contribution of soft momenta can be neglected since they scale as $\cO(\lambda^2)$. Clearly then, these observables are not dependent on our choice of jet algorithm and so the assumptions we made about factorization of the algorithm in deriving \eq{Njetcs} are trivially satisfied.

However, for observables that scale as $\cO(\lambda^2)$ such as angularities, soft contributions become important and so the details of the algorithms we consider become relevant.
We now estimate how closely the phase space region included in the soft function in the SCET factorization theorem approximates the region of soft particles actually included by the jet algorithm. We will argue that unless $R\gtrsim \cO(\lambda)$ for the anti-$\kt$ and infrared-safe cone algorithms, and $R\gg\lambda$ for the $\kt$ algorithm, the mismatch in areas is sufficiently large so as to cause a leading-order power correction to the measured jet angularities. For $R$ larger than these bounds, the corrections are negligible.
This miscounting arises due to the fact that factorization requires that collinear particles be combined first, and that the soft function only knows about the parent collinear direction. When algorithms do not obey this ordering, factorization may be violated.

\FIGURE[t]{
\includegraphics[width = 1.0\textwidth]{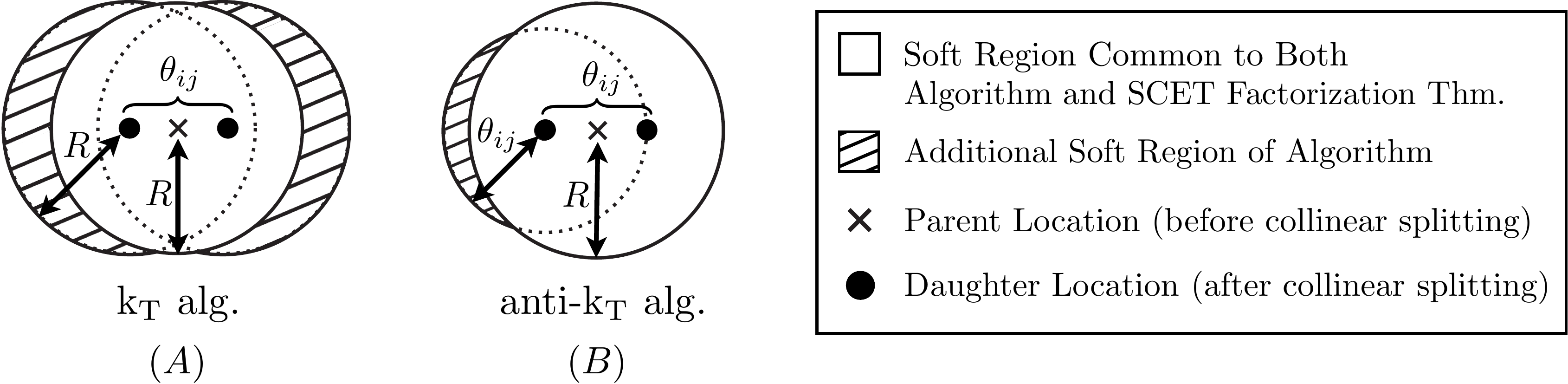}
\vspace{-2em}
{ \caption[1]{Difference between regions of soft radiation included in the SCET factorization theorem and the actual (A) $\kt$ and (B) anti-$\kt$ algorithms. We illustrate how the regions of soft radiation included by the algorithms change when a single, energetic parent particle splits into two collinear daughters.
Both the algorithm and the soft function merge soft particles contained in the large white circle. The algorithm also merges the hatched area and hence contains a region of phase space which is different than that included in the SCET soft function (since, as explained in the text, the shape and size of the region used in SCET cannot depend on the details of collinear splittings). }
\label{fig:kt-error}}
}

  To determine the size of the soft particle phase space region for each jet algorithm that is not included by the factorization theorem, we consider the situation depicted in \fig{fig:kt-error}.  A parent collinear particle splits into two daughter collinear particles.  In the factorization theorem, since collinear particles are combined first, the region of phase space where soft particles are combined into the jet is a circle of radius $R$ about the parent particle direction.  However, in a jet algorithm, soft particles in an additional region outside of this may also be combined into the jet (the hatched regions in \fig{fig:kt-error}).  If the area of this region is of the same order as the area included by the factorization theorem, then the power corrections to jet angularities induced by the jet algorithm will be leading-order.

Because soft particles have momenta that are parametrically smaller than collinear particle momenta, we determine the omitted region of soft particle phase space by considering the dominant action of the jet algorithm.  The $\kt$ algorithm serves as a useful example.  The $\kt$ metric between a pair of soft particles is $\cO(\lambda^2)\theta_{ss}$, the metric between a soft particle and collinear particle is $\cO(\lambda^2)\theta_{cs}$, and the metric between a pair of collinear particles is $\cO(\lambda^0)\theta_{cc}$.  Therefore, collinear-collinear recombinations only occur if the angle $\theta_{cc}$ between the collinear particles is smaller than the separation between any soft particle and its nearest neighbor by a factor of $\cO(\lambda^2)$.  Given that the  typical angle between collinear particles is $\cO(\lambda)$, the dominant action of the jet algorithm is to first merge all soft particles with their nearest neighbors,  while collinear-collinear recombinations occur late in the operation of the algorithm.  This description will suffice to accurately determine the area of the omitted soft phase space for the $\kt$ algorithm.
Since collinear particles are combined last, on average soft particles within circles of radius $R$ about the collinear daughter particles are included in the $\kt$ algorithm jet,\footnote{Soft particles in this region can also be removed from this region by merging with other soft particles outside of the region and vice-versa, but this average area suffices for our discussion.} as shown in \fig{fig:kt-error}A. The area that the $\kt$ algorithm includes that the soft function does not is represented by the  hatched region, which is an area of $\cO(\theta_{ij}R)$. This area must be parametrically smaller than the area included by the soft function (of $\cO(R^2)$) for the associated power corrections to be small. We thus require that $R \gg \theta_{ij} \sim \lambda$ in the SCET power counting.

The anti-$\kt$  algorithm combines particles in a manner that is closer to respecting factorization. It finds the hardest particle first and merges particles at successively larger distances from this particle. For the example of two collinear daughters, it will merge all soft particles with the hardest daughter that are closer than the distance to the softer daughter before merging the two daughters and then merging all soft particles a distance $R$ from the merged daughters (i.e., the parent particle), as shown in \fig{fig:kt-error}B. As the Figure illustrates, the  hatched area of the anti-$\kt$ jet tends to be smaller than that of the $\kt$ jet. In fact, for $R > 3 \theta_{ij}/2$, this region vanishes completely (and this case of having only two collinear daughters is a worst-case scenario). 
This leads us to expect that, for any number of collinear splittings, for $R \gtrsim \lambda$ (i.e., not necessarily parametrically larger), power corrections due the action of the anti-$\kt$ algorithm vanish.

Cone algorithms such as the SISCone algorithm can also include regions that differ from the lowest-order region at higher orders in perturbation theory. We now argue that an arithmetic bound $R\gtrsim\lambda$ is sufficient to minimize the power correction from these differences, as for the anti-$\kt$ algorithm. This situation arises due to the fact that stable solutions may exist with overlapping cones when collinear splittings are larger than the cone radius, i.e., $R < \theta_{ij}$. In these cases, the amount of radiation that falls into the overlapping region is used to decide whether the cones are split or merged. In either case, the boundary of the resulting jet(s) has roughly the appearance of \fig{fig:kt-error}A and the difference between the region of soft radiation assumed in SCET and that by the actual algorithm is $\cO(1)$.
However, for $R >  \theta_{ij} \sim \lambda$ for the case of a single collinear splitting, all of the collinear radiation lies within a region of size $R$ and there will always be a stable cone that includes this radiation and thus the algorithm and the SCET soft function will be sensitive to soft particles in the same region of phase space. 

In summary, we have argued that for all the algorithms we consider ($\kt$, anti-$\kt$, infrared-safe cone), power corrections are negligible for sufficiently large $R$. While anti-$\kt$ and cone allow simply $R\gtrsim\lambda$ instead of $R\gg\lambda$ as for $\kt$, we will in fact always consider $R \gg \lambda$ (for the $\lambda$ in a measured jet sector) in the remainder of this paper, guaranteeing small power corrections for all these algorithms.  ($R$ still determines the scale $\lambda$ in an unmeasured jet sector.)
Our focus will remain on resumming logs of jet shapes such as angularities $\tau_a$ in the presence of jet algorithms, without worrying about resumming logs of $R$ themselves.\footnote{Because small $R$ ($\lesssim$ 0.3) jets cannot be well resolved in current experiments, resummation of logarithms of $R$ is not of primary practical importance in the near future.}

\section{Jet Functions at $\mathcal{O}(\as)$ for Jet Shapes}
\label{sec:jet}

In this section, we calculate the quark and gluon jet functions for jet shapes at next-to-leading order in perturbation theory.  The jet functions can be divided into two categories: those for measured jets, which are fixed to have a specific angularity $\tau_a$, and those for unmeasured jets, which are not.  We will denote the quark jet function by $J_{\omega}^q$, the gluon jet function by $J_{\omega}^g$, where $\omega$ is the label momentum, and a jet function $J^{q,g}(\tau_a)$ with an argument of $\tau_a$ denotes a measured jet.  We will calculate the jet functions for the two classes of jet algorithms, $\kt$-type and cone-type algorithms.

In the course of these calculations, we will demonstrate the crucial role of zero-bin subtractions \cite{Manohar:2006nz} from collinear jet functions in obtaining the consistent anomalous dimensions and the correct finite parts. In this case zero-bin subtractions are not merely scaleless integrals converting IR to UV divergences, but in fact contribute part (sometimes the most important part) of the correct nonzero result, as was already pointed out by \cite{Cheung:2009sg,Chiu:2009yx}. The relation of zero-bin subtractions in SCET to eikonal jet subtractions from soft functions in traditional methods of QCD factorization was explored in \cite{Lee:2006nr,Idilbi:2007ff,Idilbi:2007yi}. In addition, we find that the zero-bin subtraction removes the contribution of collinear emissions that escape a jet, leaving only power-suppressed pieces in $\Lambda/\omega_i$.

\subsection{Phase Space Cuts}

\FIGURE{
\includegraphics[totalheight = .08\textheight]{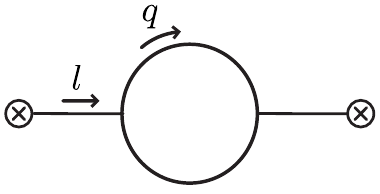}

{ \caption[1]{A representative diagram for the NLO quark and gluon jet functions. The incoming momentum is $l = \frac{n}{2}\omega + \frac{\bar n}{2} l^+$ and particles in the loop carry momentum $q$ (``particle 1'') and $l-q$ (``particle 2'').}
\label{fig:generic-jet}}
}

To calculate the jet functions for a particular algorithm, we must impose phase space restrictions in the matrix element.  From the jet function definitions, \eq{jetfuncdefs}, these cuts take two forms.  One kind, imposed by the operator $\delta_{N(\hat{\mathcal{J}}),1}$ in \eq{jetfuncdefs}, is common to every jet function.  It is the set of phase space restrictions related to the jet algorithm, and requires exactly one jet to arise from each collinear sector of SCET.  The other, imposed by the operator $\delta({\tau}_a - \hat \tau_a)$, is implemented only on measured jets and restricts the kinematics of the cut final states to produce a fixed value of the jet shape.  In this section we describe these phase space cuts in detail.

The typical form of the NLO diagrams in the jet functions is shown in \fig{fig:generic-jet}.  As shown in the figure, the momentum flowing through the graph has label momentum $l^- \equiv \bn\cdot l = \omega$ and residual momentum $l^+ \equiv n\cdot l$, and the loop momentum is $q$.  We will label ``particle 1'' as the particle in the loop with momentum $q$ and ``particle 2'' as the particle in the loop with momentum $l - q$.  For the quark jet, we take particle 1 as the emitted gluon and particle 2 as the quark.

As usual, the total forward scattering matrix element can be written as a sum over all cuts. Cutting through the loops corresponds to the interference of two real emission diagrams, each with two final state particles, whereas cutting through a lone propagator that is connected to a current corresponds to the interference between a tree-level diagram and a virtual diagram, each with a single final state particle. Thus, the phase space restrictions and measurements we impose act differently depending on where the diagrams are cut. In addition, since we will be working in dimensional regularization (with $d = 4-2\epsilon$), which sets scaleless integrals to zero, the only diagrams that contribute are the cuts through the loops. This means that we only need to focus on the form of phase-space restrictions and angularities in the case of final states with two particles.

The regions of phase space for two particles created by cutting through a loop in the jet function diagrams can be divided into three contributions:
\begin{enumerate}
\item Both particles are inside the jet.
\item Particle 1 exits the jet with energy $E_1<\Lambda$.
\item Particle 2 exits the jet with energy $E_2<\Lambda$.
\end{enumerate}
In contributions (2) and (3), the jet has only one particle, which is the remaining particle with $E > \Lambda$.

It is well known that collinear integrations of jet functions can be allowed to extend over all values of loop momenta so long as a ``zero-bin subtraction'' is taken from the result to avoid double counting the soft region already accounted for in the soft function \cite{Manohar:2006nz}.  We will demonstrate that contributions (2) and (3) are power suppressed by $\cO(\Lambda/\omega)$, which scales as $\lambda^2$, after the zero-bin subtraction.

The phase space cuts that enforce both particles to be in the jet depend on the jet algorithm.  There are two classes of jet algorithms that we consider, cone-type algorithms and (inclusive) $\kt$-type algorithms, and all the algorithms in each class yield the same phase space cuts.  We label the phase space restrictions as $\Theta_\text{cone}$ and $\Theta_{\kt}$, generically $\Theta_\text{alg}$.  For the cone-type algorithms,
\be
\Theta_\text{cone} \equiv \Theta_\text{cone}(q, l^+) = \Theta \left( \tan^2{\frac{R}{2}} >  \frac{q^+}{q^- }\right) \Theta \left( \tan^2{\frac{R}{2}} > \frac{l^+ - q^+}{\omega - q^-} \right) .
\ee
These $\Theta$ functions demand that both particles are within $R$ of the label direction.  For the $\kt$-type algorithms, the only restriction is that the relative angle of the particles be less than $R$:
\begin{align}
\Theta_{\kt} \equiv \Theta_{\kt}(q, l^+)  &= \Theta \left( \cos{R} < \frac{\vec{q} \cdot \vec{l} - q^2}{q \sqrt{l^2 + q^2 - 2 \vec{q} \cdot \vec{l}}} \right) \nn\\
&= \Theta \left( \tan^2{\frac{R}{2}} >  \frac{q^+\omega^2}{q^- \left(\omega - q^-\right)^2} \right)\, .
\end{align}
In the second line we took the collinear scaling of $q$ ($q^+ \ll q^-$). While this is not strictly needed, it makes the calculations significantly simpler.

For the phase space restrictions of zero-bin subtractions, we take the soft limit of the above restrictions. The zero-bin subtractions are the same for all the algorithms we consider. For the case of particle 1, which has momentum $q$, the zero-bin phase space cuts are given by
\begin{align}
\Theta_\text{alg}^{(0)} = \Theta_\text{cone}^{(0)} =
\Theta_{\kt}^{(0)} =  \Theta \left( \tan^2{\frac{R}{2}} >  \frac{q^+}{q^- }\right)
\,.\end{align}
The zero bin of particle 2 is given by the replacement $q \to l - q$.

For all the jet algorithms we consider, the zero-bin subtractions of the unmeasured jet functions are scaleless integrals.\footnote{Note that algorithms do exist that give nonzero zero-bin contributions to unmeasured jet functions \cite{Cheung:2009sg}.} However, for the measured jet functions, the zero-bin subtractions give nonzero contributions that are needed for the consistency of the effective theory.

In the case of a measured jet, in addition to the phase space restrictions we also demand that the jet contributes to the angularity by an amount $\tau_a$ with the use of the delta function $\delta_R = \delta(\tau_a - \hat \tau_a) $, which is given in terms of $q$ and $l$ by
\begin{equation}
\delta_R \equiv \delta_R(q, l^+) = \delta\left(\tau_a - \frac{1}{\omega}(\omega - q^-)^{a/2}(l^+ - q^+)^{1-a/2} - \frac{1}{\omega}(q^-)^{a/2}(q^+)^{1-a/2}\right)
\,.\end{equation}
In the zero-bin subtraction of particle 1, the on-shell conditions can be used to write the corresponding zero-bin $\delta$-function as
\begin{equation}
\delta^{(0)}_R = \delta\left(\tau_a - \frac{1}{\omega}(q^-)^{a/2}(q^+)^{1-a/2}\right)
\,,\end{equation}
(and for particle 2 with $q \to l - q$).

\subsection{Quark Jet Function}
The diagrams corresponding to the quark jet function are shown in Fig.~\ref{fig:quark-jet-function}.  The fully inclusive quark jet function is defined as
\be
\int\!d^4 x \, e^{i l\cdot x} \bra{0}\chi_{n, \omega}^{a \alpha}(x) \bar\chi_{n, \omega}^{b \beta} (0)\ket{0} \equiv \delta^{a b}\left(\frac{\nslash}{2}\right)^{\alpha\beta} J^q_\omega(l^+)\, ,
\ee
and has been computed to NLO (see, e.g., \cite{Bauer:2003pi, Bosch:2004th}) and to NNLO \cite{Becher:2006qw}.  Below we compute the quark jet function at NLO with phase space cuts for the jet algorithm for both the measured jet, $J^q_{\omega}(\tau_a)$, and the unmeasured jet, $J^q_{\omega}$.  As discussed above, we will find that the only nonzero contributions come from cuts through the loop when both cut particles are inside the jet.

\FIGURE[t]{
\includegraphics[totalheight = .08\textheight]{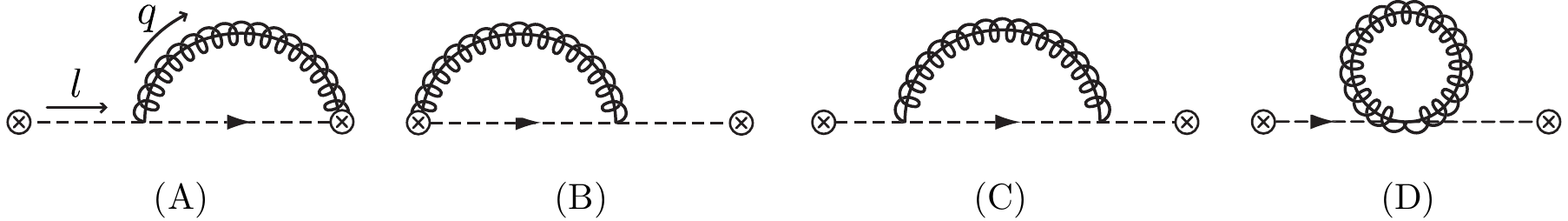}

{ \caption[1]{Diagrams contributing to the quark jet
function. (A) and (B) Wilson line emission diagrams; (C) and (D) QCD-like diagrams. The momentum assignments are the same as in \fig{fig:generic-jet}.}
\label{fig:quark-jet-function}}
}

\subsubsection{Measured Quark Jet}
\label{ssec:quark-meas}

The measured quark jet function includes contributions from naive Wilson line graphs (A) and (B) and QCD-like graphs (C) and (D) in \fig{fig:quark-jet-function}.  The sum of these contributions is
\begin{equation}
\label{Jnaivein}
\begin{split}
\qjetnaive^q_{\omega}(\tau_a) &= g^2\mu^{2\epsilon} C_F \int\frac{d l^+}{2\pi}\frac{1}{(l^+)^2}\int \frac{d^d q}{(2\pi)^d}\left(4\frac{l^+}{q^-} + (d-2)\frac{l^+ - q^+}{\omega - q^-}\right) 2\pi\delta(q^- q^+ - q_\perp^2)\\
&\quad \times\Theta(q^-)\Theta(q^+) 2\pi\delta\left(l^+ - q^+ - \frac{q_\perp^2}{\omega - q^-}\right)\Theta(\omega - q^-)\Theta(l^+ - q^+) \, \Theta_\text{alg} \delta_R
\,.\end{split}
\end{equation}
The contribution proportional to $d-2$ comes from the QCD-like graphs (C) and (D) in \fig{fig:quark-jet-function}.  Only the Wilson line graphs have a nonzero zero-bin limit, which comes from taking the scaling limit $q\sim\lambda^2$ of the naive contribution:
\begin{equation}
\label{Jzeroin}
\begin{split}
J^{q(0)}_{\omega}(\tau_a) &= 4g^2\mu^{2\epsilon} C_F \int\frac{dl^+}{2\pi}\frac{1}{l^+}\int \frac{d^d q}{(2\pi)^d}\frac{1}{q^-} 2\pi\delta(q^- q^+ - q_\perp^2)\Theta(q^-)\Theta(q^+) \\
&\quad \times 2\pi\delta\left(l^+ - q^+\right)\Theta(l^+ - q^+) \,  \Theta_\text{alg}^{(0)} \delta_R^{(0)}
\,.\end{split}
\end{equation}
All jet algorithms that we use yield the same zero-bin contribution, since the phase space cuts are the same.

To evaluate these integrals, we can analytically extract the coefficient of $\delta(\tau_a)$ by integrating over $\tau_a$ and using the fact that the remainder is a plus distribution, as defined in \eq{mitplus}. We find the naive contribution is
\begin{equation}
\label{Jnaiveinresult}
\begin{split}
\qjetnaive^q_{\omega}(\tau_a) &= \frac{\alpha_s C_F}{2\pi} \frac{1}{\Gamma(1-\epsilon)} \left(\frac{4\pi\mu^2}{\omega^2\tan^2\frac{R}{2}}\right)^\epsilon\left( \frac{1}{\epsilon^2}  + \frac{3}{2\epsilon} \right) \delta(\tau_a)
 +  \frac{\alpha_s}{2\pi} \qjetnaive^q_\text{alg} (\tau_a)
\,.\end{split}
\end{equation}
The only difference between the jet algorithms that we consider resides in the finite distribution $\qjetnaive^q_\text{alg}(\tau_a)$, which is a complicated function of $\tau_a$ that we give in Appendix~\ref{app:jetcalc}. Note that the divergent part of the naive contribution is proportional to $\delta(\tau_a)$. This is due to the fact that the jet algorithm regulates the distribution for $\tau_a >0$. The divergent plus distributions come entirely from the zero-bin subtraction, which is given by
\begin{equation}
\label{Jzeroinresult}
\begin{split}
J^{q(0)}_{\omega}(\tau_a) &= \frac{\alpha_s C_F}{\pi} \frac{1}{\Gamma(1-\epsilon)} \left(\frac{4\pi\mu^2\tan^{2(1-a)}\frac{R}{2}}{\omega^2}\right)^\epsilon \frac{1}{(1-a)\epsilon} \frac{1}{\tau_a^{1+2\epsilon}}\, .
\end{split}
\end{equation}

Adding the leading-order contribution to all of the NLO graphs and expanding in powers of $\epsilon$, adopting the $\overline{\text{MS}}$ scheme, we find the total quark jet function,
\begin{align}
\label{Jtotal}
J^q_\omega(\tau_a) = \delta (\tau_a) + \qjetnaive^q_{\omega}(\tau_a) - J^{q(0)}_{\omega}(\tau_a)  &= \Biggl\{ 1 + \frac{\alpha_s C_F}{\pi}\Biggl[ \frac{1-\frac{a}{2}}{1-a}\frac{1}{\epsilon^2} + \frac{1-\frac{a}{2}}{1-a}\frac{1}{\epsilon}\ln\frac{\mu^2}{\omega^2} + \frac{3}{4\epsilon}
 \Biggr]\Biggr\}\delta(\tau_a) \nn \\
 & \qquad - \frac{\alpha_s C_F}{\pi}\Biggl[ \frac{1}{\epsilon}\frac{1}{1-a} \frac{\Theta(\tau_a)}{\tau_a} \Biggr]_+  + \frac{\as} {2\pi} J^q_\text{alg}(\tau_a)\,.
\end{align}
This agrees with the standard jet function $J(k^+)$ given in  \cite{Bauer:2003pi, Bosch:2004th} by setting $a=0$ and $k^+ = \omega\tau_a$.  We have shown the divergent terms explicitly, and collect the finite pieces in $J_\text{alg}^q(\tau_a)$, which we  give in \eq{eq:quark_meas_finite}.  Note that there is no jet algorithm dependence in the divergent parts of the jet function at this order in perturbation theory.

\subsubsection{Gluon Outside Measured Quark Jet}
\label{ssec:quark-out}

In this section we calculate the contribution to the quark jet function from the region of phase space in which the gluon exits the jet carrying an energy $E_g<\Lambda$.  This cut causes the contribution to be power suppressed by $\Lambda/\omega$, which scales as $\lambda^2$.  However, we elect to evaluate this case explicitly as it provides a clear example of the zero-bin subtraction giving the proper scaling to the total contribution.  We only evaluate this contribution for the cone algorithm; the details of the $\kt$ algorithm calculation are similar.  Note that the contribution when the quark is out of the jet is power suppressed at the level of the Lagrangian given in \ssec{SCET}, in which soft quarks do not couple to collinear partons at leading order in $\lambda$.

For the cone algorithm, the gluon exits the jet when the angle between the jet axis, $\vect{n}_1$, and the gluon is greater than $R$.  When the gluon is not in the jet, the cone axis is the quark direction, and so it makes no contribution to the angularity.  Therefore, this region of phase space contributes only to the $\delta(\tau_a)$ part of the angularity distribution.

For the naive contributions, requiring the gluon to be outside the jet and have energy less than $\Lambda$, we have the integral
\begin{align}
\label{Jnaiveout}
\qjetnaive^{q, \rm out}_{\omega}(\tau_a) &= g^2\mu^{2\epsilon} C_F \int\frac{d l^+}{2\pi}\frac{1}{(l^+)^2}\int \frac{d^d q}{(2\pi)^d}\left(4\frac{l^+}{q^-} + (d-2)\frac{l^+ - q^+}{\omega - q^-}\right) 2\pi\delta(q^- q^+ - q_\perp^2) \nn \\
&\quad \times\Theta(q^-)\Theta(q^+) 2\pi\delta\left(l^+ - q^+ - \frac{q_\perp^2}{\omega - q^-}\right)\Theta(\omega - q^-)\Theta(l^+ - q^+) \nn \\
&\quad\times \Theta\left(\frac{q^+}{q^-} - \tan^2\frac{R}{2}\right)\Theta\left(2\Lambda - q^-\right)\delta(\tau_a) \,.
\end{align}
Note that the theta function requiring $q^- < 2\Lambda$ is more restrictive than $q^- < \omega$.  Evaluating \eq{Jnaiveout} yields a contribution that scales with $\Lambda$ only below the leading term in $1/\epsilon$:
\begin{equation}
\label{Jnaiveoutresult}
\qjetnaive^{q, \rm out}_{\omega}(\tau_a) = -\frac{\alpha_s C_F}{2\pi} \frac{1}{\Gamma(1-\epsilon)} \left(\frac{4\pi\mu^2}{(2\Lambda\tan\frac{R}{2})^2}\right)^{\epsilon} \delta(\tau_a) \left(\frac{1}{\epsilon^2} + \frac{1}{\epsilon}\left(\frac{4\Lambda}{\omega} - \frac{2\Lambda^2}{\omega^2}\right) + \frac{8\Lambda}{\omega}\right)
\end{equation}
The zero-bin subtraction of Eq.~\ref{Jnaiveout} is
\begin{align}
\label{Jzeroout}
\qjetnaive^{q, {\rm out}(0)}_{\omega}(\tau_a) &= g^2\mu^{2\epsilon} C_F \int\frac{d l^+}{2\pi}\frac{1}{(l^+)^2}\int \frac{d^d q}{(2\pi)^d}\left(4\frac{l^+}{q^-} + (d-2)\frac{l^+ - q^+}{\omega - q^-}\right) 2\pi\delta(q^- q^+ - q_\perp^2)\nn \\
&\quad \times\Theta(q^-)\Theta(q^+) 2\pi\delta\left(l^+ - q^+\right) \Theta\left(\frac{q^+}{q^-} - \tan^2\frac{R}{2}\right)\Theta\left(2\Lambda - q^-\right)\delta(\tau_a) \,.
\end{align}
Evaluating \eq{Jzeroout}, we find the zero bin will exactly remove the leading term in $1/\epsilon$:
\begin{equation}
\label{Jzerooutresult}
\qjetnaive^{q, {\rm out}(0)}_{\omega}(\tau_a) = -\frac{\alpha_s C_F}{2\pi} \frac{1}{\Gamma(1-\epsilon)} \left(\frac{4\pi\mu^2}{(2\Lambda\tan\frac{R}{2})^2}\right)^{\epsilon} \delta(\tau_a) \frac{1}{\epsilon^2}
\end{equation}
Therefore, the difference is power suppressed only after the zero bin is included.  Because other contributions when one particle is outside of the jet are similarly power suppressed, we will drop them in our remaining discussion of the jet functions.

\subsubsection{Unmeasured Quark Jet}
\label{ssec:quark-unmeas}

When the angularity of a jet is not measured, the jet function has no $\tau_a$ dependence.  The naive and zero-bin contributions are the same as \eqs{Jnaivein}{Jzeroin} except for the factor of $\delta_R$.  The zero-bin contribution is
\begin{equation}
\label{J2zeroin}
\begin{split}
J^{q(0)}_{\omega} &= 2g^2\mu^{2\epsilon} C_F n\mcdot\bn \int\frac{dl^+}{2\pi}\frac{1}{l^+}\int \frac{d^d q}{(2\pi)^d}\frac{1}{q^-} 2\pi\delta(q^- q^+ - q_\perp^2)\Theta(q^-)\Theta(q^+) \\
&\quad \times 2\pi\delta\left(l^+ - q^+\right)\Theta(l^+ - q^+) \,  \Theta_\text{alg}^{(0)}
\,.\end{split}
\end{equation}
This integral is scaleless and therefore equal to 0 in dimensional regularization.  This implies that the NLO part of the quark jet function for an unmeasured jet is just the naive result.
We find, making the divergent part explicit, in the $\overline{\text{MS}}$ scheme,
\begin{equation}
\label{Jq}
J^q_\omega = 1+ \qjetnaive^q_\omega = 1 + \frac{\alpha_s C_F}{2 \pi} \Biggl\{ \frac{1}{\epsilon^2} + \frac{3}{2\epsilon} + \frac{1}{ \epsilon} \ln \left(\frac{\mu^2}{\omega^2\tan^2\frac{R}{2}}\right) \Biggr\} + \frac{\alpha_s}{2 \pi} J^q_\text{alg} \, ,
\end{equation}
where the finite part $J^q_\text{alg}$ is given in \eq{eq:quark_unmeas_finite}.\footnote{The unmeasured jet function \eq{Jq} is not simply obtained by integrating the measured jet function \eq{Jtotal} over $\tau_a$. This is due to the different relative scaling of $R$ with the SCET expansion parameter $\lambda_i$ in a measured and unmeasured jet sector, as noted earlier. Namely, $R\sim \lambda_i^0$ in a measured jet sector (where $\lambda\sim\sqrt{\tau_a}$) while $\lambda_k\sim\tan(R/2)$ in an unmeasured jet sector.}

\subsection{Gluon Jet Function}
The diagrams needed for the gluon jet function at NLO are shown in \fig{fig:gluon-jet-function}. The fully inclusive jet function, defined as
\begin{equation}
\label{Gjetdef}
\int\!d^4 x \, e^{i l\cdot x} \bra{0} B_{\perp, \omega}^{\mu A} (x) B_{\perp, \omega}^{\nu B} (0)\ket{0} \equiv - \frac{1}{\omega} g_\perp^{\mu\nu} \delta^{AB} J^g_\omega(l^+)
\,,\end{equation}
(with $  J^g_\omega(l^+)$ normalized to $ 2\pi\delta(l^+)$ at tree-level) has been calculated to NLO in Feynman gauge in \cite{Bauer:2006qp,Bauer:2001rh, Fleming:2003gt} and was reported to give the same result in both $R_\xi$ and light-cone gauges in \cite{Becher:2009th}. Since our phase space restrictions and the observables act differently on cuts through loops and on cuts through external propagators, it is useful to reproduce these results by explicitly cutting the diagrams.

\FIGURE[t]{
\includegraphics[totalheight = .15\textheight]{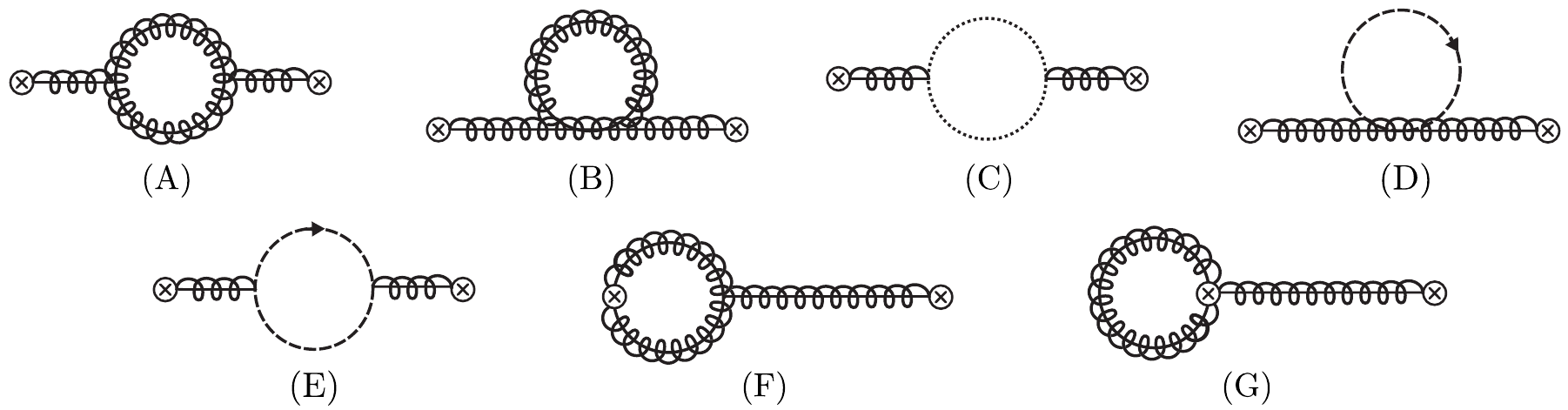}

{ \caption[1]{Diagrams contributing to the gluon jet
function. (A) sunset and (B) tadpole gluon loops; (C) ghost loop; (D) sunset and (E) tadpole collinear quark loops; (F) and (G) Wilson line emission loops. Diagrams (F) and (G) each have mirror diagrams (not shown). The momentum assignments are the same as in \fig{fig:generic-jet}.}
\label{fig:gluon-jet-function}}
}

After some algebra, we find that the sum of all cuts through the loops of the na\"{i}ve collinear graphs gives
\begin{align}
\int\!\frac{dl^+}{2\pi} \tilde J^g_\omega(l^+) &= \mu^{2 \epsilon} \frac{2  g^2}{(2 \pi)^{d-1}}\int\!\frac{dl^+}{l^+} \int\!d^d q \, \delta(q^2) \delta((l-q)^2) \, \Theta(\omega -q^-)\nn\\
& \quad \times \left[ \TR\NF \left( 1 - \frac{2}{1-\epsilon} \frac{q^+q^-}{\omega l^+}\right) - \CA \left( 2 - \frac{\omega}{q^-}  - \frac{\omega}{\omega - q^-}  - \frac{q^+ q^-}{\omega l^+} \right)\right] \, .
\label{gluon-jet-unconstrained}
\end{align}
We also record the zero bin that needs to be subtracted from \eq{gluon-jet-unconstrained}. To leading-power it is given by
\begin{align}
\int\!\frac{dl^+}{2\pi} J^{g(0)}_\omega(l^+) &=
\mu^{2 \epsilon} \CA \frac{2  g^2}{(2 \pi)^{d-1}}\int\!\frac{dl^+}{l^+} \int\!d^d q \,\bigg[ \delta(q^2)  \delta(l^+-q^+) \Theta(q^-)  \frac{1}{q^-}  \nn\\
& \qquad \qquad +  \delta((l-q)^2) \delta( q^+)  \Theta(\omega-q^-)  \frac{1}{\omega - q^-}  \bigg]
\label{gluon-zerobin-unconstrained}
\,.\end{align}
Without inserting any additional constraints, this integral is scaleless and zero in dimensional regularization. Therefore, in the absence of phase-space restrictions, the na\"{i}ve integral \eq{gluon-jet-unconstrained} gives the standard (inclusive) gluon jet function
\begin{align}
\label{Gjetincl}
\frac{J^g_\omega(l^+) }{2\pi \omega} &= \frac{\as}{4 \pi} \mu^{2\epsilon}(\omega l^+)^{-1-\epsilon} \left[ \TR\NF\left( \frac{4}{3} + \frac{20}{9}\epsilon\right) -\CA \left(\frac{4}{\epsilon} + \frac{11}{3} + \left( \frac{67}{9}-\pi^2\right)\epsilon \right)\right]\,
,\end{align}
in the $\overline{\text{MS}}$ scheme.
The measured and unmeasured jet functions are obtained by inserting $\Theta_\text{alg} \delta_R$ and $\Theta_\text{alg}$, respectively, into \eqs{gluon-jet-unconstrained}{gluon-zerobin-unconstrained}.

\subsubsection{Measured Gluon Jet}
\label{ssec:gluon-meas}

The naive contribution to the measured gluon jet can be written as
\begin{align}
\label{Gnaivein}
\gjetnaive^g_\omega(\tau_a) &= \frac{\as}{2\pi} \frac{1}{\Gamma(1-\epsilon)} \left( \frac{4 \pi \mu^2}{\omega^2} \right)^\epsilon \frac{1}{1-\frac{a}{2}} \left( \frac{1}{\tau_a} \right)^{1+ \frac{2\epsilon}{2-a}} \int_0^1\!dx \, (x^{a-1} + (1-x)^{a-1})^{\frac{2\epsilon}{2-a}} \\
& \qquad \times \left[ \TR\NF \left( 1 - \frac{2}{1-\epsilon} x(1-x) \right) - \CA \left( 2 -\frac{1}{x(1-x)}  - x(1-x) \right)\right]  \, \Theta_\text{alg}(x)
\, ,\nn
\end{align}
where $x \equiv q^-/\omega$. This gives
\begin{align}
\label{Gnaiveinresult}
\gjetnaive^g_\omega(\tau_a) &= \frac{\as}{2\pi} \frac{1}{\Gamma(1-\epsilon)} \left( \frac{4 \pi \mu^2}{\omega^2 \tan^2 \frac{R}{2}} \right)^\epsilon \delta(\tau_a)
\Bigg[  \CA \left( \frac{1}{\epsilon^2} +\frac{11}{6}\frac{1}{\epsilon}  \right)   -  \frac{2}{3\epsilon} \TR\NF  \Bigg] + \frac{\as}{2\pi} \gjetnaive^g_\text{alg}(\tau_a)
\,,\end{align}
where, as for the quark jet function, the finite distributions $\gjetnaive_\text{alg}^g(\tau_a)$ differ among the algorithms we consider. They are given in Appendix~\ref{app:jetcalc}.

The zero-bin result is
\begin{align}
\label{Gzeroinresult}
J^{g(0)}_\omega(\tau_a) &= \frac{\as \CA}{\pi} \frac{1}{\Gamma(1-\epsilon)}  \left( \frac{4 \pi \mu^2 \tan^{2(1-a)} \frac{R}{2}}{\omega^2 } \right)^\epsilon \left( \frac{1}{\tau_a}\right)^{1+2\epsilon}\frac{1}{(1-a)\epsilon}
\,.\end{align}
Subtracting the zero-bin from the naive integral and adding the leading-order contribution, we obtain in $\overline{\text{MS}}$
\begin{align}
\label{Gjet}
J^g_\omega(\tau_a) &= \delta(\tau_a) + \gjetnaive^g_\omega(\tau_a) - J^{g(0)}_\omega(\tau_a) \nn\\
&= \Biggl\{ 1+ \frac{\as \CA}{ \pi } \Bigg[ \frac{1-a/2}{1-a} \left(\frac{1}{\epsilon^2}+ \frac{1}{\epsilon} \ln{\frac{\mu^2}{\omega^2}} \right)  + \frac{11}{12}\frac{1}{\epsilon}  \Bigg]  - \frac{\as}{3\pi}\TR\NF \frac{1}{\epsilon}  \Biggr\}\delta(\tau_a) \nn\\
 & \quad -\frac{\alpha_s C_A}{\pi}  \frac{1}{1-a}\frac{1}{\epsilon} \left(\frac{\Theta(\tau_a)}{\tau_a}  \right)_+ + \frac{\as}{2\pi} J^g_\text{alg}(\tau_a)
\,.\end{align}
The finite distribution $J_\text{alg}^g(\tau_a)$ is given in \eq{eq:quark_meas_finite}.

\subsubsection{Unmeasured Gluon Jet}
\label{ssec:gluon-unmeas}

As for the quark jet function, for unmeasured jets the naive and zero-bin contributions are given by the measured jet contributions without the $\delta_R$ constraint.  Also, as it was for the quark jet function, the zero-bin contribution to the unmeasured jet function is a scaleless integral. Thus, the final answer is just the naive result, which is given by
\begin{align}
\label{G2result}
J^g_\omega = 1+ \frac{\as}{2 \pi}
\Bigg[  \CA \left( \frac{1}{\epsilon^2} +\frac{11}{6}\frac{1}{\epsilon} + \frac{1}{\epsilon} \ln \frac{\mu^2}{\omega^2 \tan^2 \frac{R}{2}} \right)    -  \frac{2}{3\epsilon} \TR\NF \Bigg] + \frac{\as}{2\pi} J^g_\text{alg}
\,,\end{align}
with the finite part $J^g_\text{alg}$ given in \eq{eq:gluon_unmeas_finite} in the Appendix.

\section{Soft Functions at $\mathcal{O}(\as)$ for Jet Shapes}
\label{sec:soft}

In this section we compute the soft function for a generic $N$ jet event. In \sec{sec:soft-phase-space}, we describe the phase space cuts that we impose on soft particles emitted into the final state. We then give an outline of how we disentangle the various contributions to the $N$-jet soft function in \sec{sec:soft-outline} and proceed to calculate these contributions in the remaining subsections.

\subsection{Phase Space Cuts}
\label{sec:soft-phase-space}

Soft particles in the final state must satisfy the phase space cuts required by the operator $\delta_{N(\hat{\mathcal{J}}),0}$ in \eq{softfuncdef}, which constrains the soft particles to not create an extra jet.  A soft particle is allowed in the final state if it is emitted into one of the jets as defined by the jet algorithm, or if it is not in a jet but has energy less than a cutoff $\Lambda$.  At NLO, only a single soft gluon can be emitted.  Therefore, for both cone-type and $\kt$-type algorithms, the constraint that the soft gluon be in a jet is simply that the angle of the gluon with respect to the jet axis satisfies $\theta_{gJ} < R$.  Thus, our requirement on soft gluons is that they obey one of the two following conditions:
\begin{align}
\textrm{in jet $i$: }& \theta_{gJ_i} < R \nn \\
\textrm{out of all jets: }& E_g < \Lambda \textrm{ and } \theta_{gJ_i} > R \textrm{ for all } i
\,.\end{align}
These conditions can be written in terms of theta functions on the gluon momentum $k$. We denote the energy restriction for out-of-jet gluons as
\begin{align}
\Theta_\Lambda \equiv \Theta (k^0 < \Lambda)
\,,\end{align}
and we denote the requirement that a gluon be in jet $i$ in terms of the light-cone components $k^{\pm}$ about the direction of jet $i$, $n_i$, as
\begin{align}
\Theta_R^i \equiv \Theta \left( \frac{k^+}{k^-} < \tan^2{\frac{R}{2}}\right)
\,.\end{align}

For the case when the soft gluon is in a measured jet, we demand that it contributes an amount $\tau_a$ to the angularity of a jet with label momentum $\omega$ with the use of the delta function
 \begin{equation}
\delta_R \equiv \delta\left(\tau_a - \frac{1}{\omega}(k^-)^{a/2}(k^+)^{1-a/2}\right)
\,.\end{equation}

\subsection{Calculation of contributions to the $N$-Jet Soft Function}
\label{sec:soft-outline}
\FIGURE[t]{
\includegraphics[totalheight = .15\textheight]{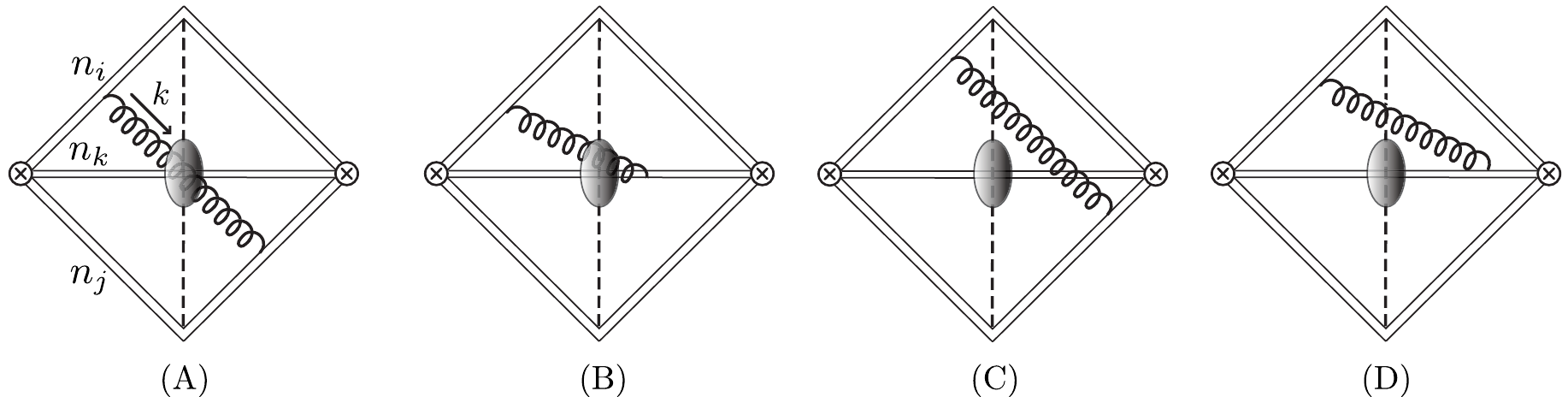}

{ \caption[1]{Soft function real-emission diagrams. Diagrams (A) and (C) are interference diagrams of Wilson line emission from lines $i$ and $j$ and (B) and (D) are from lines $i$ and $k$. The shaded region in the center represents the region of phase space corresponding to jet $k$ defined by the jet algorithm, and so the gluons in diagrams (A) and (B) are inside jet $k$ and those in (C) and (D) are not. Each diagram has a corresponding mirror diagram (not shown).}
\label{fig:soft-function}}
}

The topology of the various graphs that we need to compute the soft function is shown in \fig{fig:soft-function}. The next-to-leading order part $S_{(1)}$ of the soft function $S$ is the sum of the interference of soft gluon emissions from Wilson lines in directions $i$ and $j$, $S_{i j}$, over all lines $i$ and $j$ with $i \neq j$ (since for $i=j$, the diagram is proportional to $n_i^2 = 0$),
\be
S_{(1)} =\sum_\ijpair S_{ij}
\,.\ee
It is conceptually straightforward to see that $S_{ij}$ is just the sum of the following three classifications of the final state cut gluon:
\begin{itemize}
\item The gluon is in a measured jet and thus contributes to the jet angularity.
\item The gluon is outside of all the jets and has energy $E_g < \Lambda$.
\item The gluon is in an unmeasured jet and has any energy.
\end{itemize}
However, the second contribution is technically difficult to compute due to the complicated shape of the space with all jets cut out of it, like Swiss cheese.  A division of phase space leading to a simpler calculation is the following:
\begin{itemize}
\item $S_{ij}^{\meas}(\tau^k_a)$: The gluon is in measured jet $k$ and contributes to the jet's angularity $\tau^k_a$.
\item $S_{ij}^k$: The gluon is in jet $k$ with energy $E_g > \Lambda$ (and does not contribute to $\tau^k_a$).
\item $\bar{S}_{ij}^k$: The gluon is in jet $k$ with energy $E_g < \Lambda$ (and does not contribute to $\tau^k_a$).
\item $S_{ij}^{\incl}$: The gluon is anywhere with $E_g < \Lambda$ (and does not contribute to any angularity).
\end{itemize}
In terms of these pieces, the NLO soft function with $M$ measured jets and $N - M$ unmeasured jets is given by
\begin{equation}
\begin{split}
S_{(1)}(\tau^1_a, \tau^2_a, \dots, \tau_a^M) &=  \sum_{i\neq j}\left[\sum_{k \in \meas} S_{ij}^{\meas}(\tau^k_a)  \prod_{\substack{l = 1 \\ l \neq k}}^M \delta(\tau^l_a)\right] \\
& \left.\qquad +   \sum_{i\neq j}\left[\left( S_{ij}^\incl - \sum_{k \in \meas} \bar{S}_{ij}^k + \sum_{k \notin \meas}S_{ij}^k \right ) \prod_{l=1}^M \delta(\tau^l_a)\right.  \right]
\label{eq:contruct-soft-gen}
\,.
\end{split}
\end{equation}
From the definitions above, it is easy to see that the term in large parentheses on the second line is equivalent to the sum of the last two contributions on the original list above, i.e., the contributions from  a gluon not in any jet with $E_g < \Lambda$ and from a gluon in an unmeasured jet with any energy.

We can simplify this expression by noting that the contribution from a gluon in jet $k$ with no energy restriction involves a scaleless integral over the energy that vanishes in dimensional regularization and thus
\be
\label{eq:s-plus-sbar}
S_{ij}^k + \bar{S}_{ij}^k= 0
\,.\ee
Using this relation, the soft function simplifies to
\begin{equation}
\label{eq:construct-soft}
S_{(1)}(\tau^1_a, \dots, \tau_a^M) = S_{(1)}^\unmeas \prod_{l=1}^M \delta(\tau_a^l) + \sum_{k \in \meas}S_{(1)}^\meas(\tau^k_a)  \prod_{\substack{l=1 \\ l \neq k}}^M \delta(\tau_a^l)
\,,\end{equation}
where the first term in \eq{eq:construct-soft} is a universal contribution that is needed for every $N$-jet observable, defined as
\be
\label{eq:Sunmeas}
S_{(1)}^\unmeas \equiv \sum_\ijpair \Big( S_{ij}^\incl + \sum_{k=1}^N S_{ij}^k \Big)
\,.\ee
The second term, defined as,
\be
\label{eq:Smeas}
S_{(1)}^\meas( \tau_a^k) \equiv \sum_\ijpair S_{ij}^\meas(\tau^k_a)
\,,\ee
depends on our choice of angularities as the observable.

We now proceed to set up the one-loop expressions for the contributions in \eq{eq:construct-soft}. The integrals involved are highly nontrivial and so in this section we simply quote the result of each integral, referring the reader to Appendix~\ref{app:softcalc} for details. Most of these integrals are most easily written in terms of the variable $t_{ij}$, defined in \eq{tdef} as $t_{ij} \equiv \tan\frac{\psi_{ij}}{2}/\tan\frac{R}{2}$, where $\psi_{ij}$ is the angle between jet directions $i$ and $j$. (That is, $n_i \cdot n_j = 1 - \cos \psi_{ij}$.) In \tab{soft}, we summarize the divergent parts of the soft function.

The Feynman rules for the emission of a soft gluon from fundamental- and adjoint-representation Wilson lines (corresponding to quark and gluon jets, respectively) have the same kinematic structure. The difference is entirely encoded in the color charge of the Wilson lines which, using the color space formalism of \cite{Catani:1996jh, Catani:1996vz}, we denote as $\vect T_i$ for emission from Wilson line $i$. Thus, we can consider the $N$-jet soft function without specifying the color representation of the final-state jets.

\subsubsection{Inclusive Contribution: $S_{ij}^{\incl}$}
\label{sec:Sij-incl}

The contribution to the soft function from a gluon going in any direction with energy $E_g < \Lambda$ is given by the integral
\begin{align}
S_{ij}^{\incl} = - g^2 \mu^{2 \epsilon} \, \vect{T}_i \cdot \vect{T}_j  \int\!\frac{d^dk}{(2 \pi)^d} \frac{n_i \cdot n_j}{(n_i \cdot k) (n_j \cdot k)} 2\pi\delta(k^2)\Theta(k^0) \,\Theta_\Lambda
\label{eq:Sincl-integral}
\,.\end{align}
We evaulate this integral in  \sec{sec:app-Sinclij} of the Appendix and find
\begin{align}
S_{ij}^{\incl} = -\frac{\as}{2 \pi} \frac{ \vect{T}_i \cdot \vect{T}_j }{\Gamma(1-\epsilon)} \left( \frac{4\pi \mu^2}{4 \Lambda^2 } \right)^\epsilon   \left( \frac{1}{\epsilon^2} - \frac{1}{\epsilon} \ln{\frac{n_i \cdot n_j}{2}}  - \frac{\pi^2}{6} - \Li_2 \bigg(1-\frac{2}{n_i \cdot n_j}\bigg)\right)
\,.\end{align}

Note that this calculation is related to the inclusive, timelike soft function that has applications elsewhere (see, e.g., \cite{Korchemsky:1993uz, Belitsky:1998tc, Becher:2007ty}), generalized for non back-to-back jets:
\begin{align}
 \frac{d S_{ij}^{\incl}}{d \Lambda} =  - g^2 \mu^{2 \epsilon} \, \vect{T}_i \cdot \vect{T}_j  \int\!\frac{d^dk}{(2 \pi)^d} \frac{n_i \cdot n_j}{(n_i \cdot k) (n_j \cdot k)} 2\pi\delta(k^2)\Theta(k^0) \delta(k^0 - \Lambda)
\,.\end{align}

\subsubsection{Soft gluon inside jet $k$ with $E_g > \Lambda$: {$S_{ij}^k$}}
\label{sec:Sijk}

Using \eq{eq:s-plus-sbar}, the contribution $S_{ij}^k$ from a gluon emitted into jet $k$ from lines $i$ and $j$ is given by the integral
\be
\label{eq:Sijk-integral}
\begin{split}
S_{ij}^k = g^2\mu^{2\epsilon} \, \vect T_i\cdot \vect T_j &\int\frac{d^dk}{(2\pi)^d}\frac{n_i\cdot n_j}{(n_i\cdot k)(n_j\cdot k)}2\pi\delta(k^2)\Theta(k^0) \, \Theta_R^k\Theta_\Lambda
\end{split}
\,.\ee
Much like for the $S_{ij}^{\meas}$ contribution, if $k = i,j$, there is an additional divergence (arising from $n_k \cdot k \to 0$) relative to the case $k \neq i,j$, and so we evaluate these two cases separately below.

\paragraph{Case 1: $k=i$ or $j$} The integrals for this case are performed in \sec{app:Siji-and-Sijmeas} of the Appendix, with the result that $S_{ij}^j$ is
\begin{align}
S_{ij}^j = S_{ij}^i &= \frac{\alpha_s \vect T_i \cdot \vect T_j}{4\pi} \Bigg[\frac{1}{\epsilon^2}\frac{1}{\Gamma(1-\epsilon)} \left(\frac{4 \pi\mu^2}{4\Lambda^2} \right)^\epsilon \left(\frac{t_{ij}^2}{t_{ij}^2 - 1}\tan^2\frac{R}{2}\right)^{-\epsilon}  + \Li_2 \bigg( \frac{-1}{t_{ij}^2 - 1}\bigg) \nn\\
& \qquad \qquad \qquad + 2 \Li_2 \bigg( \frac{-1}{\cos^2{\frac{\psi_{ij}}{2}}(t_{ij}^2 - 1)}  \bigg) \Bigg]
\label{eq:Sijj-ans}
\,.\end{align}

\paragraph{Case 2: $k \neq i,j$}

These contributions are at most $1/\epsilon$ divergent since the matrix element does not have the $n_k\cdot k \to 0$ singularity. We show in Appendix~\ref{app:Sijk} that the result takes the form
\begin{align}
S_{ij}^k \big\vert_{k \neq i,j} &=  -\frac{\alpha_s }{4\pi} \vect T_i\cdot \vect T_j \,\bigg[ \frac{1}{\epsilon}\ln\bigg(\frac{t_{ik}^2 t_{jk}^2 - 2 t_{ik} t_{jk} \cos\beta_{ij}+1}{(t_{ik}^2 - 1)(t_{jk}^2 - 1)} \bigg)+ F(t_{ik}, t_{jk}, \beta_{ij}) \bigg]
\label{eq:Sijk-form}
\,,\end{align}
where $\beta_{ij}$ is the angle between the $i$-$k$ and $j$-$k$ planes and the finite function $F$ is given in \eq{eq:F-defn} and is $\cO(1/t^2)$.


\subsubsection{Soft gluon inside measured jet $k$: $S_{ij}^{\meas}(\tau_a^k)$}
\label{sec:Sij-meas}

The contribution of a gluon emitted into jet 1 (the measured jet) from  lines $i$ and $j$ is given by the integral
\be
\label{eq:Sijmeas-integral}
S_{ij}^{\meas}(\tau^k_a) = -g^2\mu^{2\epsilon} \, \vect T_i\cdot \vect T_j \int\frac{d^dk}{(2\pi)^d}\frac{n_i\cdot n_j}{(n_i\cdot k)(n_j\cdot k)}2\pi\delta(k^2)\Theta(k^0)\,  \Theta^k_R \delta_R
\,.\ee
The singularity structure of this integral depends on whether or not $k$ = $i$ or $j$.
Thus, we evaluate the case $k$ = $i$ or $j$ and the case $k \neq i,j$ separately below.

\paragraph{Case 1: $k = i$ or $j$} We consider first $S_{ij}^{\meas}(\tau^i_a)$.  Using the results of  \sec{app:Siji-and-Sijmeas} of the Appendix, we obtain the result in terms of $t_{ij}$,
\begin{align}
S_{ij}^{\meas}(\tau_a^i) &= S_{ji}^{\meas}(\tau_a^i) \nn\\
&= -\frac{\alpha_s}{2\pi} \vect T_i\cdot \vect T_j \biggl[\frac{1}{\epsilon}\frac{1}{1-a} \bigg(\frac{1}{ \tau_a^i}\bigg)^{1+2\epsilon} \frac{1}{\Gamma(1-\epsilon)} \bigg(\frac{4 \pi\mu^2}{\omega^2}\bigg)^\epsilon \bigg(\frac{t_{ij}^2}{t_{ij}^2-1}\tan^2\frac{R}{2}\bigg)^{\epsilon(1-a)} \nn\\
& \qquad \qquad \qquad \qquad \qquad+ \frac{1+a}{2} \delta(\tau^i_a)\Li_2\bigg(\frac{-1}{t_{ij}^2-1}\bigg)\biggr]
\label{eq:Smeasiji-ans}
\,.\end{align}

\paragraph{Case 2: $k\neq i,j$} The remaining contributions to the observed jet angularity are $S_{ij}^{\meas}$ for $k \neq i, j$.
Using the results from \sec{app:Smeas-ijk} in the Appendix, this contribution is
\begin{align}
S_{ij}^{\meas}(\tau^k_a)\big\vert_{i,j \neq k} &=  - \frac{\alpha_s }{2\pi} \vect T_i\cdot \vect T_j \bigg[\bigg( \frac{1}{\tau_a^k}\bigg)^{1+2\epsilon} \ln\bigg(\frac{t_{ik}^2 t_{jk}^2 - 2 t_{ik} t_{jk} \cos\beta_{ij}+1}{(t_{ik}^2 - 1)(t_{jk}^2 - 1)} \bigg)\nn\\
& \qquad + \delta(\tau_a^k) \, G(t_{ik}, t_{jk}, \beta_{ij}) \bigg]
\label{eq:Sijk-meas-form}
\,,\end{align}
where $G$ is $\cO(1/t^2)$ and is given in \eq{eq:G-defn} and, again, $\beta_{ij}$ is the angle between the $i$-$k$ and $j$-$k$ planes.

\subsection{Total $N$-Jet Soft Function in the large-$t$ Limit}
\label{sec:large-t-njet}

In this section, we add together the necessary ingredients calculated above
to obtain the $N$-jet soft function from \eq{eq:construct-soft}. The results for the divergent pieces are summarized in \tab{soft}. In \sec{sec:cons} we use \tab{soft} to show that the consistency of factorization is explicitly violated by terms of order $1/t^2$,
and so in this section we give the full soft function (including the finite terms) to $\cO(1/t^2)$.

\TABLE[t]{
\begin{tabular}{||c | c @{}c ||}
\hline\hline
\tabrule
contribution & divergent terms & \\
\hline\hline
\tabrule
$S_{ij}^{\incl}$ & $-\frac{1}{\epsilon}\frac{\as}{2\pi}\vect{T}_i\cdot\vect{T}_j \,\Big( \frac{1}{\epsilon}  -  \ln \frac{n_i\cdot n_j}{2} + \ln \frac{\mu^2}{4 \Lambda^2}\Big)$ & \\
\tabrule
$S_{ij}^i$ & $\frac{1}{\epsilon} \frac{\as}{4\pi}\vect{T}_i\cdot\vect{T}_j \Big(\frac{1}{\epsilon} - \ln\frac{t_{ij}^2\tan^2(R/2)}{t_{ij}^2-1} + \ln \frac{\mu^2}{4 \Lambda^2}\Big)$ & \\
\rule{-2pt}{3ex} \rule[-2ex]{0pt}{0pt}
$S_{ij}^k$ &  $- \frac{1}{\epsilon}\frac{\as}{4\pi}\vect{T}_i\cdot\vect{T}_j \ln\frac{t_{ik}^2t_{jk}^2- 2t_{ik}t_{jk}\cos\beta_{ij}+1}{(t_{ik}^2 - 1)(t_{jk}^2 - 1)}$   &\\
\hline
\tabrule
$S_{(1)}^{\unmeas}$ & $\frac{1}{\epsilon}\frac{\as}{2\pi} \Big[  \sum_{i=1}^N \vect{T}_i^2 \ln\tan^2(R/2)  + \sum_{i\not =j}\vect{T}_i\mcdot\vect{T}_j \ln(n_i\cdot n_j/2)\Big]+ \mathcal{O}(1/t^2)$ & \\
\hline
\hline
\tabrule
$S_{ij}^{\meas}(\tau_a^i)$ & $\frac{1}{\epsilon}\frac{\as}{4\pi} \vect{T}_i\cdot\vect{T}_j \Big[ \Big(\frac{1}{1-a}\big(\frac{1}{\epsilon} + \ln \frac{\mu^2}{\omega_i^2}\big)+ \ln\frac{t_{ij}^2\tan^2(R/2)}{t_{ij}^2-1} \Big)\delta(\tau_a^i)  - \frac{2}{1-a} \Big(\frac{1}{\tau_a^i} \Big)_{\!\!+} \Big]$  &\\
\tabrule
$S_{ij}^{\meas}(\tau_a^k)$ & $ \frac{1}{\epsilon}\frac{\as}{4\pi}\vect{T}_i\cdot\vect{T}_j \ln\frac{t_{ik}^2t_{jk}^2- 2t_{ik}t_{jk}\cos\beta_{ij}+1}{(t_{ik}^2 - 1)(t_{jk}^2 - 1)} \delta(\tau_a^k)$ &  \\
\hline
\tabrule
$S_{(1)}^{\meas}(\tau_a^i)$ & $ - \frac{1}{\epsilon}\frac{\as}{2\pi} \vect{T}_i^2 \Big[ \Big(\frac{1}{1-a}\big(\frac{1}{\epsilon} + \ln \frac{\mu^2}{\omega_i^2}\big)+ \ln\tan^2(R/2) \Big)\delta(\tau_a^i)  - \frac{2}{1-a} \Big(\frac{1}{\tau_a^i} \Big)_{\!\!+} \Big] + \mathcal{O}(1/t^2)$  &\\
\hline\hline
\end{tabular}
\caption{Summary of the divergent parts of the soft function at NLO. The first block contains the the observable-independent contributions:  soft gluons emitted by jets $i$ and $j$ in any direction with energy $E_g<\Lambda$ in row 1; soft gluons entering jet $k$ with $E_g>\Lambda$ (with $k=i$ or $j$ in the second row and $k \neq i,j$ in the third). In the last row of the first block, we summed over $i$ and $j$ and took the large-$t$ limit to get the total $S_{(1)}^\unmeas$. Similarly, in the second block we give the contributions to the angularities $\tau_a^k$ (with $k=i$ or $j$ in the first row and $k \neq i,j$ in the second) and summed over $i$ and $j$ and took the large-$t$ limit to get $S_{(1)}^\meas$ in the third row.}
\label{tab:soft} }

Using color-conservation ($\sum_i \vect{T}_i = 0$), we find that the observable-independent part, $S_{(1)}^\unmeas$,  defined in \eq{eq:Sunmeas}, is given for large $t$ by
\begin{align}
\label{eq:Sunmeas-result}
S_{(1)}^{\unmeas}
&= \frac{\as}{2\pi}   \sum_i \vect{T}_i^2 \Biggl[ \frac{1}{\epsilon}  \ln\left(\frac{\mu^2}{4\Lambda^2}\right) - \frac{1}{\epsilon}\ln\left(\frac{\mu^2}{4\Lambda^2\tan^2\frac{R}{2}}\right) \\
&\quad  + \frac{1}{2}\ln^2\left(\frac{\mu^2}{4\Lambda^2}\right) - \frac{1}{2} \ln^2\left(\frac{\mu^2}{4\Lambda^2\tan^2\frac{R}{2}}\right)
- \frac{\pi^2}{6}  \Biggr] \nn\\
& \quad +  \frac{\as}{2\pi}  \sum_\ijpair  \vect{T}_i \cdot \vect{T}_j \Biggl[  \frac{1}{\epsilon}  \ln{\frac{n_i \cdot n_j}{2}}+ \ln\left(\frac{\mu^2}{4\Lambda^2}\right)\ln\left(\frac{n_i\cdot n_j}{2}\right) \nn \\
&\qquad \qquad \qquad \qquad + \Li_2 \bigg(1-\frac{2}{n_i \cdot n_j}\bigg) \Biggr]  + \cO(1/t^2) \nn
\,.\end{align}
We see that the finite parts of this contribution are sensitive to two scales, $2\Lambda$ and $2\Lambda\tan\frac{R}{2}$. For simplicity, in this paper, since we take $\tan(R/2)\sim\cO(1)$, we will choose only a single scale $\mu_S^\Lambda$ to attempt to minimize logs in \eq{eq:Sunmeas-result}, where
\be
\label{eq:mu_lambda-defn}
\mu_S^\Lambda  \equiv 2 \Lambda \tan^{1/2} \frac{R}{2} 
\,,\ee
chosen as the geometric mean of the two.

The remaining part of the soft function that is dependent on angularities as our choice of jet observable is the sum over $S_{(1)}^\meas(\tau_a^i)$ (defined in \eq{eq:Smeas}) for each jet $i$, where $S_{(1)}^\meas(\tau_a^i)$ is given by
\begin{align}
\label{eq:Smeas-result}
S_{(1)}^\meas(\tau_a^i) &= - \frac{\as}{2\pi} \vect{T}_i^2 \frac{1}{1-a}\Biggl\{  \bigg[ \frac{1}{\epsilon^2} + \frac{1}{\epsilon} \ln\left( \frac{\mu^2}{\omega_i^2}\tan^{2(1-a)}\frac{R}{2}\right) - \frac{\pi^2}{12} \nn\\
& \qquad \qquad \qquad \qquad \qquad + \frac{1}{2} \ln^2 \bigg( \frac{\mu^2}{\omega_i^2} \tan^{2(1-a)} \frac{R}{2}\bigg)\bigg]  \delta(\tau_a^i)  \\
& \qquad \qquad \qquad- 2 \Biggl[\Bigg(\frac{1}{\epsilon} +\ln  \bigg( \frac{\mu^2  \tan^{2(1-a)}\frac{R}{2}}{(\omega_i \tau_a^i)^2} \bigg)\Bigg) \frac{\Theta(\tau_a^i)}{\tau_a^i} \Biggr]_+  \Biggr\}+ \cO(1/t^2) \nn
\,.\end{align}
The finite part of this contribution is sensitive to the scale $\mu_S^i$, where
\be
\label{eq:mu_S-defn}
 \mu_S^i \equiv \frac{\omega_i \tau_a^i}{ \tan^{1-a} \frac{R}{2}}
\,,\ee
which, in principle, differs for each jet $i$ and from the scale $\mu_S^\Lambda$ in the unmeasured part of the soft function \eq{eq:Sunmeas-result}. After discussing resummation of large logarithms through RG evolution, we will describe in \sec{ssec:refactorization}  a framework  to ``refactorize'' the soft function into pieces depending on multiple separated soft scales. This framework will enable us to resum logarithms of all of these potentially disparate scales.

\section{Resummation and Consistency Relations at NLL}
\label{sec:cons}
The factorized cross section  \eq{Njetcs} is written in terms of hard, jet, and soft functions evaluated at a factorization scale $\mu$. Since the physical cross section is independent of $\mu$, the anomalous dimensions of these functions are closely related. We derive and verify this relation in \sec{ssec:consistency}. In \sec{ssec:genericRGE} and \sec{ssec:allRGE}, we work out the form of the renormalization-group equations (RGEs) satisfied by the hard, jet, and soft functions, and obtain their solutions so that we can express each function at the scale $\mu$ in terms of its value at a scale $\mu_0$ where logarithms in it are minimized.  In \sec{ssec:refactorization}, we present a framework to refactorize the soft function and give the total resummed distribution in \sec{ssec:totalNLLdist}.

\subsection{General Form of Renormalization Group Equations and Solutions}
\label{ssec:genericRGE}

We will build solutions for the hard, jet, and soft functions from two forms of RGEs. The first form is for a function which does not depend on the observable  $\tau_a$ and is multiplicatively renormalized,
 \be
 F^{\text{bare}} = Z_F(\mu)F(\mu)\,,
 \ee
 and satisfies the RGE,
\be
\label{FRGE}
\mu \frac{d}{d\mu}F(\mu) =  \gamma_F(\mu) F(\mu) \,,\ee
where the anomalous dimension $\gamma_F$ is found from the $Z$-factor by the relation
\be
\label{eq:anom-from-Z-mult}
\gamma_F(\mu) = - \frac{1}{Z_F(\mu)}\mu \frac{d}{d \mu} Z_F(\mu)
\,,\ee
and takes the general form,
\begin{equation}
\label{gammaF}
\gamma_F(\mu) = \Gamma_F[\alpha] \ln\frac{\mu^2}{\omega^2} + \gamma_F[\alpha]\,.
\end{equation}
We call $\Gamma_F[\alpha]$ the ``cusp part'' of the anomalous dimension and $\gamma_F[\alpha]$ the ``non-cusp part''. They have the perturbative expansions
\begin{align}
\label{eq:Gammaexpansion}
\Gamma_{\!F}[\as] = \left( \frac{\as}{4\pi}\right) \Gamma_{\!F}^0 + \left( \frac{\as}{4\pi}\right)^2 \Gamma_{\!F}^1 + \cdots
\end{align}
and
\begin{align}
\label{eq:gammaexpansion}
\gamma_F[\as] = \left( \frac{\as}{4\pi}\right) \gamma_{F}^0 + \left( \frac{\as}{ 4\pi}\right)^2 \gamma_{F}^1 + \cdots
\,.\end{align}
The RGE \eq{FRGE} has the solution
\begin{align}
F(\mu)
& = U_F(\mu,\mu_0)F(\mu_0) \,,
\end{align}
where the evolution kernel $U_F$ is given by
\begin{equation}
\label{kernel}
U_F(\mu,\mu_0)= e^{K_F(\mu,\mu_0)} \left(\frac{\mu_0}{\omega}\right)^{\omega_F(\mu,\mu_0)}\,,
\end{equation}
where $K_F$ and $\omega_F$ will be defined below in \eq{eq:kernelparams}.

The second form of RGE is for a function dependent on the jet shape $\tau_a$ and is renormalized through a convolution,
\begin{equation}
F^{\text{bare}}(\tau_a) = \int d\tau_a' Z_F(\tau_a - \tau_a';\mu)F(\tau_a',\mu)\,,
\end{equation}
and satisfying the RGE
\begin{align}
   \label{RGEtau}
  & \mu \frac{d}{d\mu} F(\tau_a; \mu)= \int_{}^{}d \tau_a' \, \gamma_F (\tau_a-\tau_a'; \mu) F(\tau_a'; \mu)
\,,\end{align}
with an anomalous dimension calculated from
\be
\label{eq:anom-from-Z-convolve}
\gamma_F(\tau_a; \mu) = - \int\! d\tau'\, Z_F^{-1} (\tau_a - \tau_a'; \mu) \, \mu \frac{d}{d \mu} Z_F(\tau_a'; \mu)
\,,\ee
and taking the general form
\begin{align}
\label{gammaFtau}
\gamma_F (\tau_a; \mu) = -  \Gamma_F [\as] \left( \frac{2}{j_F} \left[ \frac{\Theta(\tau_a)}{\tau_a}\right]_+ -\ln \frac{\mu^2}{\omega^2} \,\delta(\tau_a)\right)  + \gamma_F[\as] \delta(\tau_a)
\,.\end{align}
The solution of an RGE of the form \eq{RGEtau} has the solution \cite{Korchemsky:1993uz,Becher:2006mr,Balzereit:1998yf,Neubert:2005nt,Fleming:2007xt}
\begin{align}
F(\tau_a; \mu) = \int\!d \tau' \, U_F(\tau_a-\tau_a'; \mu, \mu_0) F(\tau_a'; \mu_0)
\label{Fconvol}
\,,\end{align}
where the evolution kernel $U_F$ is given to all orders in $\alpha_s$  by the expression
\begin{align}
 U_F(\tau_a; \mu, \mu_0)= \frac{e^{{K}_F + \gamma_E{\omega}_F}}{\Gamma(-{\omega}_F)} \left(\frac{\mu_0}{\omega}\right)^{j_F\omega_F} \left[\frac{\Theta(\tau_a)}{(\tau_a)^{1+{\omega}_F}}\right]_+
 \label{kernelF}
\,,\end{align}
where $\gamma_E$ is the Euler constant.

In \eqs{kernel}{kernelF}, the exponents ${\omega}_F$ and ${K}_F$ are given in terms of the cusp and non-cusp parts of the anomalous dimensions by the expressions
\begin{subequations}
   \label{eq:kernelparams}
\begin{align}
\label{omegaF}
  {\omega}_F(\mu,\mu_0) & \equiv \frac{2}{j_F}\int_{\as(\mu_0)}^{\as(\mu)}\frac{d\alpha}{\beta[\alpha]} \Gamma_{\!F}[\alpha] \,, \\
   \label{KF}
   {K}_F(\mu,\mu_0)& \equiv \int_{\as(\mu_0)}^{\as(\mu)}\frac{d\alpha}{\beta[\alpha]} \gamma_F[\alpha]+2\int_{\as(\mu_0)}^{\as(\mu)}\frac{d\alpha}{\beta[\alpha]} \Gamma_{\!F}[\alpha]\int_{\as(\mu_0)}^{\alpha}\frac{d \alpha'}{\beta[{\alpha'}]}
\,.\end{align}
\end{subequations}
In the case of \eq{kernel} or if $\Gamma_F[\alpha]$ happens to be zero, we define $j_F$ to be 1. To achieve NLL accuracy in the evolution kernels $U_F$,  we need the cusp part of the anomalous dimension to two loops and the non-cusp part to one loop, in which case the parameters $\omega_F,K_F$ in \eq{eq:kernelparams} are given explicitly by
\begin{subequations}
  \label{eq:kernelparamsNLL}
\begin{align}
\label{omegaFNLL}
  \omega_F(\mu, \mu_0) &=-\frac{\Gamma_{\!F}^0}{j_F\, \beta_0} \left[\ln{r}+\left(\frac{\Gamma_{\cusp}^1}{\Gamma_{\cusp}^0}-\frac{\beta_1}{\beta_0}\right)\frac{\as(\mu_0)}{4\pi}(r-1)\right] \,,\\
  \label{KFNLL}
  K_F(\mu_,\mu_0) &=-\frac{\gamma_{F}^0}{2\beta_0}\ln {r} - \frac{2\pi\Gamma_{\!F}^0}{(\beta_0)^2}\bigg[\frac{r-1-r\ln{r}}{\as(\mu)} \nn\\
  & \qquad \qquad \qquad\qquad  +\left(\frac{\Gamma^1_{\cusp}}{\Gamma^0_{\cusp}}-\frac{\beta_1}{\beta_0}\right)\frac{1-r+\ln{r}}{4\pi}+\frac{\beta_1}{8\pi\beta_0}\ln^2{r}\bigg]
\,.\end{align}
\end{subequations}
Here $r=\frac{\as(\mu)}{\as(\mu_0)}$, and $\beta_0, \beta_1$  are the one-loop and two-loop  coefficients of the beta function,
\begin{equation}
\beta[\alpha_s] = \mu\frac{d \alpha_s}{d \mu} = -2\alpha_s\left[\beta_0\left(\frac{\alpha_s}{4\pi}\right) + \beta_1\left(\frac{\alpha_s}{4\pi}\right) ^2 +\cdots\right]\,,
\end{equation}
where (with $\TR = 1/2$)
\begin{align}
   \beta_0=\frac{11C_A}{3} - \frac{2\NF}{3} \qquad {\rm and } \qquad \beta_1= \frac{34C_A^2}{3}-\frac{10C_A \NF}{3} -2\CF \NF
\,.\end{align}
The two-loop running coupling $\as(\mu)$ at any scale $\mu$ in terms of its value at another scale $Q$ is given by
\begin{equation}
\label{eq:2loopalphas}
\frac{1}{\as(\mu)} = \frac{1}{\as(Q)} + \frac{\beta_0}{2\pi}\ln\left(\frac{\mu}{Q}\right) + \frac{\beta_1}{4\pi\beta_0}\ln\left[1+\frac{\beta_0}{2\pi}\as(Q)\ln\left(\frac{\mu}{Q}\right)\right]\,.
\end{equation}
In \eq{eq:kernelparamsNLL}, we have used the fact that, for the hard, jet, and soft functions for which we will solve, the cusp part of the anomalous dimension $\Gamma_F[\alpha_s]$ is proportional to \emph{the} cusp anomalous dimension $\Gamma_{\rm cusp}[\alpha_s]$, where
 \begin{align}
 \label{gamma-gammacusp}
 & \Gamma_{\cusp}[\as]=\left(\frac{\as}{4\pi}\right)\Gamma^0_{\cusp}+\left(\frac{\as}{4\pi}\right)^2 \Gamma^1_{\cusp}+\cdots
 \,.\end{align}
The ratio of the one-loop and two-loop coefficients of $\Gamma_{\rm cusp}$ is \cite{Korchemsky:1987wg}
 \begin{align}
 &\frac{\Gamma_{\cusp}^1}{\Gamma^0_{\cusp}}=\left(\frac{67}{9}-\frac{\pi^2}{3}\right)C_A-\frac{10 \NF}{9}
\,.\end{align}
$\Gamma_{\cusp}^1$ and $\beta_1$ are needed in the expressions of $\omega_F$ and $K_F$ for complete NLL resummation since we formally take $\as^2 \ln\tau_a \sim \mathcal{O}(\as)$.

\subsection{RG Evolution of Hard, Jet, and Soft Functions}
\label{ssec:allRGE}

\subsubsection{Hard Function}

The hard function is related to the matching coefficient $C_N$ of the $N$-jet operator in \eq{Njetmatching}. If there are multiple operators with different color structures, $C_N$ is a vector of coefficients. The hard function is then a matrix $H_{ab} = C^\dag_b C_a$\,. The hard function is contracted in the cross section \eq{Njetcs} with a matrix of soft functions.

The anomalous dimensions of the matching coefficients $C_a$ have been calculated in the literature (for example, Table III of Ref.~\cite{Chiu:2009mg}). For an operator with $N$ legs with color charges $\vect{T}_i^2$, the anomalous dimension is
\begin{equation}
\label{gammaCN}
\gamma_{C_N}(\alpha_s) = -\sum_{i=1}^N \left[\vect{T}_i^2 \Gamma(\alpha_s) \ln\frac{\mu}{\omega_i} + \frac{1}{2}\gamma_i(\alpha_s)\right] - \frac{1}{2}\Gamma(\alpha_s)\sum_{i\not = j}\vect{T}_i\mcdot\vect{T}_j \ln\left(\frac{-n_i\mcdot n_j - i0^+}{2}\right)\,,
\end{equation}
where $\vect{T}_i$ is a matrix of color charges in the space of operators, and $\gamma_i$ is given for quarks and gluons by
\be
\label{gammaqg}
\gamma_q = \frac{3\alpha_s C_F}{2\pi}\ , \quad \gamma_g = \frac{\alpha_s}{\pi}\frac{11 C_A - 2 \NF}{6}\,.
\ee
The coefficient $\Gamma(\as)$ is the \emph{cusp anomalous dimension} and is given to one-loop order by $\Gamma(\as) = \as/\pi$.
The anomalous dimension of the hard function itself is given by $\gamma_H = \gamma_{C_N}^\dag + \gamma_{C_N}$, and takes the form of \eq{gammaF}, with cusp and non-cusp parts $\Gamma_H[\alpha_s]$ and $\gamma_H[\alpha_s]$ given to one loop in \tab{anomalous-coeff}\,, with the result
\begin{equation}
\gamma_H(\as) = -\Gamma(\as) \vect{T}^2 \ln\frac{\mu^2}{\bar\omega_H^2}- \sum_{i=1}^N \gamma_i(\as)  - \Gamma(\as) \sum_{i\not=j}\vect{T}_i\mcdot\vect{T}_j\ln\frac{n_i\mcdot n_j}{2}\,,
\end{equation}
where $\vect{T}^2 = \sum_{i=1}^N \vect{T}_i^2$ is the sum of all the Casimirs, and the effective hard scale $\bar\omega_H$ appearing as the scale $\omega$ in the logarithm in \eq{gammaF} is given by the color-weighted average of the jet energies,
\begin{equation}
\label{omegaH}
\bar\omega_H = \prod_{i=1}^N \omega_i^{\vect{T}_i^2/\vect{T}^2}
\end{equation}

\TABLE[t]{
\begin{tabular}{ | c || c | @{}c | c | c |}
\hline
 & $ \Gamma_F[\alpha_s] $ & $ \gamma_F[\alpha_s] $ & $j_F$ & $\omega$ \\
\hline
\hline
$ \gamma_H $ & \tabrule $-\Gamma \sum_i \vect{T}_i^2$ & $-\sum_i  \gamma_i - \Gamma\sum_{\ijpair} \vect{T}_i \cdot \vect{T}_j   \ln \frac{n_i \cdot n_j}{2} $ & $1$ & $\bar\omega_H$ \\
\hline
$\gamma_{J_{i}}(\tau_a^i)$ &  \tabrule  $\Gamma  \vect{T}_i ^2 \frac{2-a}{1-a}$ & $\gamma_i $  & $2-a$ & $\omega_i$ \\
$\gamma_S^\meas(\tau_a^i)$\!\! & \tabrule $  - \Gamma \vect{T}_i^2 \frac{1}{1-a}$ & $0$  & $1$& $\omega_i \tan^{-1+a}\frac{R}{2}$ \\
\hline
$\gamma_{J_{i}}$ & \tabrule $\Gamma \vect{T}_i^2$  & $\gamma_i$  & $1$ & $\omega_i\tan\frac{R}{2}$ \\
\hline
$\gamma_S^\unmeas$ & \tabrule $0$ & $\Gamma\sum_i \vect{T}_i^2 \ln \tan^2\frac{R}{2} + \Gamma\sum_{\ijpair} \vect{T}_i \cdot \vect{T}_j  \ln \frac{n_i \cdot n_j}{2}$  & $1$ & --- \\
\hline
$\cO(1/t^2)$ & $0$ & \rule{-5pt}{3.5ex} \rule[-3.2ex]{0pt}{0pt} $ \Gamma \sum_{\ijpair} \vect{T}_i \cdot \vect{T}_j \Big[ \delta_{i \notin \meas}  \,2 \ln\frac{t_{ij}^2}{t_{ij}^2 - 1} $ & $1$ & --- \\
& & \rule{-5pt}{3.5ex} \rule[-3.2ex]{0pt}{0pt} $ \qquad + \Gamma \sum_{\!\!\substack{ k \neq i,j \\ k \notin \meas}} \ln\Big(\frac{t_{ik}^2 t_{jk}^2 - 2 t_{ik} t_{jk} \cos\beta_{ij}+1}{(t_{ik}^2 - 1)(t_{jk}^2 - 1)} \Big) \Big] $  &  & \\
\hline
\end{tabular}
{ \caption[1]{Anomalous dimensions for the jet and soft functions. We give the cusp and non-cusp parts of the anomalous dimensions, $\Gamma_F[\as]$ and $\gamma_F[\as]$. $\Gamma$ is the cusp anomalous dimension itself, equal to $\as/\pi$ at one loop. $\gamma_i$ is given for quark and gluon jets in \eq{gammaqg}. The third column gives the value of $j_F$ appearing in \eq{gammaFtau} or \eq{eq:kernelparams}. The last column gives the values of $\omega$ appearing in the logarithm $\ln\mu^2/\omega^2$ multiplying the cusp part of the anomalous dimension in \eqs{gammaF}{gammaFtau}. The scale $\bar\omega_H$ is the color-weighted averages of all jet energies defined in \eq{omegaH}. All rows except for the last are given in the large-$t$ limit and in the last row we give the remaining terms that are present for arbitrary $t$.
This last row explictly violates consistency at $\cO(1/t^2)$. The first group of rows are needed for measured jets and the second group for unmeasured jets. In the large-$t$ limit, for any number of measured and unmeasured jets, the consistency relation \eq{eq:consistency} is satisfied.}
\label{tab:anomalous-coeff} }}

\subsubsection{Jet Functions}

There are two forms of jet functions, those for measured and those for unmeasured jets.  Unmeasured jet functions $J_\omega^{q,g}(\mu)$ satisfy multiplicative RGEs, with anomalous dimensions given by the infinite parts of \eqs{Jq}{G2result},
\begin{equation}
\gamma_{J_i} = \Gamma(\alpha_s) \vect{T}_i^2 \ln \frac{\mu^2}{\omega_i^2\tan^2\frac{R}{2}} + \gamma_i\,,
\end{equation}
which falls into the general form \eq{gammaF}, with cusp and non-cusp parts of the anomalous dimension given in \tab{anomalous-coeff}, and the scale $\omega$ in \eq{gammaF} being $\omega_i\tan\frac{R}{2}$. The part $\gamma_i$ is given by \eq{gammaqg}.

Measured jet functions satisfy RGEs of the form \eq{RGEtau}, with anomalous dimensions given by the infinite parts of \eqs{Jtotal}{Gjet},
\begin{equation}
\gamma_{J_i}(\tau_a^i) = \left[\vect{T}_i^2\,\Gamma(\alpha_s)\frac{2-a}{1-a}\ln\frac{\mu^2}{\omega_i^2} + \gamma_i\right]\delta(\tau_a^i) - 2\Gamma(\alpha_s)\vect{T}_i^2 \frac{1}{1-a}\left[\frac{\Theta(\tau_a^i)}{\tau_a^i}\right]_+\,,
\end{equation}
which takes the form \eq{gammaFtau} with cusp and non-cusp parts of the anomalous dimension split up as in \tab{anomalous-coeff}, and the scale $\omega$ in \eq{gammaFtau} being $\omega_i$.

\subsubsection{Soft Function}

The total $N$-jet soft function is given by \eq{eq:Sunmeas-result} for unmeasured jets added to the sum over measured jets of \eq{eq:Smeas-result}. This soft function depends on the $M$ jet shapes $\tau_a^1,\dots,\tau_a^M$,  and satisfies the RGE
\begin{equation}
\label{softRGE}
\mu\frac{d}{d\mu} S(\tau_1,\dots,\tau_M;\mu) = \int d\tau_1' \cdots d\tau_M' \, \gamma_S(\tau_1 - \tau_1',\dots,\tau_M-\tau_M';\mu) S(\tau_1',\dots,\tau_M';\mu) \,.
\end{equation}
From the infinite parts of the soft function given in \tab{soft}, we find the anomalous dimension $\gamma_S(\tau_1,\dots,\tau_M;\mu)$ decomposes, as required by the consistency condition \eq{eq:consistency} given below, into a sum of terms,
\begin{equation}
\gamma_S(\tau_1,\dots,\tau_M;\mu) = \gamma_S^{\text{unmeas}}(\mu)\delta(\tau_1)\cdots\delta(\tau_M) + \sum_{k=1}^M \gamma_S^{\text{meas}}(\tau_k;\mu)\prod_{j\not = k} \delta(\tau_j)\,,
\end{equation}
where the pieces $\gamma_S^\unmeas(\mu)$ and $\gamma_S^\meas(\tau_k;\mu)$ are given in terms of their cusp and non-cusp parts in \tab{anomalous-coeff}, with the result
\begin{equation}
\gamma_S^\unmeas(\mu) = \sum_i^N \Gamma(\as)\vect{T}_i^2 \ln\tan^2\frac{R}{2} + \Gamma(\as)\sum_{i\not=j} \vect{T}_i\cdot\vect{T}_j \ln\frac{n_i\cdot n_j}{2}\,,
\end{equation}
which takes the form of \eq{gammaF} with no cusp part, and
\begin{equation}
\gamma_S^\meas (\tau_k;\mu) = -\Gamma(\as) \vect{T}_k^2\frac{1}{1-a}\biggl\{ \ln\left(\frac{\mu^2\tan^{2(1-a)}\frac{R}{2}}{\omega_k^2}\right) \delta(\tau_k) - 2\left[\frac{\Theta(\tau_k)}{\tau_k}\right]_+\biggr\}\,,
\end{equation}
which takes the form of \eq{gammaFtau} with no non-cusp part, and the scale $\omega$ in \eq{gammaFtau} being $\omega_k/\tan^{1-a}\frac{R}{2}$.

The solution of the RGE \eq{softRGE} is given by
\begin{equation}
\label{softsolution}
\begin{split}
S(\tau_1,\dots,\tau_M;\mu) &= \int\!d\tau_1'\cdots d\tau_M' \,S(\tau_1',\dots,\tau_M';\mu_0)    U_S^{\text{unmeas}}(\mu,\mu_0)\prod_{k=1}^M U_S^k(\tau_k-\tau_k';\mu,\mu_0)\,,
\end{split}
\end{equation}
where $U_S^{\text{unmeas}}$ is given by the form of \eq{kernel} and $U_S^k(\tau_a^k)$ by the form of \eq{kernelF}.

The solution \eq{softsolution} is appropriate if all the scales appearing in the soft function are similar, and thus all large logarithms in the finite part can be minimized at a single scale $\mu_0$. As we noted in \sec{sec:large-t-njet}, however, the potentially disparate scales $\omega_i\tau_a^i$, set by the jet shapes of the measured jets, and $\Lambda$, set by the cutoff on particles outside jets, appear together in the soft function, and logarithms of ratios of these scales may be large. In this case, the soft function should be ``refactorized'' into pieces depending only on one of these scales at a time. We describe a framework for doing so below in \sec{ssec:refactorization}.

But first, we verify the consistency of the anomalous dimensions for the hard, jet, and soft functions to the order we have calculated them.

\subsection{Consistency Relation among Anomalous Dimensions}
\label{ssec:consistency}

We summarize the anomalous dimensions of the hard, jet, and soft functions in \tab{anomalous-coeff}. We separate contributions to the jet and soft anomalous dimensions that arise from measured jets, from unmeasured jets, and those that are universally present. In all rows except the last row, we take the large-$t$ limit and give the additional terms that arise (from the soft function) for arbitrary $t$.

Consistency of the effective theory requires that the anomalous dimensions satisfy
\begin{align}
\label{eq:consistency}
0 &=   \Big( \gamma_H(\mu) +\gamma_{ S}^\unmeas(\mu)+ \sum_{i \notin \meas}  \gamma_{J_{i}}(\mu)  \Big) \delta(\tau_a^i)+ \sum_{i \in \meas} \Big(\gamma_{ J_{i}}(\tau^i_a; \mu)) +  \gamma^\meas_{S}(\tau_a^i; \mu)  \Big)
\,.\end{align}
From the results tabulated in \tab{anomalous-coeff}, up to corrections of $\cO(1/t^2)$, we see that,  remarkably, this relation is indeed satisfied! This is highly nontrivial, as jet and soft anomalous dimensions depend on the jet radius $R$ and the jet shape $\tau_a$, while the hard function does not; in addition, the hard and soft functions know the directions $n_i$ of all $N$ jets, while the jet functions do not. These dependencies cancel precisely between the $R$-dependent pieces of unmeasured jet contributions to the jet and soft functions, between $\tau_a$-dependent pieces of the measured jet contributions, and between $n_i\cdot n_j$-dependent pieces of the hard and soft functions. The sum of all jet and soft anomalous dimensions then precisely matches the hard anomalous dimensions, satisfying \eq{eq:consistency}.

Making the satisfaction of consistency even more nontrivial, individual contributions to the infinite part of the soft function, and therefore its anomalous dimension, given by \tab{soft} depend on the energy cut parameter $\Lambda$ as well. However, these terms cancel in the sum over the contributions $S_{ij}^{\text{incl}}$ and $S_{ij}^i$ in the first two rows of \tab{soft}. The double poles of the form $\frac{1}{\epsilon}\ln\Lambda$ arise from regions of phase space where a soft gluon can become both collinear to a jet direction (giving a $1/\epsilon$) and soft (giving a $\ln\Lambda$). These regions exist in the integral over all directions giving $S_{ij}^{\text{incl}}$ but are subtracted back out in the contributions $S_{ij}^i$.  In the surviving ``Swiss cheese'' region the regions giving these double poles are cut out.

The $\cO(1/t^2)$ terms that violate consistency arise whenever there are unmeasured jets. While this limit is not required for the contribution of   measured jets to the anomalous dimension to satisfy the consistency condition \eq{eq:consistency}, the finite parts of measured jet contributions to the soft function contain large logarithms of $\omega/\Lambda$ that can not be minimized with a scale choice but are suppressed by $\cO(1/t^2)$ (cf. \eq{eq:Sijk-Smeas-sum} of Appendix \ref{app:softcalc}). This is the manifestation of the fact that jets need to be well-separated, as explained in \sec{sec:fact}. For the remainder of the paper, we work strictly in the large-$t$ limit.

It may seem mysterious that the calculations of the hard, jet, and soft functions themselves and requiring their consistency lead to the condition of a large separation parameter $t$. Although we already specified qualitatively in the proof of factorization the requirement of well-separated jets, it may not be immediately apparent where it is implemented in the actual calculations. It enters in the definition of the collinear jet functions. In the large-$t$ limit, the $N$ jets are infinitely separated from one another according to the measure given by \eq{tdef}. And indeed, when  $N$-jet operators are constructed in SCET, each collinear jet field contains a Wilson line $W_n$, defined below in \eq{Wdef}, of collinear gluons in the direction $n$ that were emitted from the back-to-back direction $\bn$, for which the separation measure $t\to\infty$. (This is similar to  QCD factorization proofs  of hard scattering cross sections, e.g. in \cite{Berger:2003iw}, in which this direction $\bn$ is chosen to be along some arbitrary path $\xi$ that is separated by an $\mathcal{O}(1)$ amount from the jet direction $n$.) Furthermore, the $n_i$-collinear jet function knows only its own color representation, and not those of the other jets. Meanwhile, the hard and soft functions we calculate ``know'' about all $N$ jets  and their precise directions and color representations. Therefore it is no surprise that,  when we actually calculate the anomalous dimensions of these functions, we find that they are consistent  with one another only in the limit that the separation parameter $t\to \infty$.

\subsection{Refactorization of the Soft Function}
\label{ssec:refactorization}

Our results for the soft function in \sec{sec:large-t-njet} make clear that in general there are multiple scales that appear in the soft function: the $\mu_S^1, \dots, \mu_S^M$ associated with the $M$ measured jets and the scale $\mu_S^\Lambda$ associated with the out-of-jet cutoff $\Lambda$ (see \eq{eq:mu_lambda-defn}). When these scales are all comparable, we can RG evolve the soft function from the single scale $\mu_S$. However, when any of them differ considerably from the others, we need to ``refactorize'' the soft function into multiple contributions, each of which is sensitive to a single energy scale. For illustration, take the scales $\mu_S^i$ to be such that $\mu_S^1 \gg \mu_S^2 \gg \cdots  \gg \mu_S^M$ as in \fig{fig:softscales}.
We also take $\mu_S^{l-1} \ll \mu_S^\Lambda \ll \mu_S^{l}$ for our discussion, but the result is independent of whether $\mu_S^\Lambda$ lies in the range  $\mu_S^1 < \mu_S^\Lambda < \mu_S^M$ or not.

\FIGURE[r]{
\includegraphics[totalheight = .15\textheight]{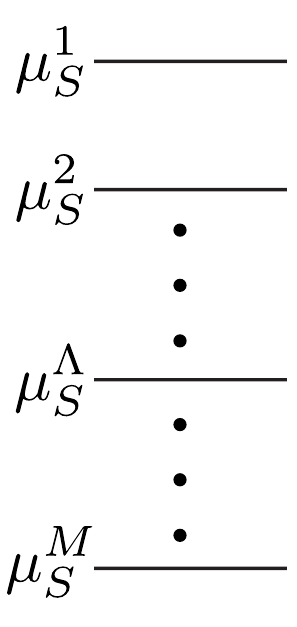}

{ \caption[1]{Soft scales.}
\label{fig:softscales}}
}

We can express the soft function appearing in \eq{Njetcs} as
\be
S(\tau_a^1, \tau_a^2, \dots, \tau_a^M;\mu) = \langle0\vert \cO_S^\dagger \Theta(\Lambda - \hat \Lambda) \prod_{i=1}^M \delta(\tau_a^i - \hat \tau_a^i ) \cO_S \vert 0 \rangle
\,,\ee
where the operator $\tau_a^i$ returns the contribution to $\tau_a$ of final-state soft particles entering jet $i$, and $\hat\Lambda$ returns the energy of final-state soft particles outside of all $N$ jets.
We have kept the dependence on the scales $\mu_S^i$ and on $\Lambda$ implicit on the left-hand side.

There are $N$ Wilson lines appearing in the operator $\cO_S$,
\be
\label{eq:soft_times_ultrasoft}
\mathcal{O}_S = Y_1 \dots Y_M Y_{M+1}\dots Y_N\,,
\ee
corresponding to the $M$ measured jets and $N-M$ unmeasured jets.
The scales associated with soft gluons entering the $M$ measured jets whose shapes are measured to be $\tau_1,\dots,\tau_M$ are $\mu_S^1,\dots,\mu_S^M$, given by \eq{eq:mu_S-defn}. The scale associated with soft gluons outside of measured jets is $\mu_S^\Lambda$ given by \eq{eq:mu_lambda-defn}. We have illustrated the ladder of these scales in \fig{fig:softscales}. Each of these soft scales can be associated with different soft fields $A_s^i$ whose momenta scale as $\lambda_i^2\omega_i$ where $\lambda_i$ is associated with the typical transverse momentum $\lambda_i\omega_i$ of the collinear modes for the $i$th jet. For measured jets, $\lambda_i$ is determined by $\tau_a^i$, while for unmeasured jets $\lambda_i\sim \tan(R/2)$.  For soft gluons outside jets, however, the soft momentum is set by the cutoff scale $\Lambda$, which is why $\mu_S^\Lambda$ appears in the ladder of \fig{fig:softscales}.

At a scale $\mu$ larger than all $\mu_S^i$ and $\mu_S^\Lambda$, the soft function is calculated as we presented in \sec{sec:soft}. In particular, we take the quantities $\tau_a^i$ and $\Lambda$ to be nonzero and finite. At a scale $\mu$ below $\mu_S^1$, we integrate out soft gluons of virtuality $\mu_S^1$ and match onto a theory with soft gluons of virtuality $\mu_S^2$. The scale $\mu_S^1$ associated with $\tau_a^1$ is taken to infinity, and phase space integrals for soft gluons entering the measured jet 1 become zero (see, e.g., \eq{eq:Sijmeas-integral-2}). Therefore, the matching coefficient from the theory above $\mu_S^1$ to the theory below is just the measured jet 1 contribution $S^\meas(\tau_a^1)$  to the soft function given by \eq{eq:Smeas-result}. The same occurs when matching from the theory above each scale $\mu_S^i$ for soft gluons entering measured jet $i$ to the scale below $\mu_S^i$, giving a matching coefficient $S^\meas(\tau_a^i)$.

When going through the scale $\mu_S^\Lambda$, in the theory above this scale, the calculation of unmeasured contributions to the soft function gives the result \eq{eq:Sunmeas-result}, by treating $\Lambda$ as a nonzero, finite cutoff. In the theory below $\mu_S^\Lambda$, we take $\Lambda$ to infinity, making all phase space integrals originally cutoff by $\Lambda$ to be scaleless and thus zero. So the matching coefficient between the theory above and below $\mu_S^\Lambda$ is just $S^\unmeas$.

After performing the above matchings all the way down to the lowest soft scale in \fig{fig:softscales},
we find that the original soft function $S(\tau_a^1,\dots,\tau_a^M;\mu)$ can be expressed to all orders as
\be
\label{refactorizedsoft}
S(\tau_a^1, \dots, \tau_a^M;\mu) = S^\unmeas(\mu)  \prod_{i=1}^M S^\meas( \tau_a^i;\mu )\bra{0}\mathcal{O}_S^\dag\mathcal{O}_S\ket{0}\,,
\,\ee
where to next-to-leading order $S^\meas$ and $S^\unmeas$ are given by
\begin{align}
S^\unmeas(\mu) &= 1 + S_{(1)}^\unmeas(\mu)\\
S^\meas(\tau_a^i;\mu) &= \delta(\tau_a^i) + S_{(1)}^\meas(\tau_a^i;\mu)  \, ,
\end{align}
where $S_{(1)}^\unmeas$ is given by \eq{eq:Sunmeas-result} and $S_{(1)}^\meas$ is given by \eq{eq:Smeas-result}.  Note that no operators restricting the jet shape or the phase space appear in the final matrix element of soft fields living at the lowest scale on the ladder in \fig{fig:softscales}.  If all the scales on the ladder are at a perturbative scale, we can now just use $\langle\cO_S^\dagger \cO_S\rangle = 1$ to eliminate the final matrix element. If any scale is nonperturbative, we should have stopped the matching procedure before that scale, and defined the surviving soft matrix element still containing additional delta function operators as a nonperturbative shape function.

Since the factors $S^\unmeas(\mu)$ and $S^\meas(\tau_a^i,\mu)$ are now matching coefficients between two theories above and below the respective scales $\mu_S^\Lambda$ and $\mu_S^i$, we can run each of the individual factors separately from their natural scale, instead of from a single soft scale $\mu_0$ as in \eq{softsolution}. The result for the RG-evolved soft function is then \eq{refactorizedsoft} where each factor at NLO is given by the solution of its RGE,
\begin{subequations}
\label{eq:Sevolve}
\begin{align}
S^\unmeas(\mu) & = U_S^\unmeas(\mu,\mu_S^\Lambda) S^\unmeas(\mu_S^\Lambda) \\
S^\meas(\tau_a^i,\mu) &= \int d\tau' U_S^i(\tau_a^i - \tau';\mu,\mu_S^i) S^\meas(\tau',\mu_S^i)\,.
\end{align}
\end{subequations}
These solutions allow us now to resum logarithms of all of the scales appearing in the ladder in \fig{fig:softscales} when these scales are widely disparate. However, the result we obtained in \eq{softRGE} when we took all scales to be of the same order and had a single soft scale has the form \eq{eq:Sevolve} at NLL accuracy. We will use equation \eq{eq:Sevolve} in all cases to interpolate between these two extremes.

\subsection{Total Resummed Distribution}

\label{ssec:totalNLLdist}

Collecting together the above results for the running of hard, jet, and soft functions in the factorized cross section \eq{Njetcs}, we obtain the RG-improved $N$-jet cross section differential in $M$ jet shapes,
\begin{equation}
\label{NjetcsRG}
\begin{split}
\frac{1}{\sigma^{(0)}}\frac{d\sigma_N}{d\tau_{a_1}^1\cdots d\tau_{a_M}^M} &=  H(\mu_H) \left(\frac{\mu_H}{\bar\omega_H}\right)^{\omega_H(\mu,\mu_H)} \!\!\!\! \prod_{k=M+1}^N \!\!\! J^k_{\omega_k}(\mu_J^k) \left(\frac{\mu_J^k}{\omega_k \tan\frac{R}{2}}\right)^{\!\!\omega_J^k(\mu,\mu_J^k)}   \!\!\! S^\unmeas(\mu_S^\Lambda) \\
&\quad\times \prod_{i=1}^M \Biggl\{ \left[ 1 + f_J^i(\tau_a^i,\mu_J^i) + f_S^i(\tau_a^i,\mu_S^i)\right] \left(\frac{\mu_S^i \tan^{1-a}\frac{R}{2}}{\omega_i}\right)^{\omega_S^i(\mu,\mu_S^i)} \\
&\quad\times \! \left(\frac{\mu_J^i}{\omega_i}\right)^{\!\!(2-a)\omega_J^i(\mu,\mu_J^i)}  \!\!\!\! \frac{1}{\Gamma[-\omega_J^i(\mu,\mu_J^i) \! - \! \omega_S^i(\mu,\mu_S^i)]} \frac{1}{(\tau_a^i)^{1+\omega_J^i(\mu,\mu_J^i) + \omega_S^i(\mu,\mu_S^i)}}\Biggr\}_+ \\
&\quad \times \exp\left[\mathcal{K}(\mu;\mu_H,\mu_J^{1,\dots,N},\mu_S^{1,\dots,M},\mu_S^\Lambda) + \gamma_E \Omega(\mu;\mu_J^{1,\dots,M},\mu_S^{1,\dots,M})\right] \,,
\end{split}
\end{equation}
where $\bar\omega_H$ is defined by \eq{omegaH}, the evolution parameters $\omega_F(\mu,\mu_F)$ and $K_F(\mu,\mu_F)$ are defined in \eq{eq:kernelparams}, and we have defined the collective parameters,
\begin{subequations}
\label{KOmega}
\begin{align}
\begin{split}
\label{K}
\mathcal{K}(\mu;\mu_H,\mu_J^{1,\dots,N},\mu_S^{1,\dots,M},\mu_S^\Lambda) &= K_H(\mu,\mu_H) + \sum_{i=1}^N K_J^i(\mu,\mu_J^i) + \sum_{j=1}^M K_S^j(\mu,\mu_S^j) \\
&\quad + K_S^\unmeas (\mu,\mu_S^\Lambda)
\end{split}
\\
\label{Omega}
\Omega(\mu;\mu_J^{1,\dots,M},\mu_S^{1,\dots,M}) &= \sum_{i=1}^M \Omega_i(\mu;\mu_J^i,\mu_S^i) \equiv  \sum_{i=1}^M [ \omega_J^i(\mu,\mu_J^i) + \omega_S^i(\mu,\mu_S^i)]\,.
\end{align}
\end{subequations}
Using results from Appendix~\ref{app:convolve}, we obtain the functions $f_{J,S}^i$ generated by the finite pieces of the measured jet and soft functions,
\begin{subequations}
\label{measuredlogs}
\begin{align}
\label{measuredlogs-jet}
f_J^i(\tau_a^i;\mu_J^i) &= \frac{\as(\mu_J^i)\vect{T}_i^2}{2\pi}  \Theta( \tau_a^{\text{max}} - \tau_a^i)\biggl\{ \frac{4-2a}{1-a} \ln^2\frac{\mu_J^i}{\omega_i(\tau_a^i)^{\frac{1}{2-a}}} + \frac{1}{1-a}\frac{1}{1-\frac{a}{2}}\left[\frac{\pi^2}{6} - \psi^{(1)}(-\Omega_i)\right]\nn \\
&\qquad  + \left[c_i + \frac{2}{1-a}H(-1-\Omega_i)\right] \left[ 2 \ln \frac{\mu_J^i}{\omega_i (\tau_a^i)^{\frac{1}{2-a}}} + \frac{1}{2-a} H(-1-\Omega_i)\right]
 \nn \\
&\qquad\qquad\qquad + \left(4-2a\right)\ln^2\tan\frac{R}{2} - 2c_i \ln\tan\frac{R}{2} \biggr\} + \frac{\as(\mu_J^i)}{2\pi} d_J(\tau_a^i)
\end{align}
\begin{align}
\label{measuredlogs-soft}
f_S^i(\tau_a^i;\mu_S^i) &= -\frac{\as(\mu_S^i)\vect{T}_i^2}{\pi}\frac{1}{1-a} \Biggl\{ \! \left[\ln \frac{\mu_S^i\tan^{1-a}\frac{R}{2}}{\omega_i\tau_a^i}  + H(-1\!-\!\Omega_i)\right]^2 \!\! + \frac{\pi^2}{6} - \psi^{(1)}(-\Omega_i) \!+\! d_S\Biggr\} \,, \nn\\
\end{align}
\end{subequations}
where $c_i =3/2$ for quark jets and $\beta_0/(2C_A)$ for gluon jets. $\tau_a^{\text{max}}$ is the upper limit on $\tau_a^i$ found in the finite part of the na\"{i}ve jet function, given in Appendix~\ref{app:jetcalc}. $H(-1-\Omega_i)$ is the harmonic number function, with $\Omega_i$ given by \eq{Omega}. $\psi^{(1)}$ is the first derivative of the digamma function, $\psi^{(1)}(z) = (d/dz) [\Gamma'(z)/\Gamma(z)]$. The  terms $d_{J,S}$  are additional contributions from the finite parts of jet and soft functions  that do not contain any logarithms, where $d_S = -\pi^2/24$, and $d_J$ is given in \eq{eq:d_J} in the Appendix. $d_{J,S}$ are not needed at NLL accuracy.
Similarly, the terms containing large logarithms in the unmeasured jet functions and unmeasured contribution to the soft function are
\begin{subequations}
\label{unmeasuredlogs}
\begin{align}
\label{unmeasuredlogs-jet}
J^i_{\omega}(\mu_J) &= 1 +  \biggl[ \Gamma(\as(\mu_J)) \vect{T}_i^2 \ln^2\frac{\mu_J}{\omega\tan\frac{R}{2}} + \gamma_k[\as(\mu_J)] \ln\frac{\mu_J}{\omega\tan\frac{R}{2}} + d^i_J \biggr ]
\end{align}
\begin{align}
\label{unmeasuredlogs-soft}
S^\unmeas(\mu_S^\Lambda)& = 1 + \Gamma(\as(\mu_S^\Lambda)) \sum_i\vect{T}_i^2 \biggl[ \ln\biggl(\frac{\mu_S^\Lambda}{2\Lambda\tan^{1/2}\frac{R}{2}}\biggr)\ln\tan^2\frac{R}{2} - \frac{\pi^2}{8}\biggr] \nn\\
&\quad + \Gamma(\as(\mu_S^\Lambda)) \sum_{i\not = j} \vect{T}_i\mcdot\vect{T}_j \biggl[ \ln\frac{\mu_S^\Lambda}{2\Lambda}\ln \frac{n_i\mcdot n_j}{2} + \Li_2 \biggl( 1- \frac{2}{n_i\mcdot n_j}\biggr)\biggr]\,,
\end{align}
\end{subequations}
where $d^i_J$ is the part of the unmeasured jet function containing no large logarithms (given in \eqs{eq:dJq}{eq:dJg} in the Appendix).

The finite parts of the measured and unmeasured jet and soft functions given in \eqs{measuredlogs}{unmeasuredlogs} show that to minimize large logarithms in the $\cO(\as)$ finite parts in the resummed distribution \eq{NjetcsRG}, we should choose  initial scales for the running to be
\begin{subequations}
\label{scalechoices}
\begin{gather}
\mu_H = \bar\omega _H \\
\mu_J^i = \omega_i(\tau_a^i)^\frac{1}{2-a}  \ , \quad
\mu_J^k = \omega_k\tan\frac{R}{2}    \\
\mu_S^i = \frac{\omega_i \tau_a^i}{\tan^{1-a}\frac{R}{2}} \ , \quad
\mu_S^\Lambda = 2\Lambda\tan^{1/2}\frac{R}{2}\,.
\end{gather}
\end{subequations}
These choices eliminate all large logarithms in the  $\cO(\as)$ hard, jet, and soft functions. They still leave logs of $\tan\frac{R}{2}$ and $n_i\cdot n_j$ in the unmeasured part of the soft function, and logs of $\tan\frac{R}{2}$ in the measured jet function, but we already take $R$ numerically of $\mathcal{O}(1)$ \footnote{We still consider $\tan(R/2)$ to be of order $\lambda_k$ in the collinear sectors describing unmeasured jets, as implied by \eq{scalechoices}. This means $\lambda_k$ is effectively much larger than the parameter $\lambda_i$ in a measured jet sector. In fact, note that \eq{scalechoices} tells us that $\tan\frac{R}{2}$ must be parametrically larger than $(\tau_a^i)^{\frac{1}{2-a}}$; otherwise, the jet scale falls below the soft scale in the measured jet sectors, invalidating the use of SCET and, thus, the validity of the factorization theorem.}  to minimize power corrections from our implementation of the jet algorithm as discussed in \sec{sec:power}, and $n_i\cdot n_j \approx 1$ since the jet separation parameter $t_{ij}$ is large compared to 1. All logs of $R$, $\Lambda$, and $\tau_a^i$ are eliminated in the unmeasured jet function and measured part of the soft function.

\section{Plots of Distributions and Comparisons to Monte Carlo}
\label{sec:plots}
Having resummed the jet shape distributions in $\tau_a$ to NLL accuracy, in this section we plot the distributions for various values of $a$ and $R$, compare to Monte Carlo simulated events, and perform scale variation on the resummed distribution.  We use the process $e^+e^- \to 3$ jets to study our predictions of jet shapes, where the jets arise from partons in the ``Mercedes-Benz'' configuration, with each parton having equal energy.  In these configurations, the partons lie in a plane and are equally separated with a pairwise angle of $2\pi/3$.  This allows us to study event shape distributions of well-separated jets where $t$ is reasonably large.  We choose three values of $R$ to study, $R = 1.0$, 0.7, and 0.4.  With these values of R, the $1/t^2$ suppression factor for corrections to the large-$t$ limit are 0.10, 0.044, and 0.014 respectively.  We will measure the angularity of only one of the three jets; the other two jets will be unmeasured.

In general, the $\vect{T}_i \cdot \vect{T}_j$ color correlations in the soft and hard functions lead to operator mixing in color space under RG evolution. This implies that the RG kernels $U_S$ and $U_H$ are matrices in color space and must be studied on a process-by-process basis (see, e.g., \cite{Chiu:2009mg,Kidonakis:1998bk,Kidonakis:1998nf,Dokshitzer:2005ek,Sjodahl:2008fz,Sjodahl:2009wx}). For the case of $N = 2,3$ jets there is only one color-singlet operator and hence no mixing. This can be seen, for example, by noting that all color correlations reduce to the Casimir invariants ($\CF$ and $\CA$) in this case (cf. Appendix~\ref{app:color}).  We have restricted the example process we use in this work to $N = 3$ jets, avoiding the additional complication of color-correlations that comes with a larger number of jets.

The NLL resummed distribution for one quark or gluon jet shape (jet 1) in a three-jet final state, written as the derivative of the radiator \eq{radiator}, is
\begin{align}
\label{3jetNLL}
\frac{1}{\sigma^{(0)}} \frac{d\sigma_3}{d\tau_a^1} &=   \left(\frac{\mu_H}{\bar\omega_H}\right)^{\omega_H(\mu,\mu_H)}\left(\frac{\mu_J^1}{\omega_1}\right)^{(2-a)\omega_J^1(\mu,\mu_J^1)} \left(\frac{\mu_J^2}{\omega_2 \tan\frac{R}{2}}\right)^{\omega_J^2(\mu,\mu_J^2)} \left(\frac{\mu_J^3}{\omega_3\tan\frac{R}{2}}\right)^{\omega_J^3(\mu,\mu_J^3)} \nn \\
&\quad \times \left(\frac{\mu_S^1 \tan^{1-a}\frac{R}{2}}{\omega_1}\right)^{\omega_S^1(\mu,\mu_S^1)}
 \exp\left[\mathcal{K}(\mu;\mu_H,\mu_J^1,\mu_J^2,\mu_J^3,\mu_S^1,\mu_S^\Lambda) + \gamma_E\Omega(\mu;\mu_J^1,\mu_S^1)\right] \nn \\
&\quad\times [1 + \hat f_J(\tau_a^1) + \hat f_S(\tau_a^1)] \frac{1}{\Gamma[-\Omega(\mu;\mu_J^1,\mu_S^1)] }\left[\frac{1}{(\tau_a^1)^{1+\Omega(\mu;\mu_J^1,\mu_S^1)}}\right]_+ \,,
\end{align}
where the various evolution parameters $\omega_{J,S}^i, \, \Omega, \, \mathcal{K}$ are all defined in \eqs{eq:kernelparams}{KOmega}, and $\hat f_{J,S}$ are given by $f_{J,S}$ in \eq{measuredlogs} with the $d_{J,S}$ terms set to zero (accurate to NLL).
The best scale choices \eq{scalechoices} for this case are
\begin{subequations}
\label{3jetscalechoices}
\begin{gather}
\mu_H = \left(\omega_1^{\vect{T}_1^2} \omega_2^{\vect{T}_2^2} \omega_3^{\vect{T}_3^2}\right)^{\frac{1}{2C_F+C_A}} \\
 \mu_J^1 = \omega_1(\tau_a^1)^{\frac{1}{2-a}} \ , \  \mu_J^{2,3} = \omega_{2,3}\tan\frac{R}{2} \\
  \mu_S^1 = \frac{\omega_1\tau_a^1}{\tan^{1-a}\frac{R}{2}}\ , \ \mu_S^\Lambda = 2\Lambda \tan^{1/2}\frac{R}{2}\,.
  \end{gather}
\end{subequations}
In \eq{3jetNLL} we have used tree-level initial conditions for the hard, jet, and soft functions at these scales. \eq{3jetNLL} evolves these functions to the arbitrary scale $\mu$ at NLL accuracy.

\FIGURE[t]{
\resizebox{.95\textwidth}{!}{\includegraphics{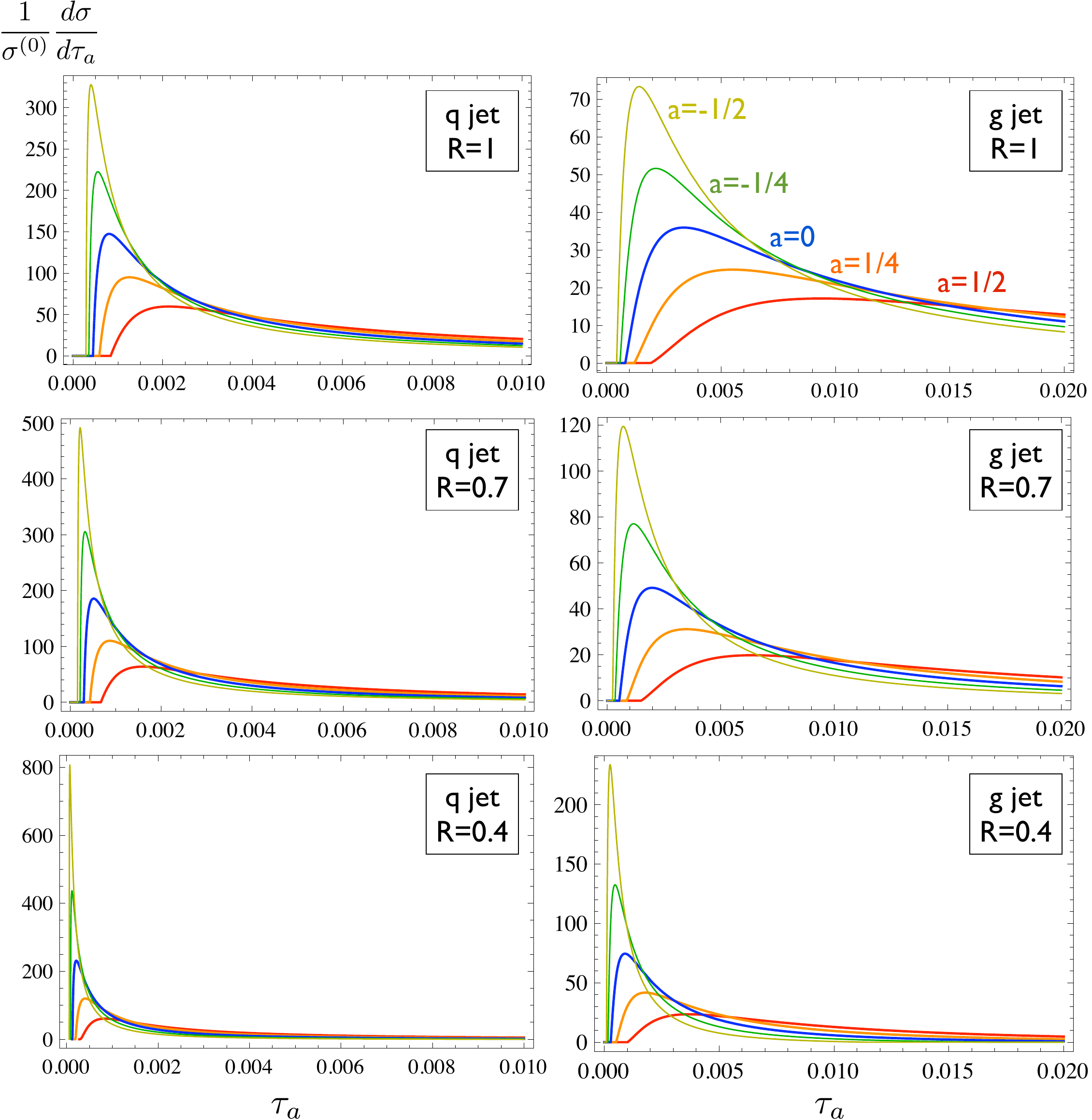}}
\vspace{-5pt}
{\caption[1]{Quark and gluon jet shapes for several values of $a$ and $R$. The NLL resummed distribution in \eq{3jetNLL} is plotted for $ a = -\frac{1}{2},-\frac{1}{4},0,\frac{1}{4}, \frac{1}{2}$ for quark and gluon jets with $R = 1, 0.7,0.4$. The plots are for jets produced in $e^+ e^-$ annihilation at center-of-mass energy $Q = 500\GeV$, with three jets produced in  a Mercedes-Benz configuration with equal energies  $E_J = 150\GeV$, and minimum threshold $\Lambda = 15\GeV$ to produce a jet.}
\label{fig:qgaplots}}
}

With these choices, we plot \eq{3jetNLL} in \fig{fig:qgaplots} for several values of $a$ and $R$ for a quark or gluon jet shape in a three-jet final state in $e^+e^-$ annihilation at center-of-mass energy $Q = 500\GeV$.\footnote{The distributions plotted with the $\hat f_{J,S}$ terms included in \eq{3jetNLL} exhibit a small negative dip near $\tau_a=0$ (not shown) that can be cured by convolving with a nonperturbative shape function with a renormalon-free gap parameter \cite{Hoang:2007vb,Hornig:2009vb}.  This is beyond the scope of the present work, so we only plot the perturbative distributions where they are positive.} The jets are chosen to be in a Mercedes-Benz configuration, where all jets have equal energies of $150\GeV$. We choose the jet energy cutoff $\Lambda$ to be $15\GeV$. We choose the factorization scale to be $\mu=\mu_H$.

\FIGURE[t]{
\resizebox{\textwidth}{!}{\includegraphics{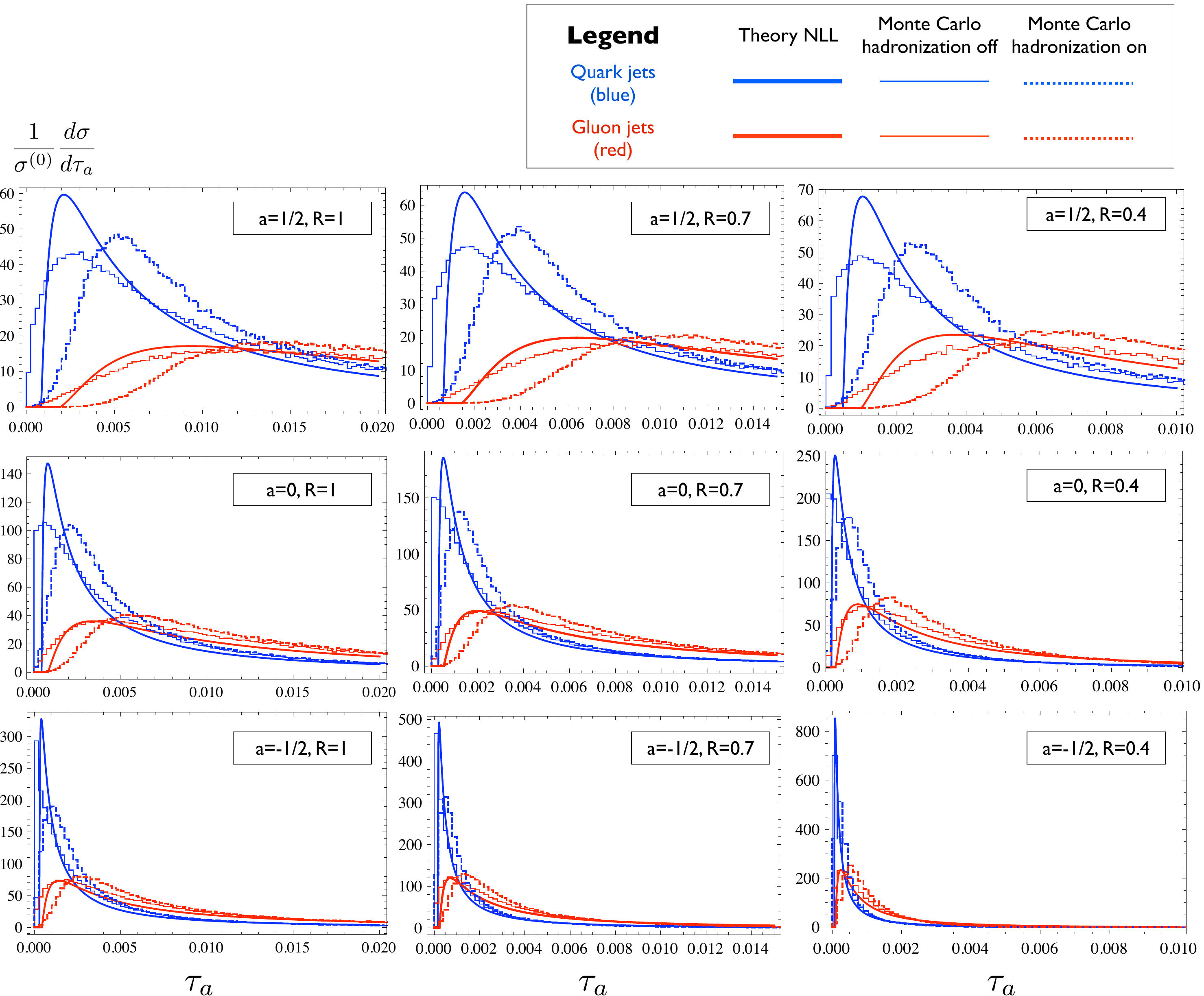}}
\vspace{-18pt}
{ \caption[1]{Quark vs. gluon jet shapes with comparison to Monte Carlo.  Solid, straight curves represent the resummed jet shape distribution in \eq{3jetNLL}, and jagged curves are histograms from the Monte Carlo, normalized as described in the text.  The solid histogram has no hadronization, while the dashed histogram includes the effects of hadronization.  The distributions are plotted for $ a = -\frac{1}{2},0,\frac{1}{2}$ with quark (blue) and gluon (red) jets compared on the same plot, for jets of size $R=1.0,0.7,0.4$.  Gluon jet shape distributions peak at larger values of $\tau_a$ than quark jets, indicative of their broader shape. The plots are for jets in $e^+ e^-$ annihilation at center-of-mass energy $Q = 500\GeV$, with three jets produced with equal energies  $E_J = 150\GeV$, angular separation $\psi = 2\pi/3$ between all pairs of jets, and minimum threshold $\Lambda = 15\GeV$ to produce a jet.}
\label{fig:qgsingleaRplots}}
}

We compare the results of a jet algorithm implemented on Monte Carlo simulated events with our NLL resummed predictions for a variety of $a$ and $R$ values in \fig{fig:qgsingleaRplots}.  Because the resummed NLL distribution we choose to study applies to an exclusive process, three-jet events in the Mercedes-Benz configuration, we implement cuts on the simulated events to obtain a sample that matches onto this configuration.  We use MadGraph/MadEvent v.4.4.21 \cite{Alwall:07.1} to generate parton-level $e^+e^-\to q\bar{q}g$ events at a center-of-mass energy $Q = 500\GeV$, with cuts imposed to obtain partons in the Mercedes-Benz configuration.  We shower and hadronize the events with Pythia v.6.414~\cite{Sjostrand:06.1} using $p_T$-ordered parton showers.  The process of hadronization will induce a shift in the angularity distribution, which we do not model in our resummed distribution.  Therefore, we produce two samples: one sample with only QCD final-state showering, no initial-state radiation, and no hadronization, and another sample with complete showering, initial-state radiation, and hadronization.  The anti-$\kt$ jet algorithm is run on the final state particles from Pythia, and we use FastJet~\cite{Cacciari:05.1} to implement the jet algorithm.  The anti-$\kt$ algorithm is particularly well suited for this comparison, as very few particles at an angle $\theta > R$ to the jet axis are included in the jet.  With anti-$\kt$, the phase space cut on an individual particle matches well with the phase space cuts in the next-to-leading order calculation.

To select a sample of events to compare to our resummed distributions, we make cuts on the final state jets, requiring each of the three hard, well-separated partons from MadGraph to be associated with a jet.  This involves a cut on the jet direction and momentum:
\be
\frac{\vect{p}_{\rm parton}\cdot\vect{p}_{\rm jet}}{\left|\vect{p}_{\rm parton}\right|\left|\vect{p}_{\rm jet}\right|} > 0.9 \quad \textrm{ and } \quad \frac{\left|\left|\vect{p}_{\rm parton}\right| - \left|\vect{p}_{\rm jet}\right|\right|}{\left|\vect{p}_{\rm parton}\right|} < 0.15 \, .
\ee
We analyze events passing these cuts, and tag each associated jet as coming from a quark or a gluon based on which parton it matches onto.  The angularity value for each jet is computed from the constituent particles of the jet, using the matching parton direction as the jet axis.  The jet direction only differs from the parton direction by a power correction (see \ssec{jetshapedist3jets}).  In \fig{fig:qgsingleaRplots}, we isolate some of the quark and gluon jet shapes in \fig{fig:qgaplots} and compare to Monte Carlo events.

\FIGURE[b]{
\qquad\qquad\resizebox{.7\textwidth}{!}{\includegraphics{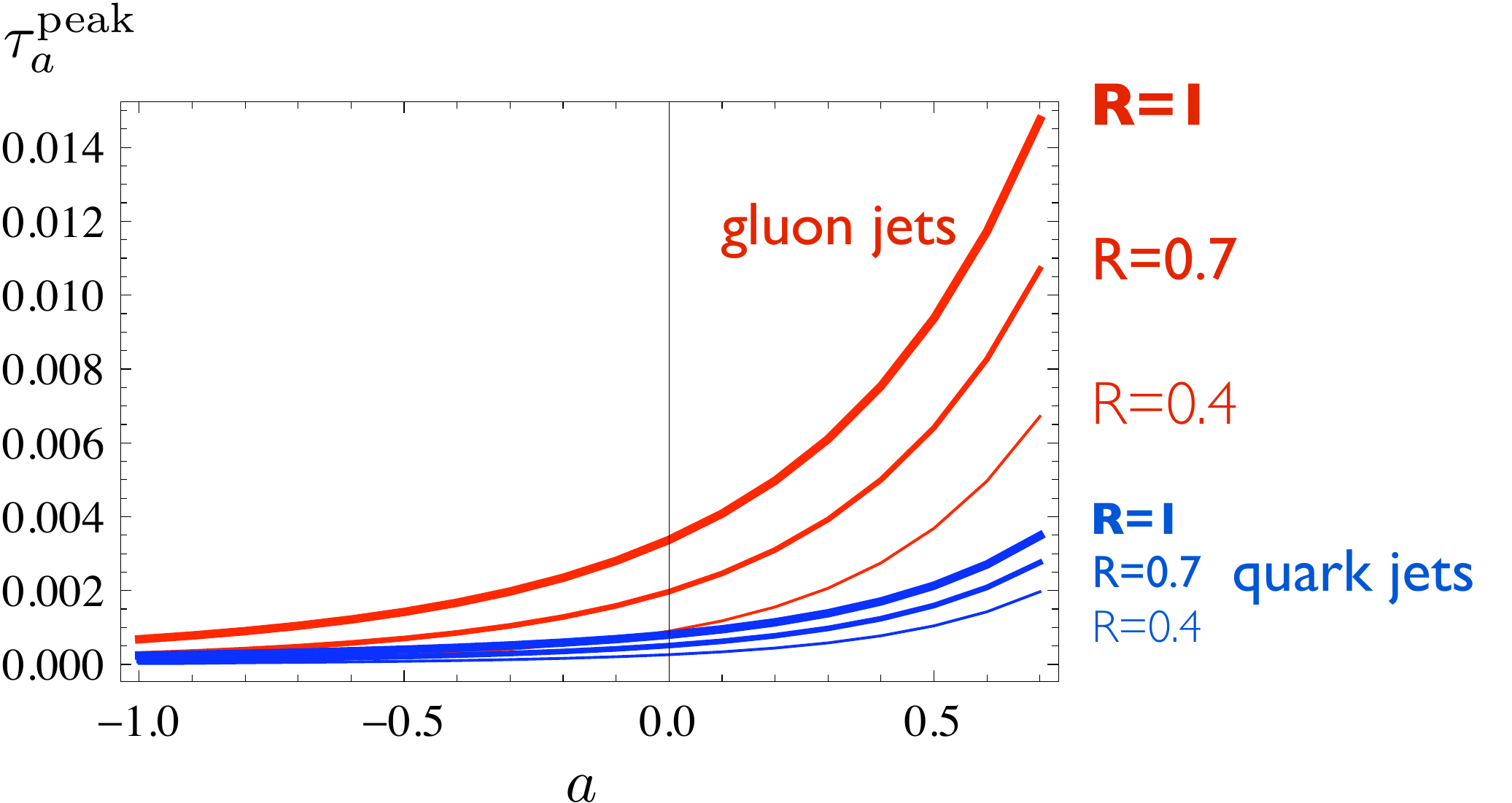}}
\vspace{-7pt}
{ \caption[1]{Location of peak of jet shape distribution as a function of $a$ for quark and gluon jets. We plot the value of $\tau_a$ at the peak of the jet shape distribution for $a$ between -1.0 and 0.8 for quark vs. gluon jets, using $R = 1,0.7,0.4$. }
\label{fig:taupeak}}
}

\FIGURE[t!]{
\centerline{\includegraphics[height=.83\textheight]{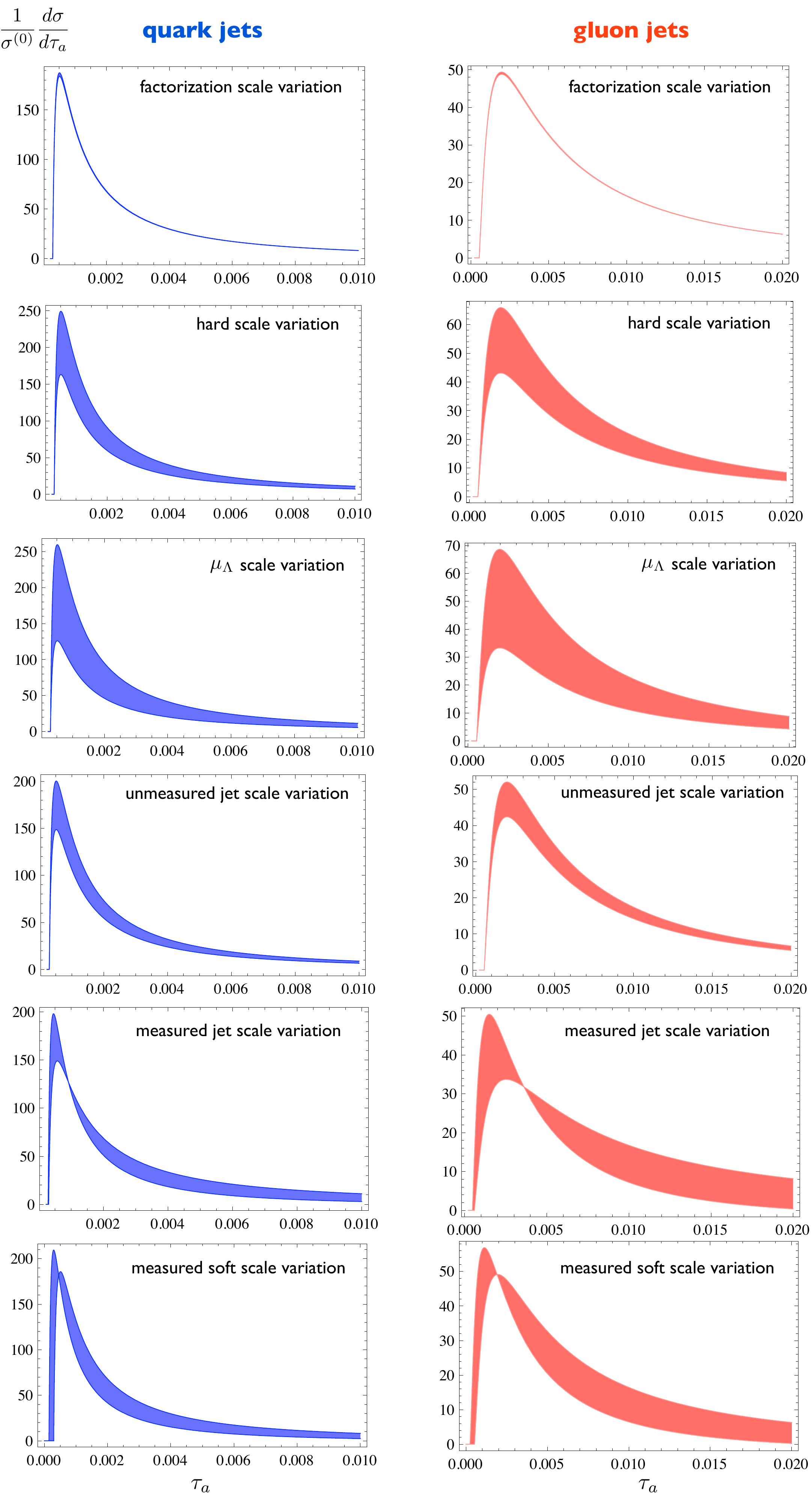}}
\vspace{-20pt}
{ \caption[1]{Scale variation of quark and gluon jet shapes. For $a=0$ and $R=0.7$, we display the variation of the NLL resummed jet shape distributions with the hard scale $\mu_H$, the jet cutoff scale $\mu_S^\Lambda$, the unmeasured jet scales $\mu_J^{2,3}$, the measured jet scale $\mu_J^1(\tau_a)$, and the measured soft scale $\mu_S(\tau_a)$. In each case we vary the scale between $1/2$ and $2$ times the natural choices  in \eq{scalechoices}, except for the measured soft scale, which we varied between $1$ and $2$ times the choice in \eq{scalechoices}. We keep the factorization scale fixed at the default hard scale given by \eq{3jetscalechoices}, $\mu=\omega_i$.
}

\label{fig:scalevar}}
}

The relative normalization between the distribution of Monte Carlo events and the NLL resummed angularity distribution requires some explanation.
Both our calculation and the Monte Carlo simulation are most accurate in the region that includes the peak of the distribution and the larger-$\tau$ side of the peak, but both are inaccurate as $\tau\to0$ and in the tail region.  Therefore, each will differ in the relative normalization between the peak region and the tail region. An accurate prediction of the tail region requires matching onto a calculation at fixed-order in $\alpha_s$ in full QCD as in \cite{Catani:1992ua,Becher:2008cf,Hornig:2009vb}.
In \fig{fig:qgsingleaRplots}, we choose to normalize the area of the Monte Carlo distribution to the total area of the NLL resummed theory distribution. We find the area under the theory curves for quark and gluon jets to be approximately 0.3 for $R=0.4$, 0.5 for $R=0.7$, and 0.7 for $R=1$. A more accurate prediction of the normalizations may require summing remaining unsummed logs of the phase space cuts in the theory and Monte Carlo predictions. These plots should be interpreted as comparisons of the predictions of the shapes in $\tau_a$ and these shapes' scaling as we vary $a$ and $R$, rather than the overall normalization.

The shapes of the theory and Monte Carlo distributions are largely similar, though they display noticeable differences at the leftmost endpoint near $\tau_a=0$ and in the ``sharpness'' of the peak. These may be due to the different ways the two approaches deal with the growth of the strong coupling for small $\tau_a$, the different orders of log resummation (LL vs. NLL) and the need to match the tails onto fixed-order QCD predictions. Since neither the Monte Carlo nor theory partonic predictions without inclusion of hadronization effects is yet a prediction of a physically observable quantity, we use this comparison as an intermediate diagnostic rather than a conclusive test of either method. Nevertheless, comparing the way the shapes of the distributions  and locations of the peaks vary over the range of values of $a$ and $R$ that we sample, the behavior agrees very well between the theory distributions and the Monte Carlo distributions without hadronization for both quark and gluon jets.

In \fig{fig:taupeak} we plot the location of the peak of the jet shape distributions as a function of $a$ for three values of $R$, displaying the different variation of the peak of quark and gluon jet shape distributions.  The peak value increases with increasing $R$ and $a$, as wide angle radiation is included (increasing $R$) and less suppressed (increasing $a$).  Although the difference in the peak value between the quark and gluon jet angularity distributions is large, the width of each distribution creates substantial overlap in angularity values between quark and gluon jets.  Distinguishing between quark and gluon jets using jet angularities is a complex task which we will explore in future work; for now, we note only that the NLL resummed distributions indicate that discrimination between quark and gluon jets using jet angularities is possible.

\FIGURE[htb!]{
\centerline{\includegraphics[width=\textwidth]{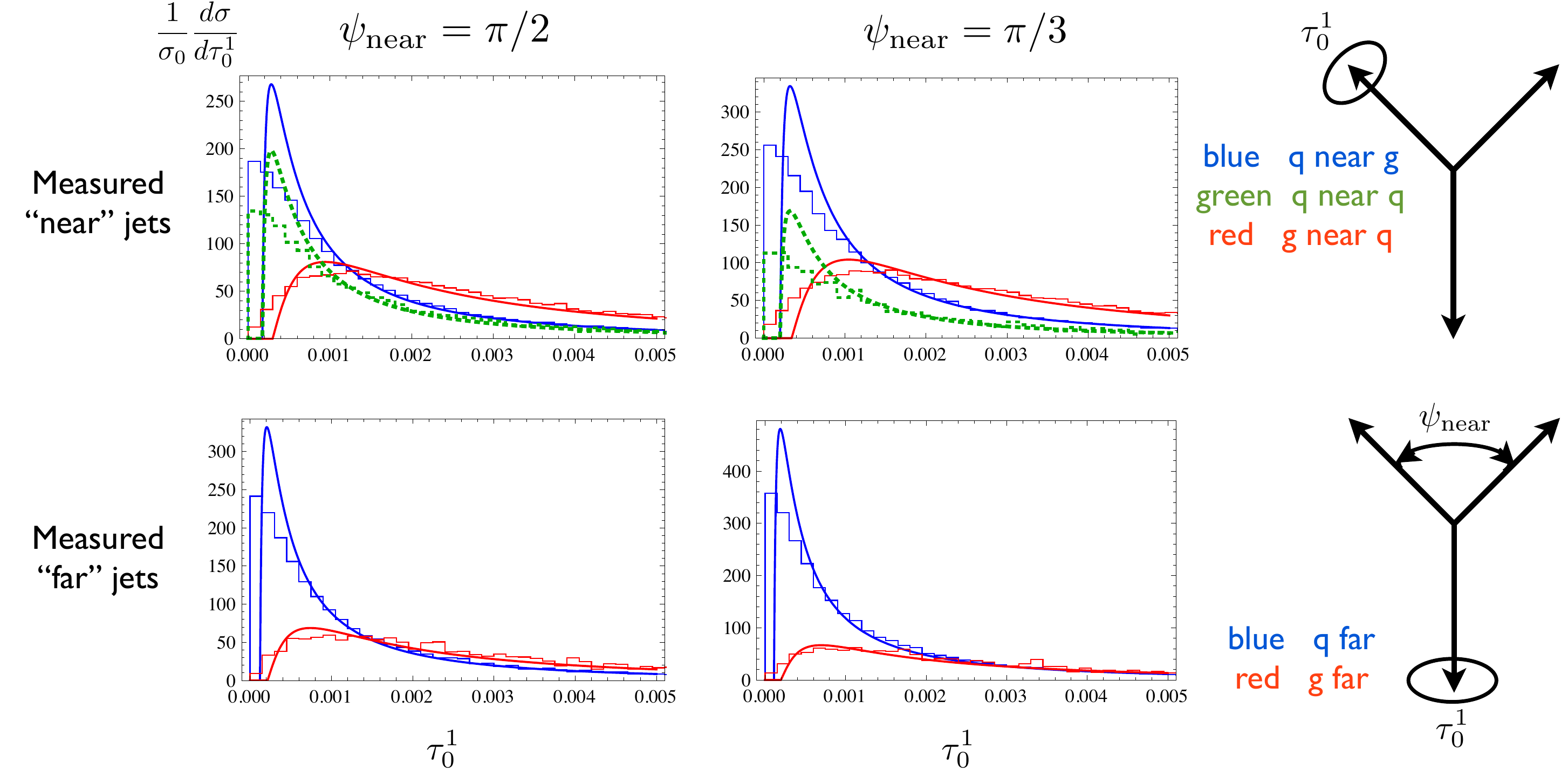}}
\vspace{-20pt}
{ \caption[1]{Jet shapes for other kinematic configurations. We compare our theoretical predictions to Monte Carlo simulations for the shape $\tau_0^1$ ($a=0$) for a quark or gluon jet found in a three-jet configuration where the two jets with narrowest separation angle $\psi_{\text{near}}$ have equal energy. We consider the two cases $\psi_{\text{near}} = \pi/2$ and $\pi/3$. In the first row, we plot shapes of one of the jets in the ``near'' pair. The blue solid curve is the shape of quark jet found near a gluon jet, the green dotted curve is a quark found near an antiquark, and the red solid curve is a gluon found near a quark. In the second row, we compare shapes of a quark or gluon jet found far from the near pair.
}

\label{fig:nonMB}}
}

As a rough estimate of the theoretical uncertainty in our NLL resummed predictions, we show  in \fig{fig:scalevar} the change in the $a=0$ quark and gluon $\tau_a$ jet shape distributions for $R=0.7$ when the various scales that appear in the resummed cross section \eq{3jetNLL} are varied. These are the initial scales at which the hard, jet, and soft functions are evaluated to minimize logarithms in the NLO fixed-order part, from which the evolution kernels run them to the common factorization scale $\mu$.  In the top  row of \fig{fig:scalevar}, we vary $\mu$ between $\bar\omega_H/2$ and $2\bar\omega_H$. The tiny variation is a sign of the consistency condition satisfied by the anomalous dimensions in \eq{eq:consistency}. In the next four rows, we vary the hard scale $\mu_H$, the soft jet energy cutoff scale $\mu_S^\Lambda$, the unmeasured jet scales $\mu_J^{2,3}$, and  the measured jet scale $\mu_J^1(\tau_a^1)$ between half and twice the natural  values given in \eq{3jetscalechoices}. In the last row, we vary the measured soft scale $\mu_S^1(\tau_a^1)$ between one and two times the value in \eq{3jetscalechoices}. This is because too low a value of $\mu_S^1(\tau_a^1)$ as $\tau_a\to 0$ brings it into the nonperturbative region where $\as(\mu_S^1)$ blows up, so that the perturbative estimate of uncertainty is not so meaningful. We note that, while the uncertainty in the vertical scale of the distributions is considerable in some cases, the location of the peak is much more stable.

Finally, in \fig{fig:nonMB} we give a sense for how robust our theoretical predictions are for other kinematic configurations.  We consider $e^+e^-\to q\bar{q}g$ events where the angle $\psi_{\rm near}$ between two partons is either $\pi/2$ or $\pi/3$, and these partons have equal energy.  We find jets using the anti-kT algorithm with $R=0.4$, and plot jet shapes for $a=0$.  The selection cuts to choose events from the Monte Carlo are the same as the Mercedes-Benz configuration.  In these events there are five distinct characterizations for a single parton.  If the event has the quark (or antiquark) as the ``far'' (most well separated) parton, then each parton in the event is distinct: there is the far quark, the near quark, and the near gluon.  If the event has the gluon as the far parton, then there are only two distinct types of partons: the far gluon and the near quark (antiquark).   In \fig{fig:nonMB}, we plot all these configurations for both $\psi_{\rm near} = \pi/2$ and $\psi_{\rm near} = \pi/3$.  The agreement between the theory predictions and the Monte Carlo are as good as in the Mercedes-Benz case, a good indication that our calculation applies to a broad range of kinematic configurations of multijet events.  Additionally, we observe features consistent with our intuition about the relative differences between the jet shape distributions between different jets in these configurations.  As one would expect, the distribution of near jet shapes is weighted more heavily towards larger $\tau_a$ than the far jet shapes, due to the enhanced soft radiation in the near jet system.  When the near quark is near a gluon, the distribution is weighted more heavily towards larger $\tau_a$ than when the near quark is near an antiquark, due to the enhanced radiation coming from a gluon rather than a quark.  These distributions serve as further evidence that jet shapes may be an effective discriminant between quark and gluon jets.

\section{Conclusions}
\label{sec:conclusions}

In this work, we have factorized an $N$-jet exclusive cross section differential in $M\leq N$ jet observables and resummed global logarithms of the jet observable $\tau_a$ to NLL accuracy, leaving summation of non-global logarithms to future work. We demonstrated that the anomalous dimensions of the hard, jet, and soft functions in the factorization theorem satisfy the nontrivial consistency condition \eq{eq:consistency} to $\cO(\as)$, for any number of quark and gluon jets, any number of jets whose shapes are measured, and any size $R$ of the jets, as long as the jets are well-separated, meaning $t\gg 1$. This condition ensures the validity of an effective theory with $N$ collinear directions that are assumed to be distinct.  We identified and estimated important power corrections to the factorized form of the cross section. We also illustrated that zero-bin subtractions give nonzero contributions to the anomalous dimensions crucial for consistency.

Armed with consistent factorization and the fixed-order jet and soft functions, we resummed large logarithms in the jet shape distribution by running each individual function from the scale where logs in it are minimized to the common factorization scale $\mu$. We thereby resummed, to NLL accuracy, global logs of the jet shape $\tau_a$ and logs of the scale $\Lambda/E_J$ of soft radiation outside of jets, but leaving some non-global logs and logs of the angular cut $R$ (but we took $R$ to be numerically of order 1). This is the first such calculation of a resummed jet shape distribution in an exclusive multijet cross section.

We constructed a framework to deal with all the scales that appear in the multijet soft function which depends on the values $\tau_a^i$ of all $M$ jet shapes and the phase space cuts $\Lambda,R$. By refactorizing the full soft function into individual pieces depending on one of these scales at a time, we were able to sum logs of ratios of these scales.

We demonstrated the accuracy of our results by comparing our resummed prediction for quark and gluon jet shapes in $e^+ e^-\to 3$ jets to the output of Monte Carlo event generators, MadGraph/MadEvent and Pythia. We compared our predictions with the Monte Carlo output without hadronization. The changes in shape and location of the peak value as functions of $a$ and $R$ match quite well between the theory and Monte Carlo.

Our results provide a basis for future studies of other jet observables at both $e^+e^-$ and hadron colliders, requiring recalculation of those parts of our jet and soft functions that depend on the choice of observable. Studying jets at hadron colliders requires constructing observables appropriate for that environment and the switching of two of our outgoing jets to incoming beams, which can be described by beam functions in SCET \cite{Stewart:2009yx}.

Precision calculations of jet shapes will allow improved discrimination of jets of different origins. We are applying the results of our predictions of light quark and gluon jet shapes to distinguish quark and gluon jets with greater efficiency than achieved before. Extensions to the shapes of heavy jets and calculations of other types of jet shapes such as the $\Psi(r/R)$ shape introduced in \cite{Ellis:1991vr,Ellis:1992qq,Abe:1992wv} can also be performed.

\paragraph{\it Note added in final preparation:}
As this paper was being completed, Ref.~\cite{Jouttenus:2009ns} appeared reporting the calculation of a quark jet function for a jet defined with a Sterman-Weinberg algorithm and whose invariant mass $s$ is measured. This jet function is related to our measured jet function $J^q_\omega(\tau_a)$ for a cone jet at $a=0$ given in \eq{Jtotal}, since $s=\omega^2\tau_0$. We have checked that the corresponding results between the two papers agree.

\acknowledgments
We are grateful to C. Bauer for valuable discussions and review of the draft.  The authors at the Berkeley CTP and in the Particle Theory Group at the University of Washington thank one another's groups for hospitality during portions of this work.  AH was supported in part by an LHC Theory Initiative Graduate Fellowship, NSF grant number PHY-0705682. The work of AH and CL was supported in part by the U.S. Department of Energy under Contract DE-AC02-05CH11231, and in part by the National Science Foundation under grant  PHY-0457315.  The work of SDE, CKV, and JRW was supported in part by the U.S. Department of Energy under Grants DE-FG02-96ER40956.

\appendix

\section{Jet Function Calculations}
\label{app:jetcalc}

\subsection{Finite Pieces of the Quark Jet Function}

\paragraph{Measured Quark Jet Function} The finite pieces the jet functions, which depend on the jet algorithm, share common features.  For cone-type algorithms, the finite piece of the naive part of the quark jet function, $\qjetnaive^q_\text{alg}(\tau_a)$, is given by
\begin{align}
\qjetnaive^{q}_\text{cone}(\tau_a) &=\CF\left(\frac{7}{2} +  3\ln 2 - \frac{\pi^2}{3}\right) \delta(\tau_a)   + \frac{\CF}{1-\frac{a}{2}} \left[ \mathcal{I}^{q}_\text{cone} \frac{\Theta(\tau_a)\Theta(\tau_a^{\max} - \tau_a)}{\tau_a}\right]_\dotplus
\end{align}
where in this Appendix, plus distributions are defined by \cite{Stewart:2009yx}
\begin{equation}
\label{mitplus}
[\Theta(x)g(x)]_+ = \lim_{\epsilon\to 0}\frac{d}{dx}[\Theta(x-\epsilon) G(x)],\qquad \text{with} \qquad G(x) = \int_1^x dx' g(x')\,,
\end{equation}
defined so as to satisfy the boundary condition $\int_0^1 dx[\Theta(x)g(x)]_+ = 0$.
The quantity $\mathcal{I}^{q}_\text{cone}$ depends implicitly on $\tau_a$ and $R$ and is given by
\be
\label{eq:xcone-defn}
\mathcal{I}^{q}_\text{cone} = \int_{\xcone}^{1 - \xcone} dx\,\frac{2(1-x) + x^2}{x} =  2\log\frac{1-\xcone}{\xcone} - \frac{3}{2} + 3\xcone \, .
\ee
The parameter $\xcone = \xcone(\tau_a)$ is the lower limit on the $x= q^-/\omega$ scaled gluon momentum integral imposed by the cone restriction. It is given by the solution of the equation
\begin{align}
\label{eq:fcone-defn}
f_{\rm cone}(\xcone) = \frac{\tau_a}{\tan^{2-a}\frac{R}{2}}  \, ,
\end{align}
where $f_{\rm cone}(x)$ is defined as
\be
\label{eq:fcone-defn2}
f_{\rm cone}(x) \equiv  x^{2-a}[x^{-1+a} + (1-x)^{-1+a}]
\ee
in the range $0 < x< 1/2$, which is plotted in \fig{fig:jetfunctions-integrations}A.  The limit $\tau_a^{\max}$ is given by the maximum value over $x$ of \eq{eq:fcone-defn2}.  Similarly, for $\kt$-type algorithms, $\qjetnaive^q_{\kt}(\tau_a)$ is given by
\begin{align}
\qjetnaive^{q}_{\kt}(\tau_a) &= \CF\left(\frac{13}{2} - 2\frac{\pi^2}{3} \right) \delta (\tau_a) + \frac{\CF}{1-\frac{a}{2}} \left[\mathcal{I}^{q}_{\kt}\,  \frac{\Theta(\tau_a)\Theta(\tau_a^{\max} - \tau_a)}{\tau_a}\right]_\dotplus \, .
\end{align}
$\mathcal{I}^{q}_{\kt}$ is given by
\be
\mathcal{I}^{q}_{\kt} = \int_{\mathcal{R}}dx\,\frac{2(1-x) + x^2}{x}
\ee
where $\mathcal{R}$ is the region in $x$ where the constraint
\be
\label{kTtauconstraint}
f_{\kt}(x) \equiv x^{2-a} (1-x)^{2-a} [x^{-1+a} + (1-x)^{-1+a}] \ge \frac{ \tau_a}{\tan^{2-a}\frac{R}{2}}
\ee
is satisfied. We plot this region in \fig{fig:jetfunctions-integrations}B and C for the cases $a>-1$ and $a<-1$, repsectively.  The boundaries of this region are the points $x_{1,2}$ illustrated in the figure, and are given by the equation
\begin{equation}
f_{\kt}(x_{1,2}) = \frac{ \tau_a}{\tan^{2-a}\frac{R}{2}}\, ,
\end{equation}
where we take $x_2>x_1$ if $x_2$ exists.
The upper limit $\tau_a^{\text{max}}$ is given by the maximum value over $x$ of the right-hand side of \eq{kTtauconstraint}.  In general, the constraint \eq{kTtauconstraint} is symmetric about $x=\frac12$, and so the region $\mathcal{R}$ is symmetric about the same point.  In general, if $a > -1$ or $\tau_a <  2^{a-2}\tan^{(2-a)}\frac{R}{2}$, then $\mathcal{R}$ is a single range in $x$.  Otherwise, $\mathcal{R}$ is two disjoint ranges in $x$. Since  $\tau_a \ge 2^{a-2}\tan^{(2-a)}\frac{R}{2}$ can only occur for $a<-1$, we can write $\mathcal{I}^{q}_{\kt}$ as
\begin{align}
\mathcal{I}^{q}_{\kt} &= \int_{x_1}^{1-x_1}dx\,\frac{2(1-x) + x^2}{x} - \Theta \left(\tau_a > 2^{a-2}\tan^{(2-a)}\frac{R}{2} \right)  \int_{x_2}^{1-x_2}dx\,\frac{2(1-x) + x^2}{x}
\end{align}

\FIGURE[t ]{
\includegraphics[width = \textwidth]{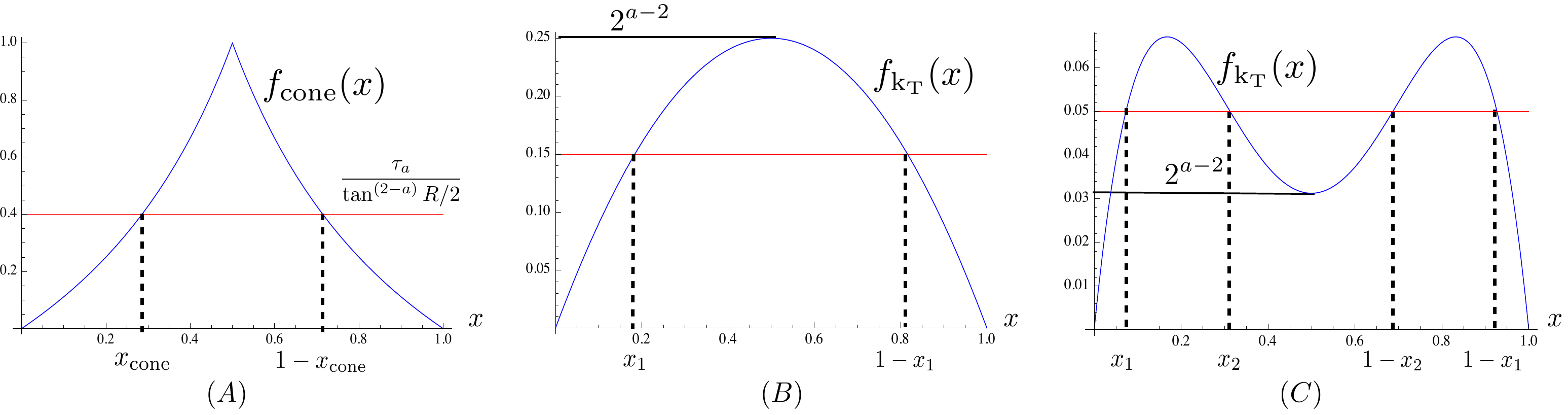}
\vspace{-.3in}
{ \caption[1]{Regions of integration for the (A) cone and $\kt$-type algorithms for (B) $a>-1$ and (C) $a<-1$. The allowed region of $x$ is when the (blue) functions $f_{{\rm cone}, \, \kt}(x)$ lie above the (red) lines of constant $\tau_a/\tan^{(2-a)}{R/2}$. When $a<-1$ for the $\kt$ algorithm, there are two regions of integration when $\tau_a> 2^{a-2} \tan^{(2-a)}{R/2}$.}
\label{fig:jetfunctions-integrations}}
}

Note that $\mathcal{I}^{q}_\text{cone}$ and $\mathcal{I}^{q}_{\kt}$ involve the same integrand, but for each algorithm the integral is over different ranges. In addition, both $\xcone$ and $x_{1}$ approach the same limiting value for small $\tau_a$,
\be
x \xrightarrow{\tau_a \to 0} \frac{\tau_a}{\tan^{(2-a)}\frac{R}{2}}
\,.\ee
Thus, we can extract the small $\tau_a$ behavior of both distributions by writing
\be
\label{eq:tau-extract}
\left[ \frac{1}{\tau_a} \ln \left( \frac{1-x}{x}\right) \right]_\dotplus  =  \left[\frac{1}{\tau_a} \ln \left( \frac{\tau_a}{\tan^{(2-a)} \frac{R}{2}} \frac{1-x}{x} \right)\right]_\dotplus - \left[ \frac{1}{\tau_a} \ln \left( \frac{\tau_a}{\tan^{(2-a)} \frac{R}{2}}\right)\right]_+
\,,\ee
where $x = \xcone$ or $x_{1}$ for the cone and $\kt$ algorithms, respectively.
Defining
\be
\label{eq:rq}
r_q(x) = 3 x + 2 \ln \frac{1-x}{x}
\,,\ee
using \eq{eq:tau-extract}, and including the zero-bin subtraction in \eq{Jzeroinresult}, we find that the finite distributions of the full measured quark jet functions are
\begin{subequations}
\label{eq:quark_meas_finite}
\begin{align}
\label{eq:quark_meas_finite_cone}
J^{q}_\text{cone}(\tau_a) &=   \CF\Biggl[  \frac{3}{2} \ln\frac{\mu^2}{\omega^2\tan^2\frac{R}{2}} + \frac{1-\frac{a}{2}}{1-a} \ln^2\frac{\mu^2}{\omega^2} + \left(1-\frac{a}{2}\right) \ln^2\tan^2\frac{R}{2} + \frac{7}{2} + 3\ln 2  \nn\\
&\quad  - \frac{\pi^2}{6} \left(2 + \frac{1-\frac{a}{2}}{1-a}\right) \Biggr] \delta(\tau_a) -  \CF\Biggl[\Biggl( \frac{4}{1-a}\ln\frac{\mu\tan^{1-a}\frac{R}{2}}{\omega \tau_a }   \Biggr)\frac{\Theta(\tau_a - \tau_a^{\text{max}})}{\tau_a} \Biggr]_+ \nn\\
&\quad -  \frac{\CF}{1-\frac{a}{2}}\Biggl[\frac{\Theta(\tau_a)\Theta(\tau_a^{\max} - \tau_a)}{\tau_a} \biggl( \frac{3}{2} + \frac{2-a}{1-a}\ln\frac{\mu^2}{\omega^2\tau_a^{\frac{1}{1-a/2}}} \nn \\
&\qquad \qquad \qquad \qquad\qquad\qquad\qquad   -  r_q(\xcone) - 2\ln\frac{\tau_a}{\tan^{2-a}\frac{R}{2}}\biggr) \Biggr]_\dotplus
\end{align}
and
\begin{align}
\label{eq:quark_meas_finite_kt}
J^{q}_{\kt}(\tau_a) &=   \CF\Biggl[ \frac{3}{2}\ln\frac{\mu^2}{\omega^2\tan^2\frac{R}{2}} + \frac{1-\frac{a}{2}}{1-a} \ln^2\frac{\mu^2}{\omega^2} + \left(1-\frac{a}{2}\right) \ln^2\tan^2\frac{R}{2}  + \frac{13}{2}  \nn\\
& \quad - \frac{\pi^2}{6} \left(4 + \frac{1-\frac{a}{2}}{1-a}\right) \Biggr] \delta(\tau_a) -  \CF\Biggl[ \Biggl(  \frac{4}{1-a}\ln\frac{\mu\tan^{1-a}\frac{R}{2}}{\omega \tau_a }  \Biggr) \frac{\Theta(\tau_a -\tau_a^{\text{max}})}{\tau_a} \Biggr]_+ \nn\\
&\quad -  \frac{\CF}{1-\frac{a}{2}} \bigg\{ \frac{\Theta(\tau_a)\Theta(\tau_a^{\max} - \tau_a)}{\tau_a} \bigg[ \frac{3}{2} + \frac{2-a}{1-a}\ln\frac{\mu^2}{\omega^2\tau^{\frac{1}{1-a/2}}}  \nn\\
&\qquad\qquad\qquad  -r_q( x_1)  - 2\ln\frac{\tau_a}{\tan^{2-a}\frac{R}{2}} +\Theta \left(\tau_a^{\frac{1}{2-a}} > 2\tan\frac{R}{2} \right)\biggl( r_q(x_2) - \frac{3}{2} \biggr)\bigg]\bigg\}_\dotplus
\, .
\end{align}
\end{subequations}
For $a=0$, these expressions for the jet functions can be simplified further to give
\begin{subequations}
\label{eq:quark_meas_finite_a0}
\begin{align}
\label{eq:quark_meas_finite_cone_a0}
J^{q}_\text{cone}(\tau_0) &=  J^{q}_{\text{incl}}(\tau_0)  + C_F\left[3 \frac{\Theta(\tau_0)\Theta\left(\tan^2\frac{R}{2}\!-\!\tau_0\right)}{\tau_0 + \tan^2\frac{R}{2}} + \frac{\Theta\left(\tau_0 \!-\! \tan^2\frac{R}{2}\right)}{\tau_0}\left(2\ln\frac{\tau_0}{\tan^{2}\frac{R}{2}} + \frac{3}{2}\right) \right] \,,
\end{align}
for the cone jet function, and
\begin{align}
\label{eq:quark_meas_finite_kt_a0}
J^{q}_{\kt}(\tau_0) &=  J^{q}_{\text{incl}}(\tau_0) +  C_F\Biggl\{\frac{\Theta(\tau_0)\Theta\left(\frac{1}{4}\tan^2\frac{R}{2} - \tau_0\right)}{\tau_0}\left[3x_1 + 2\ln\left(\frac{1-x_1}{x_1}\frac{\tau_0}{\tan^2\frac{R}{2}}\right)\right] \nn \\
&\quad \qquad\qquad\qquad + \frac{\Theta\left(\tau_0  -  \frac{1}{4}\tan^2\frac{R}{2}\right)}{\tau_0}\left(2\ln\frac{\tau_0}{\tan^{2}\frac{R}{2}} + \frac{3}{2}\right) \Biggr\}\,,
\end{align}
\end{subequations}
for the $\kt$ jet function. In \eq{eq:quark_meas_finite_kt_a0}, $x_1$ is given by its value for $a=0$,
\begin{equation}
\label{x1_a0}
x_1 = \frac{1}{2}\left(1-  \sqrt{1-\frac{4\tau_0}{\tan^2\frac{R}{2}}}\right)\,.
\end{equation}
In \eq{eq:quark_meas_finite_a0}, we have divided the cone and $\kt$ jet functions into the  contribution $J^{q}_{\text{incl}}(\tau_0)$ to the inclusive jet function \cite{Bauer:2003pi, Bosch:2004th}, given by
\begin{equation}
J^{q}_{\text{incl}}(\tau_0) =  C_F\left\{ \delta(\tau_0) \left( \frac{3}{2} \ln\frac{\mu^2}{\omega^2} + \ln^2\frac{\mu^2}{\omega^2}+ \frac{7}{2} - \frac{\pi^2}{2}\right)   - \left[\frac{\Theta(\tau_0)}{\tau_0} \left(\frac{3}{2} + 2\ln\frac{\mu^2}{\omega^2\tau}\right)\right]_+ \right\} \,,
\end{equation}
and algorithm-dependent parts. The algorithm-dependent part of the $a=0$ cone jet function \eq{eq:quark_meas_finite_cone_a0} agrees with \cite{Jouttenus:2009ns}.  Note that if one takes $R$ to be parametrically larger than $\tau_0$ (cf. \sec{sec:power} and \eq{scalechoices}), the algorithm-dependent parts of \eq{eq:quark_meas_finite_a0} are power suppressed, and the cone and $\kt$ jet functions reduce to the inclusive jet function.

\paragraph{Unmeasured Quark Jet Function}
The finite pieces for the unmeasured quark jet function are
\be
\label{eq:quark_unmeas_finite}
J^q_\text{alg} =  \frac{3 \CF}{2} \ln \left(\frac{\mu^2}{\omega^2 \tan^2\frac{R}{2}} \right) + \frac{\CF}{2}\ln^2 \left(\frac{\mu^2}{\omega^2 \tan^2\frac{R}{2}} \right) +d^{q, \, \text{alg}}_J
\,,\ee
where the constant term $d^{q, \, \text{alg}}_J$ is given by
\begin{equation}
\label{eq:dJq}
d^{q, \, \text{cone}}_J = \CF\left( \frac{7}{2} + 3\ln 2 - \frac{5\pi^2}{12}   \right) \ , \quad
d^{q, \, \kt}_J =  \CF\left( \frac{13}{2} - \frac{3 \pi^2}{4}  \right)
\,,
\end{equation}
for the cone and $\kt$ algorithms, respectively.

\subsection{Finite Pieces of the Gluon Jet Function}
\paragraph{Measured Gluon Jet Function}

The finite distributions of the naive gluon jet function are given by
\begin{align}
\label{eq:gluon_meas_finite_cone_naive}
\gjetnaive^{g}_\text{cone}(\tau_a) &= \delta(\tau_a)\left[ \CA\left(  \frac{137}{36} + \frac{11}{3} \ln{2}  - \frac{\pi^2}{3}  \right)  - \TR\NF \left( \frac{23}{18} + \frac{4}{3} \ln{2} \right) \right] \nn\\
& \qquad \qquad \qquad  + \frac{1}{1-\frac{a}{2}} \left[ \mathcal{I}^{g}_\text{cone} \frac{\Theta(\tau_a)\Theta(\tau_a^{\max} - \tau_a)}{\tau_a}\right]_\dotplus
\,,\end{align}
and
\begin{align}
\label{eq:gluon_meas_finite_kt_naive}
\gjetnaive^{g}_{\kt}(\tau_a) &=  \delta(\tau_a)\left[\CA\left(  \frac{67}{9}  - \frac{2\pi^2}{3}  \right)  - \TR\NF \left( \frac{23}{9} \right) \right]   + \frac{1}{1-\frac{a}{2}} \left[ \mathcal{I}^{g}_{\kt} \frac{\Theta(\tau_a)\Theta(\tau_a^{\max} - \tau_a)}{\tau_a}\right]_\dotplus \
\,,\end{align}
where the integrals $\mathcal{I}^g_\text{alg}$ are given by
\be
\mathcal{I}^g_\text{alg} = \int\!dx \left[ \CA \left( \frac{1}{x(1-x)} + x(1-x) -2 \right) + \TR\NF \left( 1 - 2x(1-x)\right) \right]
\,,\ee
with the cone and $\kt$ regions of integration the same as for the quark jet functions.  The value $\tau_a^{\max}$ is the same as in the measured quark jet function, for the respective jet algorithm.

Going through similar steps as for the quark jet function, defining
\be
\label{eq:rg}
r_g(x) = 2 \CA \ln \left( \frac{1-x}{x}\right) + \CA \,x \left( \frac{2}{3}x^2 - x+4 \right) - \TR \NF \,x \left(\frac{4}{3}x^2 - 2x+2 \right)
\,,\ee
and using \eq{eq:tau-extract} to make all logarithmic dependence on $\tau_a$ explicit, we find for the cone and $\kt$-type jet function finite distributions
\begin{subequations}
\label{eq:gluon_meas_finite}
\begin{align}
\label{eq:gluon_meas_finite_cone}
J^{g}_\text{cone}(\tau_a) &= \delta(\tau_a)\Bigg[\frac{\beta_0}{2} \ln \frac{\mu^2}{\omega^2 \tan^2 \frac{R}{2}} + \CA \frac{1-\frac{a}{2}}{1-a}\ln^2\frac{\mu^2}{\omega^2} + \CA \left(1-\frac{a}{2}\right)\ln^2 \tan^2\frac{R}{2}  \\
& \qquad \qquad +  \CA\left(  \frac{137}{36} + \frac{11}{3} \ln{2}  - \frac{\pi^2}{6} \left(2 + \frac{1-\frac{a}{2}}{1-a}\right)  \right)  - \TR\NF \left( \frac{23}{18} + \frac{4}{3} \ln{2} \right)\Bigg]\nn\\
& \quad - \Bigg[\Biggl( \frac{4 \CA}{1-a} \ln \frac{\mu\tan^{1-a}\frac{R}{2}}{\omega \tau_a}\Biggr)\frac{\Theta(\tau_a)\Theta(\tau_a - \tau_a^{\text{max}})}{\tau_a} \Bigg]_+   -  \frac{1}{1-\frac{a}{2}} \Biggl[ \frac{\Theta(\tau_a)\Theta(\tau_a^{\max} - \tau_a)}{\tau_a}\nn\\
& \qquad\qquad \qquad \times \Biggl(\frac{\beta_0}{2} + \frac{2-a}{1-a}\CA\ln\frac{\mu^2}{\omega^2\tau^{\frac{1}{1-a/2}}}    -r_g(\xcone) - 2\CA\ln\frac{\tau_a}{\tan^{2-a}\frac{R}{2}}   \Biggr) \Biggr]_+ \nn
\,,\end{align}
and
\begin{align}
\label{eq:gluon_meas_finite_kt}
J^{g}_{\kt}(\tau_a) &=\delta(\tau_a)\Bigg[\frac{\beta_0}{2} \ln \frac{\mu^2}{\omega^2 \tan^2 \frac{R}{2}} + \CA \frac{1-\frac{a}{2}}{1-a}\ln^2\frac{\mu^2}{\omega^2} + \CA \left(1-\frac{a}{2}\right)\ln^2 \tan^2\frac{R}{2}  \\
& \qquad \qquad +  \left. \CA\left(  \frac{67}{9}  - \frac{\pi^2}{6} \left(4 + \frac{1-\frac{a}{2}}{1-a}\right)   \right)  - \TR\NF \left( \frac{23}{9} \right) \right]\nn\\
& \quad - \Bigg[ \Bigg( \frac{4 \CA}{1-a} \ln \frac{\mu\tan^{1-a}\frac{R}{2}}{\omega \tau_a}\Bigg)\frac{\Theta(\tau_a)\Theta(\tau_a - \tau_a^{\text{max}})}{\tau_a} \Bigg]_+ \nn\\
& \quad -  \frac{1}{1-\frac{a}{2}} \Biggl\{ \frac{\Theta(\tau_a)\Theta(\tau_a^{\max} - \tau_a)}{\tau_a} \Biggl[  \frac{\beta_0}{2}+ \frac{2-a}{1-a}\CA\ln\frac{\mu^2}{\omega^2\tau^{\frac{1}{1-a/2}}}  \nn\\
&\qquad\quad  - r_g(x_1) - 2\CA\ln\frac{\tau_a}{\tan^{2-a}\frac{R}{2}}   +\Theta \left(\tau_a^{\frac{1}{2-a}} > 2\tan\frac{R}{2} \right) \bigg(r_g(x_2) - \frac{\beta_0}{2}\bigg) \Biggr] \Bigg\}_\dotplus \nn
\,,\end{align}
\end{subequations}
where $\xcone$ and $x_{1, \, 2}$ are given in \eqs{eq:xcone-defn}{kTtauconstraint}.

For $a=0$, the simplified result for the gluon cone jet function is
\begin{equation}
\label{eq:gluon_meas_finite_cone_a0}
\begin{split}
J^{g}_\text{cone}(\tau_0) =  J^{g}_\text{incl}(\tau_0) &+ \frac{\Theta(\tau_0)\Theta\left(\tan^2\frac{R}{2} - \tau_0\right)}{\tau_0+\tan^2\frac{R}{2}}f\left(\frac{\tau_0}{\tau_0+\tan^2\frac{R}{2}}\right) \\
&  + \frac{\Theta\left(\tau_0 - \tan^2\frac{R}{2}\right)}{\tau_0}\left(2\CA\ln\frac{\tau_0}{\tan^2\frac{R}{2}} + \frac{\beta_0}{2}\right)
\,,
\end{split}
\end{equation}
where
\begin{equation}
f(x) \equiv \CA  \left( \frac{2}{3}x^2 - x+4 \right) - \TR \NF  \left(\frac{4}{3}x^2 - 2x+2 \right)\,.
\end{equation}
For $a=0$,  the gluon $\kt$ jet function is given by,
\begin{equation}
\label{eq:gluon_meas_finite_kt_a0}
\begin{split}
J^{g}_{\kt}(\tau_0) =  J^{g}_\text{incl}(\tau_0) &+ \frac{\Theta(\tau_0)\Theta\left(\frac{1}{4}\tan^2\frac{R}{2} - \tau_0\right)}{\tau_0} \left[ r_g(x_1) + 2\CA\ln\frac{\tau_0}{\tan^2\frac{R}{2}}\right] \\
&  + \frac{\Theta\left(\tau_0 -\frac{1}{4}\tan^2\frac{R}{2}\right)}{\tau_0}\left(2\CA\ln\frac{\tau_0}{\tan^2\frac{R}{2}} + \frac{\beta_0}{2}\right)
\,,
\end{split}
\end{equation}
where $x_1$ is given by \eq{x1_a0}, and $r_g(x_1)$ is given by \eq{eq:rg}.
In \eqs{eq:gluon_meas_finite_cone_a0}{eq:gluon_meas_finite_kt_a0}, the contribution $J^{g}_\text{incl}(\tau_0)$  to the inclusive gluon jet function \cite{Bauer:2006qp,Becher:2009th,Bauer:2001rh, Fleming:2003gt}   is
\begin{align}
J^{g}_\text{incl}(\tau_0) &= \delta(\tau_0)\Bigg[\frac{\beta_0}{2} \ln \frac{\mu^2}{\omega^2} + \CA \ln^2\frac{\mu^2}{\omega^2} +  \CA\left(  \frac{67}{18}   - \frac{\pi^2}{2}   \right)  - \frac{10 \TR\NF}{9}\Bigg] \nn \\
&\qquad  - \Bigg[\Biggl( \frac{\beta_0}{2}+ 2\CA \ln \frac{\mu^2}{\omega^2 \tau_0}\Biggr)\frac{\Theta(\tau_0)}{\tau_0} \Bigg]_+
\end{align}
As for the quark jet functions \eq{eq:quark_meas_finite_a0}, the gluon jet functions split up into the inclusive jet function and  algorithm-dependent pieces that are power suppressed for $\tau_0\ll R$.

\paragraph{Unmeasured Gluon Jet Function}
For the unmeasured gluon jet functions, the finite pieces are given by
\be
\label{eq:gluon_unmeas_finite}
J^g_\text{alg} = \frac{\CA}{2} \ln^2 \frac{\mu^2}{\omega^2 \tan^2 \frac{R}{2}} + \frac{\beta_0}{2} \ln \frac{\mu^2}{\omega^2 \tan^2 \frac{R}{2}} + d_J^{g, \, \text{alg}}
\ee
where the constant part $d_J^{g, \, \text{alg}}$ for the cone and $\kt$ algorithms is given by, respectively,
\begin{subequations}
\label{eq:dJg}
\begin{align}
\label{eq:dJg_cone}
d^{g, \, \text{cone}}_J &= \CA\left(  \frac{137}{36} + \frac{11}{3} \ln{2}  - \frac{5\pi^2}{12}  \right) - \TR\NF \left( \frac{23}{18} + \frac{4}{3}\ln 2\right)
\end{align}
and
\begin{align}
\label{eq:dJg_kt}
d^{g, \, \kt}_J &= \CA\left(  \frac{67}{9}  - \frac{3\pi^2}{4}  \right)  - \TR\NF \left( \frac{23}{9}\right)
\,.\end{align}
\end{subequations}

\section{Soft function calculations}
\label{app:softcalc}

\subsection{$S^{\incl}_{ij}$}
\label{sec:app-Sinclij}

To evaluate the expression \eq{eq:Sincl-integral}, we first define
\begin{align}
S^{\incl}_{ij}
& \equiv \frac{1}{\epsilon} \frac{1}{\Gamma(1-\epsilon)}\frac{\as}{2 \pi} \left(\frac{4 \pi \mu^2}{4 \Lambda^2}\right)^\epsilon\, \vect{T}_i \cdot \vect{T}_j \, \mathcal{I}^{\incl}(n_i \cdot n_j)
\,.\end{align}
We need $\mathcal{I}^\incl$ to $\cO(\epsilon)$. Working in a coordinate system with $\vec n_i$ aligned along the $z$-axis and $\vec n_j$ in the $xz$-plane and defining $n \equiv 1 - n_i \cdot n_j = n_j^z$, we have
\begin{align}
\mathcal{I}^{\incl}(n_i \cdot n_j)
& = \frac{ n_i \cdot n_j \,4^\epsilon \, \Gamma(1-\epsilon)}{2\sqrt{\pi} \Gamma(\frac{1}{2}-\epsilon)} \int_0^\pi d\phi\sin^{-2\epsilon}\phi \int_0^\pi d \theta \sin^{1-2 \epsilon}\theta \frac{1}{1-\cos \theta} \nn\\
& \qquad \qquad \times \frac{1}{1-n_j^x \sin \theta \cos \phi - n^z_j \cos \theta} \nn\\
&=  \frac{4^\epsilon}{2} \Gamma(1-\epsilon) \int_{-1}^{+1} \!d u \, (1-u)^{-1-\epsilon} (1+u)^{-\epsilon} \frac{1-n}{1- u n} \,  _{2}\tilde{F}_1 \bigg(\frac{1}{2}, 1; 1-\epsilon; z \bigg)
\,\end{align}
where $z = \frac{(1-n^2)(1-u^2)}{(1- u n)^2}$. The integration over $u = \cos{\theta}$ has singularities at the points $u = 1 $ and $u= n$ which correspond to $z=1$ and $z=0$, respectively. To isolate these singularities, we split the integration over $u$ into the ranges $ (-1, \delta )$ and $ ( \delta, 1 )$ where $n < \delta < 1$,
\begin{align}
\mathcal{I}^{\incl}(n_i \cdot n_j) = \mathcal{I}_1^{\incl}(n_i \cdot n_j)+\mathcal{I}_2^{\incl}(n_i \cdot n_j)
\,,\end{align}
where
\begin{align}
\mathcal{I}_1^{\incl}(n_i \cdot n_j) &\equiv  \frac{4^\epsilon}{2} \Gamma(1-\epsilon) \int_{-1}^{\delta} \!d u \, (1-u)^{-1-\epsilon} (1+u)^{-\epsilon} \frac{1-n}{1- u n} \,  _{2}\tilde{F}_1 \bigg(\frac{1}{2}, 1; 1-\epsilon; z \bigg) \nn\\
 \mathcal{I}_2^{\incl}(n_i \cdot n_j)&\equiv  \frac{4^\epsilon}{2} \Gamma(1-\epsilon) \int_{\delta}^{1} \!d u \, (1-u)^{-1-\epsilon} (1+u)^{-\epsilon} \frac{1-n}{1-u n} \,  _{2}\tilde{F}_1 \bigg(\frac{1}{2}, 1; 1-\epsilon; z \bigg)
\,.\end{align}
Over the range of integration of $u$ in $\mathcal{I}_1^{\incl}$, $z \in [0, 1)$ for $\delta < 1$. For $\mathcal{I}_2^{\incl}$, $z \in (0, 1]$.

Furthermore, the singularity at $u = n$ in $\mathcal{I}^\incl_1$ is made more explicit through the use of the identity
\begin{align}
_{2}\tilde{F}_1 \bigg(\frac{1}{2}, 1; 1-\epsilon; z \bigg) &= f_a(z) + f_b(z) \nn\\
f_a(z)  &= \frac{\sqrt{\pi}}{\cos{(\epsilon \pi)}} \left( \frac{1-nu}{\abs{u-n}} \right)^{1+2\epsilon} {}_{2}\tilde{F}_1 \bigg(\frac{1}{2}-\epsilon, -\epsilon; \frac{1}{2}-\epsilon; 1-z \bigg) \nn\\
f_b(z)  &= \frac{\pi}{\cos{(\epsilon \pi)}} \frac{\epsilon}{\Gamma(1/2-\epsilon) \Gamma(1-\epsilon)} {}_2\tilde{F}_1 \bigg(\frac{1}{2}, 1; \frac{3}{2}-\epsilon;1-z \bigg)
\,.\end{align}
$f_a(z)$ gives an $\cO(1/\epsilon)$ contribution and we proceed by using the following trick that we exploit multiple times throughout the Appendix.

To integrate a product of functions $f(x, \epsilon) g(x, \epsilon)$ where $f$ diverges at the point $x_0$ as $(x-x_0)^{-1+\cO(\epsilon)}$, we write the integation as
\be
\label{eq:plus-trick}
\int\!dx \,f(x, \epsilon) g(x, \epsilon) = \int\!dx\, f(x, \epsilon) g(x_0, \epsilon) + \int\!dx \,f(x, \epsilon) \Big( g(x, \epsilon)  - g(x_0, \epsilon)\Big)
\,.\ee
The first integral has relatively simple $x$ dependence since $g(x_0, \epsilon)$ does not depend on $x$. The term in parenthesis in the second integral vanishes as $x-x_0$ for regular functions $g$ and so the entire integrand can be expanded in $\epsilon$.

We can now evaluate $f_a(z)$ by adding and subtracting the non-singular part of the integrand (which is the hypergeometric function) evaluated at $u=n$ as in \eq{eq:plus-trick}, whereas $f_b(z)$ is $\cO(\epsilon)$ and so we can simply expand about $\epsilon = 0$. Adding these contributions, we find that
\begin{align}
\mathcal{I}_1^{\incl}(n_i \cdot n_j) & = \frac{4^\epsilon}{2} \Bigg[\frac{\sqrt{\pi }\,\Gamma (1-\epsilon ) \left(1-n^2\right)^{\epsilon }}{\cos(\pi  \epsilon ) \Gamma \left(\frac{1}{2}-\epsilon \right)}\int _{-1}^{\delta } \frac{du}{\abs{u-n}^{1+2\epsilon }} \nn\\
& \qquad - \int _{-1}^{\delta }du \frac{\sgn(n-u)}{1-u} \Bigg( 1-\epsilon  \ln\left(\frac{4 (n-u)^2}{1-n^2}\right)\Bigg)\nn\\
& \qquad + \epsilon \int _{-1}^{\delta }\frac{du}{1-u}\frac{2}{|u-n|}\tanh ^{-1}\left(\frac{|u-n|}{1-n u}\right) \Bigg]
\label{eq:I-incl1}
\,.\end{align}

For $\mathcal{I}_2^{\incl}$, the part of the integrand that is not singular at $u=1$ is everything that multiplies $(1-u)^{-1-\epsilon}$, and so we add and subtract this part as in \eq{eq:plus-trick}. This gives
\begin{align}
\mathcal{I}_2^{\incl}(n_i \cdot n_j) =  \frac{4^\epsilon}{2} \Bigg[-\frac{1}{\epsilon} 2^{-\epsilon} (1-\delta)^{-\epsilon} + &\int_\delta^1 \frac{du}{u-n} \Bigg(1+ \epsilon \frac{1-n}{1-u}\log\left(\frac{(n-1)^2 (u+1)}{4 (n-u)^2}\right) \nn\\
& \qquad - \epsilon \log(1-u)+ \epsilon\frac{u-n}{1-u} \log(2) \Bigg) \Bigg]
\label{eq:I-incl2}
\,.\end{align}

The integrals in \eqs{eq:I-incl1}{eq:I-incl2} give rise to many terms. However, we find that, after some lengthy algebra, the dependence on $\delta$ cancels in the sum as it must and that the result can be simplified to
\begin{equation}
\mathcal{I}^\incl(n_i \cdot n_j) = -\frac{1}{\epsilon} + \ln \left( \frac{n_i \cdot n_j}{2} \right) + \epsilon \left(\frac{\pi^2}{6}+ \Li_2\left(1 - \frac{2}{n_i\cdot n_j}\right)\right)
\,.\end{equation}

\subsection{$S_{ij}^i$ and $S_{ij}^\meas(\tau^i_a)$ }
\label{app:Siji-and-Sijmeas}

\subsubsection{Common Integrals}
In evaluating the soft contributions $S_{ij}^i$ and $S_{ij}^\meas(\tau^i_a)$, we find an integral of the following form:
\begin{align}
I(\alpha, \beta, t) = 2 t^2 \int_0^1\!\frac{du}{u} u^{ 2 \alpha \epsilon} f(u; \beta, t)
\label{eq:I-defn}
\,,\end{align}
where $t>1$ and
\be
f(u; \beta, t) =  \frac{(\tan^{-2}\frac{R}{2}+ u^2)^{2 \beta \epsilon}}{(u+ t)^2} \, {}_{2}F_1 \bigg(1,\frac{1}{2}-\epsilon; 1-2\epsilon; \frac{4 t u}{(u+t)^2} \bigg)
\,.\ee
To evaluate this integral, we add and subtract the part of the integrand that is not singular at $u = 0$,  namely $f(u; \beta, t)$, as in \eq{eq:plus-trick}. This allows us to write
\begin{align}
I(\alpha, \beta, t) &= 2 \tan^{-4 \beta\epsilon}\frac{R}{2} \int_0^1 du \bigg\{ u^{-1+ 2 \alpha \epsilon}  \nn\\
& \qquad + \frac{1}{t^2-u^2}\left[ u + 2\epsilon \left( \alpha u \ln u + \frac{t^2}{u} \ln \frac{t^2}{t^2-u^2}+ \frac{\beta t^2}{u} \ln \Big(1+ \tan^2\frac{R}{2}u^2\Big)\right)  \right]\bigg\}
\,,\end{align}
where we used that
\be
f(0; \beta, t) =  \frac{1}{t^2} \tan^{-4 \beta\epsilon}\frac{R}{2}
\,,\ee
and that the expansion of the hypergeometric about $\epsilon = 0$ for $t > u$ is
\be
{}_{2}F_1 \bigg(1,\frac{1}{2}-\epsilon; 1-2\epsilon; \frac{4 t u}{(u+t)^2} \bigg) = \frac{t+u}{t-u} \left(1+2\epsilon \ln \frac{t^2}{t^2-u^2} + \cO(\epsilon^2) \right)
\,.\ee
Evaluating the integals, we obtain
\begin{align}
I(\alpha, \beta, t) &=   \frac{ \tan^{-4 \beta\epsilon}\frac{R}{2}}{\alpha \epsilon} \left(\frac{t^2}{t^2-1} \right)^{\alpha \epsilon}+ \epsilon\big( \alpha + 2 \beta - 2\big)\Li_2\left(\frac{-1}{t^2-1}\right)  \nn\\
& \qquad -2 \beta \epsilon \Li_2\left( - \frac{1+ t^2 \tan^2 \frac{R}{2}}{ t^2-1}\right) + \cO(\epsilon^2)
\label{eq:I-eval}
\,.\end{align}

\subsubsection{$S_{ij}^{\meas}(\tau_a^i)$}
To evaluate \eq{eq:Sijmeas-integral} for the case that $k = i$, we use light cone coordinates in the frame of jet i, $k^{+}   = n_i \cdot k$ and $k^{-}   = \bar{n}_i \cdot k$. In terms of these variables, the on-shell condition can be used to give
\be
n_j\cdot k = k^+\cos^2\frac{\psi_{ij}}{2} + k^-\sin^2\frac{\psi_{ij}}{2} - \sqrt{k^+k^-}\sin\psi_{ij}\cos\phi ,
\ee
with $\cos\psi_{ij} =1 -  n_i \cdot n_j$, and $\phi$ the angle in $k_\perp$-space (the azimuthal angle about $\vec{n}_i$).  We can do the $k_\perp$ and $k^+$ integrals using the on-shell and $\tau_a$ delta functions respectively.
The resulting $S_{ij}^{\meas}(\tau^i_a)$ has non-trivial integrals over $k^-$ and $\phi$:
\be
\label{eq:Sijmeas-integral-2}
\begin{split}
S_{ij}^{\meas}(\tau_a^i) &= - \frac{\alpha_s}{4\pi}\left(\frac{4\pi\mu^2}{\omega^2}\right)^{\epsilon}(n_i\cdot n_j)(\vect T_i\cdot \vect T_j) \frac{1}{\sqrt{\pi}\,\Gamma({\sfrac12} - \epsilon)}\frac{2\omega}{2-a}\frac{1}{(\tau_a^i)^{2\epsilon}}\int_0^{\pi}\!d\phi \sin^{-2\epsilon}\phi  \\
&\times  \int_0^{\infty} \frac{dk^-}{(k^-)^2}\left(\frac{\omega\tau_a^i}{k^-}\right)^{-1} \Theta\left(\tan^2\frac{R}{2} - \left(\frac{\omega\tau_a^i}{k^-}\right)^{\frac{2}{2-a}}\right)\left(\frac{\omega\tau_a^i}{k^-}\right)^{2\epsilon\frac{1-a}{2-a}} \\
&\times\left[\left(\frac{\omega \tau_a^i}{k^-}\right)^{\frac{2}{2-a}}\cos^2\frac{\psi_{ij}}{2} + \sin^2\frac{\psi_{ij}}{2} - \left(\frac{\omega\tau_a^i}{k^-}\right)^{\frac{1}{2-a}}\sin\psi_{ij}\cos\phi\right]^{-1}
\,.\end{split}
\ee
Making the change of variables $u = \cot\frac{R}{2} \sqrt{k^+/k^-} = \cot\frac{R}{2}\left(\frac{\omega\tau^i_a}{k^-}\right)^{\frac{1}{2-a}} $, we find that $S_{ij}^{\meas}(\tau^i_a)$ can be written as
\be
S_{ij}^{\meas}(\tau_a^i) = - \frac{\as}{2\pi} \vect T_i\cdot \vect T_j \,\frac{1}{\Gamma(1-\epsilon)}\left( \frac{4 \pi \mu^2}{\omega^2} \tan^{2(1-a)}\frac{R}{2} \right)^\epsilon \left( \frac{1}{\tau_a^i}\right)^{1+ 2 \epsilon} I(1-a, 0, t_{ij})
\,,\ee
where $I(\alpha, \beta, t)$ is defined in \eq{eq:I-defn}. Using \eq{eq:I-eval} we find the result given in \eq{eq:Smeasiji-ans}.

\subsubsection{$S_{ij}^i$}

The $\Theta$-functions in \eq{eq:Sijk-integral} are easiest to deal with if we shift to variables where each $\Theta$-function is in a different variable.  The simplest choices are just the arguments of the $\Theta$ functions $\Theta_\Lambda$ and $\Theta_R^i$, $k^0$ and $u = \cot\frac{R}{2} \sqrt{k^+/k^-}$, respectively, where $k^{\pm}$ are defined with respect to direction $n_i$.  This gives a form similar to the integral in $S_{ij}^{\meas}(\tau^i_a)$,
\be
S_{ij}^j = - \frac{1}{\epsilon}\frac{\as}{4 \pi}  \vect T_i\cdot \vect T_j \,\frac{1}{\Gamma(1-\epsilon)} \left(\frac{4\pi\mu^2}{4\Lambda^2}\tan^2\frac{R}{2}\right)^{\epsilon} I(-1, 1, t_{ij})
\,.\ee
where $I(\alpha, \beta, t)$ is defined in \eq{eq:I-defn} and evaluates to \eq{eq:I-eval}. This gives \eq{eq:Sijj-ans}.

\subsection{$S_{ij}^{\meas}(\tau_a^k)$ and $S_{ij}^k$ for $k \neq i,j$}

We again use light cone coordinates centered on jet $k$. The integrations involved in $S_{ij}^{\meas}(\tau_a^k)$ and $S_{ij}^k$ only give rise to a $1/\epsilon$ pole as explained in the text, but integrating the eikonal factor $1/(n_i\cdot k)(n_j\cdot k)$ is more complicated than for the other cases since there is a third direction, $n_k$, involved.

For unmeasured jets when there are $n \ge 3$ total final state jets, $S_{ij}^k$ is needed. However, as we explain in the text, measured jets violate consistency at $\cO(1/t^2)$ even for $n=2$ (non back-to-back) jets and the contribution of $S_{ij}^k$ does not ameliorate this fact when $n \ge 3$. To show this, we need to evaluate the divergent contribution of $S_{ij}^k$. In addition, we give the form of the finite pieces which are $\cO(1/t^2)$.

For each measured jet when there are $n \ge 3$, the sum $S_{ij}^{\meas}(\tau_a^k) +S_{ij}^k\delta(\tau_a^k)$ is needed. However, in this case the $1/\epsilon$ pole cancels in this sum and we are left with only a single, finite integral to evaluate. This is clear from the expressions for $S_{ij}^{\meas}(\tau_a^k)$ and $S_{ij}^k$ which we derive in \sec{app:Sijk} and \sec{app:Smeas-ijk}, respectively. We evaluate the sum explicitly in \sec{app:Sijk-and-Smeas-ijk}.

\subsubsection{Common Integrals}
We find the following integral arising in both $S_{ij}^{\meas}(\tau_a^k)$ and $S_{ij}^k$:
\begin{align}
\label{eq:I-Sijk-defn}
I(u; t_{a}, t_{b}, \beta) &\equiv -\frac{2 \epsilon}{\pi} \int_0^\pi\!d\theta_1 \, \sin^{-2\epsilon}\!\theta_1 \int_0^\pi\!d\theta_2 \, \sin^{-1-2\epsilon}\!\theta_2 \,\frac{t_a^2 +t_b^2 - 2 t_a t_b \cos\beta }{u^2+t_{a}^2 - 2 u t_{a} \cos\theta_1} \nn\\ &
\qquad \times \frac{1}{u^2+t_{b}^2 - 2 u t_{b}(\cos\beta \cos\theta_1 + \sin\beta \sin\theta_1 \cos\theta_2)} \nn\\
& = I^{(0)}(u; t_{a}, t_{b}, \beta) + \epsilon I^{(1)}(u; t_{a}, t_{b}, \beta) + \cO(\epsilon^2)
\,,\end{align}
where the $\cO(\epsilon^0)$ and $\cO(\epsilon^1)$ parts of $I$ are
\begin{align}
\label{eq:I-pieces}
 I^{(0)}(u; t_{a}, t_{b}, \beta)  &= \frac{2}{\pi} \int_0^\pi\!\d\theta \frac{A}{A^2 - B^2} \frac{t_a^2 +t_b^2 - 2 t_a t_b \cos\beta}{u^2+t_{a}^2 - 2 u t_{a} \cos\theta} \\
  I^{(1)}(u; t_{a}, t_{b}, \beta)  &= - \frac{2}{\pi} \int_0^\pi\!\d\theta \left[\frac{2\ln{(\sin\theta)} A}{A^2 - B^2} +  \frac{B}{A^2 - B^2} \log\frac{A-B}{A+B}\right]\frac{t_a^2 +t_b^2 - 2 t_a t_b \cos\beta}{u^2+t_{a}^2 - 2 u t_{a} \cos\theta} \nn
\,,\end{align}
where we defined
\begin{align}
A &= u^2 + t_b^2 - 2 u t_b \cos\beta \cos\theta \nn\\
B &= 2 u t_b \sin\beta \sin\theta
\,.\end{align}

We can evaluate $I^{(0)}$ straightforwardly. For the range of our interest, $t_{a,b}>1$ and $0 < u< 1$, it gives
\be
 I^{(0)}(u; t_{a}, t_{b}, \beta) =  \frac{2(t_a^2 +t_b^2 - 2 t_a t_b \cos\beta)( t_a^2 t_b^2-u^4)}{(t_a^2-u^2)(t_b^2 - u^2)(t_a^2 t_b^2 - 2 t_a t_b u^2 \cos\beta + u^4)}
\,.\ee
 In addition, we will need the following integrals over $ I^{(0)}(u)$:
 \begin{align}
 \label{eq:f1-app}
 f_1( t_{a}, t_{b},\beta) &\equiv \int_0^1\! du\, u \, I^{(0)}(u; t_{a}, t_{b}, \beta)
  = \ln\bigg(\frac{t_a^2 t_b^2 - 2 t_a t_b \cos\beta+1}{(t_a^2 - 1)(t_b^2 - 1)} \bigg)
\,, \end{align}
 and
 \begin{align}
  \label{eq:f2-app}
  f_2( t_{a}, t_{b},\beta, r) &\equiv \int_0^1\! du\, u\, I^{(0)}(u;  t_{a}, t_{b}, \beta)  \ln(r+u^2) \nn\\
&= -\Bigg\{g(t_a, r) + g(t_b, r)+2 \ln(r+1) \ln(t_a t_b)  \nn\\
& \qquad +2 \Real\Bigg[ \Li_2\bigg(\frac{t_a t_b - e^{i \beta}}{r e^{i\beta} +t_a t_b} \bigg) - \Li_2 \bigg( \frac{t_a t_b}{r e^{i\beta} +t_a t_b}\bigg) \nn\\
 & \qquad \qquad \qquad+ \ln\bigg(\frac{t_a t_b}{t_a t_b - e^{i \beta}}\bigg) \ln(r+ t_at_b e^{-i\beta})\bigg)\Bigg] \Bigg\}
\,, \end{align}
where
\begin{align}
g(t, r) \equiv \Li_2 \bigg( \frac{t^2}{t^2+r}\bigg) -  \Li_2 \bigg( \frac{t^2-1}{t^2+r}\bigg)  + \ln(t^2 - 1) \ln(r+t^2) - \ln(t^2) \ln((r+1)(r+t^2))
\,.\end{align}
For $r=0$, this simplifies to
\be
f_2(t_a, t_b,\beta,  0) = - \Li_2 \bigg(\frac{1}{t_a^2}\bigg) - \Li_2 \bigg(\frac{1}{t_b^2}\bigg) + 2 \Real \bigg[  \Li_2\bigg (\frac{e^{i \beta}}{t_a t_b}\bigg)\bigg]
\,.\ee
Notice that both $f_1$ and $f_2$ are $\cO(1/t^2)$.

The $\cO(\epsilon^1)$ piece, $I^{(1)}$, is less trivial. However, the only property of $I^{(1)}$ that we need is that
\be
\label{eq:f3-app}
f_3(t_a, t_b,\beta) \equiv \int_0^1\!du \,u I^{(1)}(u;  t_{a}, t_{b}, \beta) = \cO(1/t^2)
\,,\ee
which can be seen by taking the large-$t$ limit of $I^{(1)}$ in \eq{eq:I-pieces}.  The integral is finite an suppressed by $1/t^2$.

\subsubsection{$S_{ij}^k$}
\label{app:Sijk}
To compute $S_{ij}^k$, we choose a coordinate system such the $\vec n_k$ is in the $z$-direction and $\vec n_i$ lies in the $xz$-plane. In terms of the light-cone coordinates about $n_k$ and the variable $u = \cot \frac{R}{2} \sqrt{k^+/k^-}$, we have
\begin{align}
n_i \cdot k &= k^+ \cos^2 \frac{\psi_{ik}}{2} + k^- \sin^2\frac{\psi_{ik}}{2}  - \sqrt{k^+ k^-} \sin \psi_{ik} \cos{\theta_1}\nn\\
&= k^- \cos^2 \frac{\psi_{ik}}{2} \tan^2 \frac{R}{2} \Big[ u^2 + t_{ik}^2 - 2 u t_{ik} \cos\theta_1\Big] \nn\\
n_j \cdot k &= k^+ \cos^2 \frac{\psi_{jk}}{2} + k^- \sin^2\frac{\psi_{jk}}{2}  - \sqrt{k^+ k^-} (n_j^x \cos\theta_1 + n_j^y \sin\theta_1 \cos\theta_2) \nn\\
&=  k^- \cos^2 \frac{\psi_{jk}}{2} \tan^2 \frac{R}{2} \Big[ u^2 + t_{jk}^2 - 2 u t_{jk} \big(\cos\beta_{ij} \cos\theta_1 + \sin\beta_{ij} \sin\theta_1 \cos\theta_2\big)\Big]
\,,\end{align}
where $\beta_{ij}$ is defined as the angle between the $ik$- and $jk$-planes. Using the relation
\be
\frac{n_i \cdot n_j}{\cos^2\frac{\psi_{ik}}{2} \cos^2\frac{\psi_{jk}}{2} \tan^2\frac{R}{2}} = 2(t_{ik}^2+t_{jk}^2 - 2 t_{ik}t_{jk} \cos\beta_{ij})
\,,\ee
we find that $S_{ij}^k$ can be written as
\be
\label{eq:Sijk-full-integral}
S_{ij}^{k}= - \frac{1}{\epsilon} \frac{\alpha_s}{4\pi}  \vect T_i \cdot \vect T_j \frac{1}{\Gamma(1-\epsilon)} \left(\frac{4\pi\mu^2}{4\Lambda^2}\right)^{\epsilon}\tan^{2\epsilon}\frac{R}{2} \int_0^1\!du \,u^{1-2\epsilon} \Big(\tan^{-2}\frac{R}{2}+u^2\Big)^{2 \epsilon} I(u; t_{ik}, t_{jk}, \beta_{ij})
\,,\ee
where $I(u; t_a, t_b, \beta_{ij})$ is defined in \eq{eq:I-Sijk-defn}. Expanding in $\epsilon$, we find
\begin{align}
\label{eq:Sijk-full-final}
S_{ij}^k &= - \frac{\alpha_s}{4\pi}  \vect T_i \cdot \vect T_j \Bigg[\frac{1}{\epsilon} f_1(t_{ik}, t_{jk}, \beta_{ij}) + F(t_{ik}, t_{jk}, \beta_{ij}) \Bigg]
\,,\end{align}
where the finite part is given by
\begin{align}
\label{eq:F-defn}
F(t_a, t_b, \beta) & \equiv\bigg[  f_1( t_a, t_b,\beta) \ln\bigg(\frac{\mu^2}{4 \Lambda^2} \tan^2\frac{R}{2}  \bigg) - f_2(t_a, t_b,\beta, 0) \nn\\
& \qquad+ 2  f_2\Big(t_a, t_{b},\beta, \tan^{-2} \frac{R}{2}\Big) + f_3(t_a, t_b,\beta) \bigg]
\,,\end{align}
and $f_1$, $f_2$, and $f_3$ are given in \eqss{eq:f1-app}{eq:f2-app}{eq:f3-app}, respectively.

\subsubsection{$S^\meas_{ij}(\tau_a^k)$}
\label{app:Smeas-ijk}

Using the same coordinate system as for $S_{ij}^k$, we find that $S_{ij}^{\meas}(\tau_a^k)$ can be written as
\begin{align}
\label{eq:SijkMeas-full-integral}
S_{ij}^{\meas}(\tau^k_a)  &= - \frac{\alpha_s}{2\pi} \vect T_i \cdot \vect T_j \frac{1}{\Gamma(1-\epsilon)} \bigg(\frac{4\pi\mu^2}{\omega^2}\tan^{2(1-a)}\frac{R}{2} \bigg)^\epsilon \bigg( \frac{1}{\tau_a^k}\bigg)^{1+2 \epsilon} \nn\\
& \qquad \times \int_0^1\!du \,u^{1+2\epsilon(1-a)}  I(u; t_{12}, t_{ik}, \beta_{ij})
\,.\end{align}
 Expanding in $\epsilon$ gives
 \begin{align}
\label{eq:SijkMeas-full-final}
S_{ij}^{\meas}(\tau^k_a) &= - \frac{\alpha_s}{2\pi}  \vect T_i \cdot \vect T_j \Bigg[  \bigg(  \frac{1}{\tau_a^k}\bigg)^{1+2\epsilon} f_1( t_{ik}, t_{jk},\beta_{ij}) +  \delta(\tau_a^k) \,G(t_{ik}, t_{jk},\beta_{ij})  \Bigg]
\,,\end{align}
where
\begin{align}
\label{eq:G-defn}
G(t_{a}, t_{b},\beta)& \equiv  -\frac{1}{2} \Bigg[ f_1(t_{a}, t_{b},\beta) \ln\bigg(\frac{\mu^2}{\omega^2} \tan^{2(1-a)}\frac{R}{2}  \bigg)  +   (1-a)  f_2(t_{a}, t_{b},\beta, 0)+  f_3(t_{a}, t_{b},\beta) \Bigg]
\,,\end{align}
and $f_1$, $f_2$, and $f_3$ are given in \eqss{eq:f1-app}{eq:f2-app}{eq:f3-app}, respectively.

\subsubsection{$S_{ij}^k + S^\meas_{ij}(\tau_a^k)$}
\label{app:Sijk-and-Smeas-ijk}

The sum of \eqs{eq:Sijk-full-final}{eq:SijkMeas-full-final} is finite. We find
\begin{align}
\label{eq:Sijk-Smeas-sum}
S_{ij}^{\meas}(\tau_a^k) +S_{ij}^k\delta(\tau_a^k)
&= \frac{\alpha_s}{4\pi}  \vect T_i \cdot \vect T_j \Bigg[ \delta(\tau_a^k) \Bigg(  f_1( t_{ik}, t_{jk},\beta_{ij}) \ln \bigg(\frac{4 \Lambda^2}{\omega^2} \tan^{-2a}\frac{R}{2}\bigg) \nn\\
&\qquad \qquad  + (2-a) f_2(t_{ik}, t_{jk},\beta_{ij}, 0) - 2  f_2\Big(t_{12}, t_{ik},\beta_{ij}, \tan^{-2} \frac{R}{2}\Big) \Bigg) \nn\\
& \qquad \qquad - 2\bigg( \frac{1}{\tau_a^k} \bigg)_{\!\!+} f_1( t_{ik}, t_{jk},\beta_{ij})  \bigg]
\,.\end{align}
where $f_1$ and $f_2$ are given in \eqs{eq:f1-app}{eq:f2-app}, respectively.

\section{Convolutions and Finite Terms in the Resummed Distribution}
\label{app:convolve}

In evaluating the final resummed distribution \eq{NjetcsRG},  each measured jet function must be convolved against a corresponding soft function piece $S^\meas$. These convolutions take the form.
\be
\int\!d\tau_J d\tau_S d\tau'_J \, d\tau'_S \, J(\tau'_J, \mu_J) S^\meas(\tau'_S, \mu_S)\left[\frac{\Theta(\tau_J  - \tau'_J)}{(\tau_J-\tau'_J)^{1+\omega_J^i}} \right]_{\!+} \left[\frac{\Theta(\tau_S  - \tau'_S)}{(\tau_S-\tau'_S)^{1+\omega_S^i}} \right]_{\!+} \delta(\tau - \tau_J - \tau_S)
\,.\ee
For the class of functions of the form $x^{-1-\omega}$ with $\omega \neq 0$ and $\omega <1$, we define the plus distribution  by
\begin{equation}
\begin{split}
   \left[\frac{\Theta(x)}{x^{1+\omega}}\right]_{+} &\equiv \lim_{\beta\to 0} \left[\frac{\Theta(x-\beta)}{x^{1+\omega}}-\frac{\beta^{-\omega}}{\omega} \delta(x-\beta)\right]  = -\frac{\delta(x)}{\omega} + \sum_{n=0}^\infty (-\omega)^n \left[\frac{\Theta(x)\ln^n x}{x}\right]_+
   \label{eq:omegaplusdef}
\,,\end{split}
\end{equation}
where the plus functions on the second line are given by \eq{mitplus},
\begin{equation}
     \left[\frac{\Theta(x)\ln^n(x)}{x}\right]_+ \equiv \, \lim_{\beta\to 0} \left[\frac{\Theta(x-\beta) \ln^n(x)}{x}+\frac{\ln^{n+1}\beta}{n+1} \delta(x-\beta)\right]
     \label{eq:lognplusdef}
\,.\end{equation}
From these definitions, we can derive the identities (see, e.g., Appendix B of \cite{Hornig:2009vb})
\begin{equation}
\int d\tau'' \left[ \frac{\Theta(\tau-\tau'')}{(\tau-\tau'')^{1+\omega_1}}\right]_+  \left[ \frac{\Theta(\tau''-\tau')}{(\tau''-\tau')^{1+\omega_2}}\right]_+ = \frac{\Gamma(-\omega_1)\Gamma(-\omega_2)}{\Gamma(-\omega_1 - \omega_2)} \left[ \frac{\Theta(\tau-\tau')}{(\tau-\tau')^{1+\omega_1+\omega_2}}\right]_+\,,
\end{equation}
and
\begin{subequations}
\begin{align}
\int d\tau' \left[ \frac{\Theta(\tau-\tau')}{(\tau-\tau')^{1+\omega}}\right]_+\delta(\tau') &=  \left[ \frac{\Theta(\tau)}{\tau^{1+\omega}}\right]_+ \\
\int d\tau' \left[ \frac{\Theta(\tau-\tau')}{(\tau-\tau')^{1+\omega}}\right]_+\left[\frac{\Theta(\tau')}{\tau'}\right]_+ &=  \left[ \frac{\Theta(\tau)}{\tau^{1+\omega}}\right]_+\Bigl[ \ln\tau - H(-1-\omega)\Bigr] \\
\int d\tau' \left[ \frac{\Theta(\tau-\tau')}{(\tau-\tau')^{1+\omega}}\right]_+\left[\frac{\Theta(\tau')\ln\tau'}{\tau'}\right]_+ &=  \left[ \frac{\Theta(\tau)}{\tau^{1+\omega}}\right]_+\frac{ \left[\ln\tau - H(-1-\omega)\right]^2 + \frac{\pi^2}{6} - \psi^{(1)}(-\omega)}{2}
\end{align}
\end{subequations}

Using the above identities, we find that the final result can be written in the form \eqs{NjetcsRG}{measuredlogs} with the functions $d_J(\tau^i_a)$ given by
\begin{subequations}
\label{eq:d_J}
\begin{align}
\label{eq:d_JS}
&d_J^{q, \rm cone}(\tau^i_a) = \CF\Biggl[ \frac{7}{2} + 3\ln 2 - \frac{\pi^2}{6}\left(2 + \frac{1-\frac{a}{2}}{1-a}\right) \Biggr]  + C_F\frac{\Theta(\tau_a^{\text{max}} - \tau_a^i)}{1-\frac{a}{2}  } (\tau_a^i)^{1+\Omega_i} \nn \\
&\qquad\qquad\qquad \times\int  d\tau_J \left[\frac{\Theta(\tau_a^i - \tau_J)}{(\tau_a^i - \tau_J)^{1+\Omega_i}} \right]_{\!+}  \left[\frac{\Theta(\tau_J)}{\tau_J} \left(3\xcone + 2\ln\biggl(\frac{1-\xcone}{\xcone}\frac{\tau_J}{\tan^{2-a}\frac{R}{2}}\biggr)\right)\right]_+\nn \\
&\qquad\qquad\quad + \Theta(\tau_a^i - \tau_a^{\text{max}}) C_F\Biggl\{ \frac{3}{2}\ln\frac{\mu^2}{\omega^2\tan^2\frac{R}{2}} + \frac{1-\frac{a}{2}}{1-a}\ln^2\frac{\mu^2}{\omega^2} + \left(1-\frac{a}{2}\right) \ln^2\tan^2\frac{R}{2} \nn \\
&\qquad \qquad\qquad  - \frac{4}{1-a} \biggl[\bigl(\ln\tau_a^i - H(-1-\Omega_i)\bigr)\ln\frac{\mu\tan^{1-a}\frac{R}{2}}{\omega_i} - \frac{1}{2}\bigl(\ln\tau_a^i - H(-1-\Omega_i)\bigr)^2 \nn\\
&\qquad\qquad\qquad\qquad\qquad    - \frac{\pi^2}{12} + \frac{1}{2}\psi^{(1)}(-\Omega_i)\biggr] \nn \\
&\qquad\qquad\qquad + \frac{(\tau_a^i)^{1+\Omega_i}}{1-\frac{a}{2}}\int_0^{\tau_a^{\text{max}}}d\tau_J \left[\frac{1}{(\tau_a^i - \tau_J)^{1+\Omega_i}} \right]_{\!+} \biggl[\frac{1}{\tau_J} \biggl(r_q(\xcone)- \frac{3}{2}\biggr)\biggr]_+\Biggr\}
\end{align}
and
\begin{align}
&d_J^{q, \kt}(\tau^i_a) = \CF\Biggl[ \frac{13}{2}  - \frac{\pi^2}{6}\left(4 + \frac{1-\frac{a}{2}}{1-a}\right) \Biggr]  + C_F\frac{\Theta(\tau_a^{\text{max}} - \tau_a^i)}{1-\frac{a}{2}  }  (\tau_a^i)^{1+\Omega_i} \nn \\
&\qquad\qquad\qquad\times \int  d\tau_J \left[\frac{\Theta(\tau_a^i - \tau_J)}{(\tau_a^i - \tau_J)^{1+\Omega_i}} \right]_{\!+} \Biggl[\frac{\Theta(\tau_J)}{\tau_J} \Biggl(3x_1 + 2\ln\biggl(\frac{1-x_1}{x_1}\frac{\tau_J}{\tan^{2-a}\frac{R}{2}}\biggr) \nn \\
&\qquad\qquad\qquad\qquad\qquad - \Theta\left(\tau_a^{\frac{1}{2-a}} - 2\tan\frac{R}{2}\right)\left(r_q(x_2)-\frac{3}{2}\right)\Biggr)\Biggr]_+\Biggr \} \nn \\
&\qquad\qquad\quad + \Theta(\tau_a^i - \tau_a^{\text{max}}) C_F\Biggl\{ \frac{3}{2}\ln\frac{\mu^2}{\omega^2\tan^2\frac{R}{2}} + \frac{1-\frac{a}{2}}{1-a}\ln^2\frac{\mu^2}{\omega^2} + \left(1-\frac{a}{2}\right) \ln^2\tan^2\frac{R}{2} \nn \\
&\qquad \qquad\qquad  - \frac{4}{1-a} \biggl[\bigl(\ln\tau_a^i - H(-1-\Omega_i)\bigr)\ln\frac{\mu\tan^{1-a}\frac{R}{2}}{\omega_i} - \frac{1}{2}\bigl(\ln\tau_a^i - H(-1-\Omega_i)\bigr)^2 \nn\\
&\qquad\qquad\qquad\qquad\qquad   - \frac{\pi^2}{12} + \frac{1}{2}\psi^{(1)}(-\Omega_i)\biggr] \nn \\
&\qquad\qquad\qquad + \frac{(\tau_a^i)^{1+\Omega_i}}{1-\frac{a}{2}}\int_0^{\tau_a^{\text{max}}}d\tau_J \left[\frac{1}{(\tau_a^i - \tau_J)^{1+\Omega_i}} \right]_{\!+} \biggl[\frac{1}{\tau_J} \Biggl(r_q(x_1) - \frac{3}{2} \nn \\
&\qquad\qquad\qquad\qquad\qquad\qquad\qquad  - \Theta\left(\tau_a^{\frac{1}{2-a}} - 2\tan\frac{R}{2}\right)\left(r_q(x_2)-\frac{3}{2}\right)\Biggr)\Biggr]_+\Biggr \}
\end{align}
for quarks and by
\begin{align}
&d_J^{g, \rm cone}(\tau^i_a) =C_A\left[\frac{137}{36} + \frac{11}{3}\ln 2 - \frac{\pi^2}{6}\left(2 + \frac{1-\frac{a}{2}}{1-a}\right)\right] - T_R N_f\left(\frac{23}{18} + \frac{4}{3}\ln 2\right) \nn \\
&\qquad\qquad\quad  + \frac{\Theta(\tau_a^{\text{max}} - \tau_a^i)}{1-\frac{a}{2}  }  (\tau_a^i)^{1+\Omega_i}\int  d\tau_J \left[\frac{\Theta(\tau_a^i - \tau_J)}{(\tau_a^i - \tau_J)^{1+\Omega_i}} \right]_{\!+} \nn \\
&\qquad\qquad\qquad\qquad\qquad \times \left[\frac{\Theta(\tau_J)}{\tau_J} \left(r_g(\xcone) + 2C_A\ln\biggl(\frac{\tau_J}{\tan^{2-a}\frac{R}{2}}\biggr)\right)\right]_+ \nn \\
&\qquad\qquad  + \Theta(\tau_a^i - \tau_a^{\text{max}})\Biggl\{ \frac{\beta_0}{2}\ln\frac{\mu^2}{\omega^2\tan^2\frac{R}{2}} + C_A\frac{1-\frac{a}{2}}{1-a}\ln^2\frac{\mu^2}{\omega^2} + C_A\left(1-\frac{a}{2}\right) \ln^2\tan^2\frac{R}{2} \nn \\
&\qquad \qquad\qquad  - \frac{4C_A}{1-a} \biggl[\bigl(\ln\tau_a^i - H(-1-\Omega_i)\bigr)\ln\frac{\mu\tan^{1-a}\frac{R}{2}}{\omega_i} - \frac{1}{2}\bigl(\ln\tau_a^i - H(-1-\Omega_i)\bigr)^2 \nn\\
&\qquad\qquad\qquad\qquad \qquad  - \frac{\pi^2}{12} + \frac{1}{2}\psi^{(1)}(-\Omega_i)\biggr] \nn \\
&\qquad\qquad\quad  + \frac{(\tau_a^i)^{1+\Omega_i}}{1-\frac{a}{2}}\int_0^{\tau_a^{\text{max}}}d\tau_J \left[\frac{1}{(\tau_a^i - \tau_J)^{1+\Omega_i}} \right]_{\!+} \biggl[\frac{\Theta(\tau_J)}{\tau_J} \biggl(r_g(\xcone) - \frac{\beta_0}{2}\biggr)\biggr]_+\Biggr\}
\end{align}
and
\begin{align}
&d_J^{g, \kt}(\tau^i_a) =    C_A\left[\frac{67}{9}  - \frac{\pi^2}{6}\left(4 + \frac{1-\frac{a}{2}}{1-a}\right)\right] - T_R N_f\left(\frac{23}{9} \right) \nn \\
&\qquad\qquad\qquad   + \frac{\Theta(\tau_a^{\text{max}} - \tau_a^i)}{1-\frac{a}{2}  }  (\tau_a^i)^{1+\Omega_i}\int  d\tau_J \left[\frac{\Theta(\tau_a^i - \tau_J)}{(\tau_a^i - \tau_J)^{1+\Omega_i}} \right]_{\!+} \nn \\
&\qquad\qquad\qquad\times \Biggl[\frac{\Theta(\tau_J)}{\tau_J} \Biggl(r_g(x_1) + 2C_A\ln\biggl(\frac{\tau_J}{\tan^{2-a}\frac{R}{2}}\biggr)  \nn \\
&\qquad\qquad \qquad\qquad - \Theta\left(\tau_a^{\frac{1}{2-a}} - 2\tan\frac{R}{2}\right)\left(r_g(x_2)-\frac{\beta_0}{2}\right)\Biggr) \Biggr]_+\Biggr \} \nn \\
&\qquad\qquad + \Theta(\tau_a^i - \tau_a^{\text{max}})\Biggl\{ \frac{\beta_0}{2}\ln\frac{\mu^2}{\omega^2\tan^2\frac{R}{2}} + C_A\frac{1-\frac{a}{2}}{1-a}\ln^2\frac{\mu^2}{\omega^2} + C_A\left(1-\frac{a}{2}\right) \ln^2\tan^2\frac{R}{2} \nn \\
&\qquad \qquad\quad - \frac{4C_A}{1-a} \biggl[\bigl(\ln\tau_a^i - H(-1-\Omega_i)\bigr)\ln\frac{\mu\tan^{1-a}\frac{R}{2}}{\omega_i} - \frac{1}{2}\bigl(\ln\tau_a^i - H(-1-\Omega_i)\bigr)^2 \nn\\
&\qquad\qquad\qquad\qquad\qquad  - \frac{\pi^2}{12} + \frac{1}{2}\psi^{(1)}(-\Omega_i)\biggr] \nn \\
&\qquad\qquad + \frac{(\tau_a^i)^{1+\Omega_i}}{1-\frac{a}{2}}\int_0^{\tau_a^{\text{max}}}d\tau_J \left[\frac{1}{(\tau_a^i - \tau_J)^{1+\Omega_i}} \right]_{\!+} \biggl[\frac{1}{\tau_J} \biggl(r_g(x_1) - \frac{\beta_0}{2} \nn\\
&\qquad\qquad\qquad\qquad\qquad\qquad\qquad  - \Theta\left(\tau_a^{\frac{1}{2-a}} - 2\tan\frac{R}{2}\right)\left(r_g(x_2)-\frac{\beta_0}{2}\right)\Biggr]_+\Biggr \}
\end{align}for gluons, where in all cases $\xcone$, $x_{1,2}$ (defined in \eqs{eq:xcone-defn}{kTtauconstraint}) are evaluated at $\tau = \tau'_J$ and $r_{q,g}$ are defined in \eqs{eq:rq}{eq:rg}.
\end{subequations}

\section{Color Algebra for $n=2,3$ Jets}
\label{app:color}

For the two and three jet cases, there are no color correlations since all color generator inner products $\vect{T}_i \cdot \vect{T}_j$ can be expressed in terms of the Casimir invariants $\CA$ and $\CF$. For $n=2$, there is a quark jet with charge $\vect{T}_q$ and an anti-quark jet with charge $\vect{T}_{\bar q}$ that each square to $\CF$. There is only one inner-product in this case and using color conservation ($\sum_i \vect{T}_i = 0$), we have that
\begin{align}
\vect{T}_q \cdot \vect{T}_{\bar q} = - \vect{T}_q^2 = - \vect{T}_{\bar q}^2 = -\CF
\,.\end{align}

For $n=3$ jets color conservation gives that, for example,
\begin{align}
\vect{T}_1 \cdot \vect{T}_2 &= \frac{1}{2} \left[ ( \vect{T}_1 +  \vect{T}_2)^2 - \vect{T}_1^2-  \vect{T}_2^2 \right] \nn\\
&= \frac{1}{2}  \left[  \vect{T}_3^2 - \vect{T}_1^2-  \vect{T}_2^2 \right]
\label{eq:3jet-colors-gen}
\,.\end{align}
Referring to the quark, anti-quark, and gluon generators as $\vect{T}_q$, $\vect{T}_{\bar q}$, and $\vect{T}_g$, respectively, using $\vect{T}_q^2 = \vect{T}_{\bar q}^2 = \CF$ and $\vect{T}_g^2 = \CA$ in \eq{eq:3jet-colors-gen} gives
\begin{align}
\vect{T}_q \cdot \vect{T}_{\bar q} &= \frac{\CA}{2} -\CF \nn\\
\vect{T}_q \cdot \vect{T}_g &=  \vect{T}_{\bar q} \cdot \vect{T}_g = - \frac{\CA}{2}
\label{eq:3jet-colors}
\,.\end{align}

\bibliography{shapes}

\end{document}